\documentclass[useAMS,usenatbib]{mn2e}
\pdfoutput=1
\usepackage{lscape}
\usepackage{rotating,booktabs}
\usepackage{arydshln}
\usepackage{amssymb,amsmath}
\graphicspath{{ps_apendix/}} 
\usepackage[section]{placeins}
\usepackage{stfloats}
\usepackage{float}
\usepackage{lipsum}
\usepackage{graphicx}

\newcommand{\zlbg}{$z\sim5.7$}

\newcommand{\zlae}{$z\sim5.7$}

\newcommand{\HI}{\hbox{H\,{\sc i}}}

\newcommand{\Lya}{\mbox{Ly$\alpha$}}

\newcommand{\Hb}{\mbox{H$\beta$}}
\newcommand{\Ha}{\mbox{H$\alpha$}}
\newcommand{\Hg}{\mbox{H$\gamma$}}
\newcommand{\Hd}{\mbox{H$\delta$}}
\newcommand{\msun}{\mbox{$M_{\sun}$}}

\newcommand{\kms}{\mbox{km s$^{{-}1}$}}
\newcommand{\cm}{cm$^{-2}$}
\newcommand{\civ}{\hbox{C\,{\sc iv}}}
\newcommand{\siiv}{\hbox{Si\,{\sc iv}}}
\newcommand{\siivoiv}{\hbox{Si\,{\sc iv}+O\,{\sc vi]}}}

\newcommand{\oi}{\hbox{O\,{\sc i}}}
\newcommand{\oii}{\hbox{O\,{\sc ii}}}
\newcommand{\oiii}{\hbox{O\,{\sc iii}}}

\newcommand{\siii}{\hbox{Si\,{\sc ii}}}
\newcommand{\cii}{\hbox{C\,{\sc ii}}}
\newcommand{\Neiii}{\hbox{Ne\,{\sc iii}}}
\newcommand{\Nv}{\hbox{N\,{\sc v}}}
\newcommand{\Niv}{\hbox{N\,{\sc iv}}}

\newcommand{\nciv}{\mbox{$N_{\text \civ}$}}
\newcommand{\vmax}{$\Delta v_{{\rm MAX}}$}

\def\ltsima{$ \buildrel < \over \sim $}
\def\simlt{\lower.5ex\hbox{\ltsima}}
\def\gtsima{$ \buildrel > \over \sim $}
\def\simgt{\lower.5ex\hbox{\gtsima}}

\title[Environment of $z{\sim}5.7$ \civ\ absorption systems II]{Large-scale environment of $z\sim5.7$ \civ\ absorption systems --II. Spectroscopy of Lyman-$\alpha$ emitters\thanks{
Based on observations obtained at the W.M. Keck Observatory,
which is operated as a scientific partnership among the California Institute of
Technology, the University of California, and NASA,
and was made possible by the generous
financial support of the W.M. Keck Foundation.}}
\author[D\'{i}az et. al.]{
C. Gonzalo D\'{i}az$^{1}$\thanks{E-mail:gdiaz@swin.edu.au}, 
Emma V. Ryan-Weber$^{1}$, Jeff Cooke$^{1}$,Yusei Koyama$^{2,3}$, 
\newauthor \, and Masami Ouchi$^{4,5}$\\
\\
$^{1}$Centre for Astrophysics and Supercomputing, Swinburne University of Technology, Hawthorn, VIC 3122, Australia\\
$^{2}$National Astronomical Observatory of Japan, Mitaka, Tokyo 181-8588, Japan\\
$^{3}$Institute of Space Astronomical Science, Japan Aerospace Exploration Agency, Sagamihara, Kanagawa 252-5210, Japan\\
$^{4}$Institute for Cosmic Ray Research, The University of Tokyo, Kashiwa, Chiba 277-8582, Japan\\
$^{5}$Kavli Institute for the Physics and Mathematics of the Universe (WPI), The University of Tokyo, Kashiwa, Chiba 277-8583, Japan}

\begin{document}
\date{Accepted. Received}
\maketitle
\label{firstpage}
\begin{abstract}

The flow of baryons to and from a galaxy, 
which is fundamental for galaxy formation and evolution, 
can be studied with galaxy-metal absorption system pairs. 
Our search for galaxies around \civ\ absorption systems at 
\zlae\ showed an excess of photometric Lyman-$\alpha$ 
emitter (LAE) candidates in the fields J1030+0524 and J1137+3549. 
Here we present spectroscopic follow-up of 33 LAEs in both fields. 
In the first field, three out of the five LAEs within 10$h^{{-}1}$ 
projected comoving Mpc from the \civ\ system are within $\pm500$\,\kms\
from the absorption at $z_{\text{\civ}}=5.7242\pm0.0001$. 
The closest candidate (LAE 103027+052419) is robustly confirmed at 
$212.8^{+14}_{-0.4}h^{-1}$ physical kpc 
from the \civ\ system. In the second field, the LAE sample is 
selected at a lower redshift ($\Delta z\sim0.04$) than the \civ\ absorption 
system as a result of the filter transmission and, thus, do not trace its environment. 

The observed properties of LAE 103027+052419 indicate
that it is near the high-mass end of the LAE distribution, 
probably having a large \HI\ column density and 
large-scale outflows. Therefore, our results suggest that 
the \civ\ system is likely produced by a star-forming galaxy which has 
been injecting metals into the intergalactic medium since $z>6$. 
Thus, the \civ\ system is either produced by LAE 103027+052419, 
implying that outflows can enrich larger volumes at $z>6$ than 
at $z\sim3.5$, or an undetected dwarf galaxy. In either case, 
\civ\ systems like this one trace the ionized intergalactic medium 
at the end of cosmic hydrogen reionization and may trace the sources 
of the ionizing flux density.
\end{abstract}


\section{Introduction}

The detection of metal absorption systems
depends on the ionization balance of the absorbing gas,
thus we can learn about the galaxies that reionized 
the Universe through the study of the environment of
metal absorption systems at the tail end of the epoch 
of reionization (EoR).

Statistical studies of \civ\ absorption systems 
across cosmic time \citep[e.g.][]{ryan-weber2009,
becker2009,simcoe2011b,dodorico2013}
suggest that some physical properties of the absorbing 
`clouds', for example the size and the number density,
are changing towards higher redshift.
\citet{dodorico2013} compared the column densities of 
\civ , \siiv\ and \cii\ in metal absorption systems across 
cosmic time and concluded that, at \zlbg,
\civ\ metal absorption systems trace
less dense gas (over-densities $\delta \sim10$)
than at $z\sim3$ ($\delta\sim100$).
In addition, \citet[][hereafter Paper I]{diaz2014} report that 
two independent lines of sight towards \zlae\ \civ\ absorption systems 
are distant from the main projected over-densities of rest-frame 
UV bright Lyman break galaxy (LBG) candidates, opposite to 
the results at $z\sim2$--3 \citep[e.g.][]{adelberger2005b,steidel2010}.
Thus, it is possible that the absorbing gas at these two epochs
is found in different environments on large-scales,
which opens the possibility that the detection of \civ\ systems 
depends on properties that are linked to larger scales
that go beyond the local density of the absorbing gas,
such as the ionizing flux density background.

Interestingly, at scales of 10$h^{-1}$ comoving Mpc,
a projected over-density of Lyman-$\alpha$ emitter 
(LAE) candidates was found in the field J1030+0524 
towards the \civ\ absorption system. Similarly in the field 
J1137+3549, the surface density of LAE candidates 
within 10$h^{-1}$ comoving Mpc from the \civ\ system
is higher than the average over the observed field of view of 
$\sim 80 {\times}60 {\it h}^{-1}$ comoving Mpc. If these 
over-densities of LAEs in the environment of \civ\ systems
are confirmed (the aim of this paper), 
the association of highly ionized gas with 
a large scale excess of faint UV star-forming galaxies
would be in agreement with the current literature
where faint galaxies are found to dominate the ionizing photon budget
at $z\sim6$ \citep[e.g.][]{cassata2011, dressler2011,
finkelstein2012b, ferrara2013, cai2014, fontanot2014}.
Moreover, it would imply that high ionization metal absorption systems 
at the end of the EoR can trace the highly ionized intergalactic medium (IGM).

At smaller scales, galaxy-metal absorption system pairs 
provide useful information on the distribution of metals. 
It has been observed that \civ\ absorption systems 
with column densities \nciv$ > 10^{14}$ at $z=2$--3 are related 
to the circumgalactic medium (CGM) of LBGs 
\citep{adelberger2005b, steidel2010}. Moreover, 
in these galaxies the red-shifted \Lya\ emission 
line and the blue-shifted interstellar absorption with 
respect to the systemic redshift determined by nebular 
emission lines (e.g. \Ha , \Hb\ and \oiii ), is evidence 
for enriched gas moving at high velocities (hundreds of \kms).
These galactic outflows are commonly observed in
star-forming galaxies across cosmic time 
\citep{rupke2005, weiner2009, steidel2010, coil2011, jones2012, 
bouche2012b, martin2012, bradshaw2013, karman2014} and represent an
important source of chemical feedback 
widely explored in theoretical studies on the redistribution of metals
from star-forming regions to the CGM and the IGM
\citep{madau2001, oppenheimer2006, oppenheimer2008, cen2011, 
murray2011, tescari2011, brook2012, hopkins2012, shen2012, 
pallottini2014}.

Although it is well accepted that star-forming galaxies can
produce outflows that will enrich their CGM and potentially 
the IGM, the abundance of metal absorption systems and 
the sizes of the enriched regions suggest that a significant
fraction of the metals observed in the CGM and IGM 
at redshift $z=1.7$--4.5 have been produced by satellites and progenitors 
of the observed galaxies 
during an early stage of galaxy formation at redshifts above 6
 \citep[e.g.][]{porciani2005, martin2010}.
This is commonly known as pre-galactic enrichment.
For example, \citet{shen2012} use a detailed cosmological 
hydrodynamic simulation to study the sources of metals in 
the CGM of a $z=3$ LBG, and find that metals observed 
at low radii ($<3R_{vir}$) are mainly produced in the host 
galaxy while metals at larger radii ($\,>3R_{vir}$)
are mainly produced in the satellite companions.
Moreover, the metals observed at $\,>2R_{vir}$ 
were released at redshift $z>5$.
Therefore, searching for star-forming galaxies close to 
the highest redshift \civ\ absorption systems known to date 
is the next step in testing theories on the enrichment of the IGM.

This work presents the spectroscopic observations
obtained with \textsc{deimos} on Keck-II for the LAE 
photometric candidates in Paper I. 
The follow-up spectroscopy demonstrates
that the \zlae\ LAE selection criteria is robust, 
in particular for bright sources in the narrow-band filter NB\civ.
We report tentative evidence that the projected over-density
of LAEs in the field J1030+0524 corresponds to an excess
of sources within ${\pm}500$\,\kms\ from the
second \civ\ absorption system at $z\,\simgt\,5.5$ in the line of sight
to the QSO ($z_{\text{\civ,b}}= 5.7242\pm0.0001$).
However, better data is needed
for at least two of the sources, to confirm the nature of 
emission line currently detected with very low signal-to-noise.
In the field J1137+3549, the sample of LAEs is not at the redshift
of the \civ\ system and therefore does not trace the environment 
of this absorption system.

Our spectroscopic campaign confirmed the redshift of 
LAE 103027+052419 whose angular separation on the 
sky implies a distance of $212.8^{+14}_{-0.4}h^{-1}$ 
physical kpc from the absorption system \civ$_b$, mentioned above.
This is evidence that the strongest \civ\ system known at $z\,\simgt\,5.5$ 
is associated with a detectable star-forming galaxy.
Considering the evidence of the high incidence of galactic
outflows at high redshift \citep[e.g.][]{vanzella2009, steidel2010, jones2012, shibuya2014b},
it is very likely that LAE 103027+052419 hosts some
kind of outflow. Therefore, the question of interest is whether 
or not the carbon produced by an earlier generation of stars in 
the LAE could have reached the distances at which the
\civ\ absorption system is observed.

The analysis shows that the impact parameter of 212.8$h^{-1}$ 
physical kpc is difficult to reconcile with a typical galactic wind 
scenario due to the short time since the Big Bang at which 
the LAE-\civ\ system pair is observed. As a result, even if the 
metals were distributed by an outflow from the LAE, the mechanisms 
should have already been in place at a time that is consistent with 
the pre-galactic enrichment scenario.
Moreover, it is possible that the metals in the absorbing gas
were born in undetected dwarf galaxies that polluted
the IGM, while the LAE only provides the ionizing radiation 
to maintain the triply ionized carbon.
Therefore, the simplest explanation for our results is that 
LAE 103027+052419 is associated with the \civ\ system 
as the source of ionizing radiation instead of the source of enrichment.
The main implication is that \civ\ absorption systems at \zlae\
trace highly ionized IGM. 
This is in agreement with predictions from cosmological simulations 
\citep[e.g.][]{oppenheimer2009}
and suggests that \civ\ systems in the post-reionization Universe
are important sources of information about the ionization state of the IGM.

This paper is organised as follows: Section \ref{s:observation-specLAE} describes
the observations, Section \ref{s:spectroscopic-LAEs} presents the spectroscopic
catalogue of LAEs and reviews the contamination in the sample, 
Section \ref{s:colmag-spec} reviews the colours and magnitudes of \zlbg\ LAEs,
Section \ref{s:redshift-dist} presents the redshift distribution of
the sample and Section \ref{s:discussion-large-scale} discusses the distribution
of LAEs in the environment of \civ\ absorption systems, both
projected and in the line of sight.
After that, a description of the LAE-\civ\ system pair in the J1030+0524 field
is presented in Section \ref{s:lae-civ-pair}. Then, the rest-frame equivalent 
width of the \Lya\ line (hereafter EW$_0$) and the velocity shift of the \Lya\ 
maximum (hereafter $\Delta v_{\text{MAX}}$) are presented in 
Sections \ref{s:eqw} and \ref{s:dv}, respectively.
Finally, the origin of the \civ\ system in the field J1030+0524 is discussed in 
Section \ref{s:discussion-origin-CIV} and the conclusions are summarised
in Section \ref{s:conclusion-spec}.
Throughout this work we use AB magnitudes and assume a 
flat universe with $H_{0}=70$\,\kms\ Mpc$^{-1}$,
 $\Omega_{\text{m}}=0.3$ and $\Omega_{\lambda}=0.7$.  
 
\section{Observations and data reduction: DEIMOS, Keck Telescope}\label{s:observation-specLAE}

This work is based on the spectroscopic observations of
\zlae\ LAE photometric candidates from Paper I.
The fields of view are centred on QSOs SDSS J103027.01+052455.0 
($z_{\text{em}} = 6.309$, $\text{RA}=10^{\text{h}}30^{\text{m}}27^{\text{s}}.01$, 
$\text{Dec.}=05^{\circ}24'55''.0$) and SDSS J113717.73+354956.9 
($z_{\text{em}}=6.01$, $\text{RA}=11^{\text{h}}37^{\text{m}}17^{\text{s}}.73$, 
$\text{Dec.}=35^{\circ}49'56''.9$) 
\citep{fan2006a}, hereafter J1030+0524 and J1137+3549.
This section describes two data sets obtained with the \textsc{DEIMOS}
spectrograph on the Keck-II telescope:
first, the main data set collected the nights of the 27th and 28th of February 
2012\footnote{Program ID: W136D}, and second, additional data obtained 
at the end of the night of the 26th of February 2014\footnote{Program ID: N121D}.

For the observing run of 2012, the seeing was in the range 
0.55--0.8 arcsec FWHM during the first night and 0.55--1.0 
arcsec FWHM during the second night. 
In order to acquire spectra of high redshift galaxies and to enable 
the identification of many key features of low redshift contaminants,
we used the OG550 order-blocking filter and the 830 line 
mm$^{-1}$ grating (830G), which provides wavelength 
coverage from $\sim6500$ to $\sim10000$\,\AA\
with a central wavelength $\lambda_c=8579$\,\AA.
For a 1 arcsec slit width, the spectral resolution is 
$\text{FWHM}=2.5$\,\AA\ (or 91.4\,\kms\,at 8200\,\AA ),
which is sampled with a scale of 0.47\,\AA\ per pixel.
The 830G grating provides sufficient resolution to identify 
the \oii\ emission doublet of galaxies at $z\sim1.19$ 
($\lambda_{\text{obs}}\sim8166$\,\AA), which are the main 
contaminants in $z\sim5.7$ LAE samples.

In each of the two fields, three multi-slit masks were used,
carefully designed with position angles that minimise the flux 
lost due to atmospheric dispersion.
Single exposure times of 1200 seconds were adopted for all 
the exposures but the number of exposures per mask varied 
due to priority and weather conditions.
In particular, for the field J1030+0524 the number of exposures 
(and total exposure time) per mask was 14 (16800 s), 7 (8400 s) 
and 3 (3600 s). For the field J1137+3549, we acquired 12 (14400 s), 
9 (10800 s) and 3 (3600 s) exposures per mask. 
Therefore, the exposure time is not uniform across the LAE sample. 

One additional mask per field was observed in February 2014 with 
the same instrument but lower spectral resolution and lower 
signal-to-noise ratio (S/N). We used the 600 line mm$^{{-}1}$ 
grating (600ZD) and the GG495 order-blocking filter covering 
the wavelength range $\sim5000$ to $\sim10500$\,\AA\
with a scale of 0.65\,\AA\ per pixel. The width of the slits 
was 1 arcsec resulting in a $\text{FWHM}=3.5$\,\AA\,(128.0\,\kms\,at 8200\,\AA ).
For the field J1030+0524, only two exposures of 800 seconds 
each were obtained, which provided the detection of an emission 
line in three LAE candidates, although with low significance.
Unfortunately, the S/N is not sufficient to distinguish lower 
redshift contamination because companion emission lines 
would be undetected and the symmetry of the line cannot 
be determined with sufficient accuracy 
(see Section \ref{s:spectroscopic-LAEs-contamination}).
Thus, the data from February 2014 for the field J1030+0524
is presented as `tentative identification' and are included in 
the analysis with a note of caution that the spectroscopic 
detections cannot robustly confirm nor rule-out the high-redshift 
nature of the sources. 
In the field J1137+3549, three exposures of 1500 seconds 
(4500 s) were obtained and provided sufficient S/N to measure 
the redshift of LAEs and to identify contaminants.
Thus, the additional data on this field are included in our analysis 
together with the main data set.

Most of the reduction process was performed in \textsc{idl} using 
the \textsc{deimos} pipeline software in the \textsc{spec2d} package 
\citep{cooper2012, newman2013}.
The pipeline processes the flats and arcs, which are rectified 
into rectangular arrays, and calculates the wavelength solution 
of the 2D data array. 
Then, it applies corrections for flatfield and fringing to the science 
data of each slit, which are also rectified into rectangular arrays.
A b-spline model of the sky is obtained for each of the individual 
science exposures, and subtracted from them.
The residuals are then combined into an inverse variance 
weighted average 2D spectrum of each slit and
cosmic-rays rejection is applied on individual pixels. 
 
\begin{figure}
\includegraphics[width=85mm]{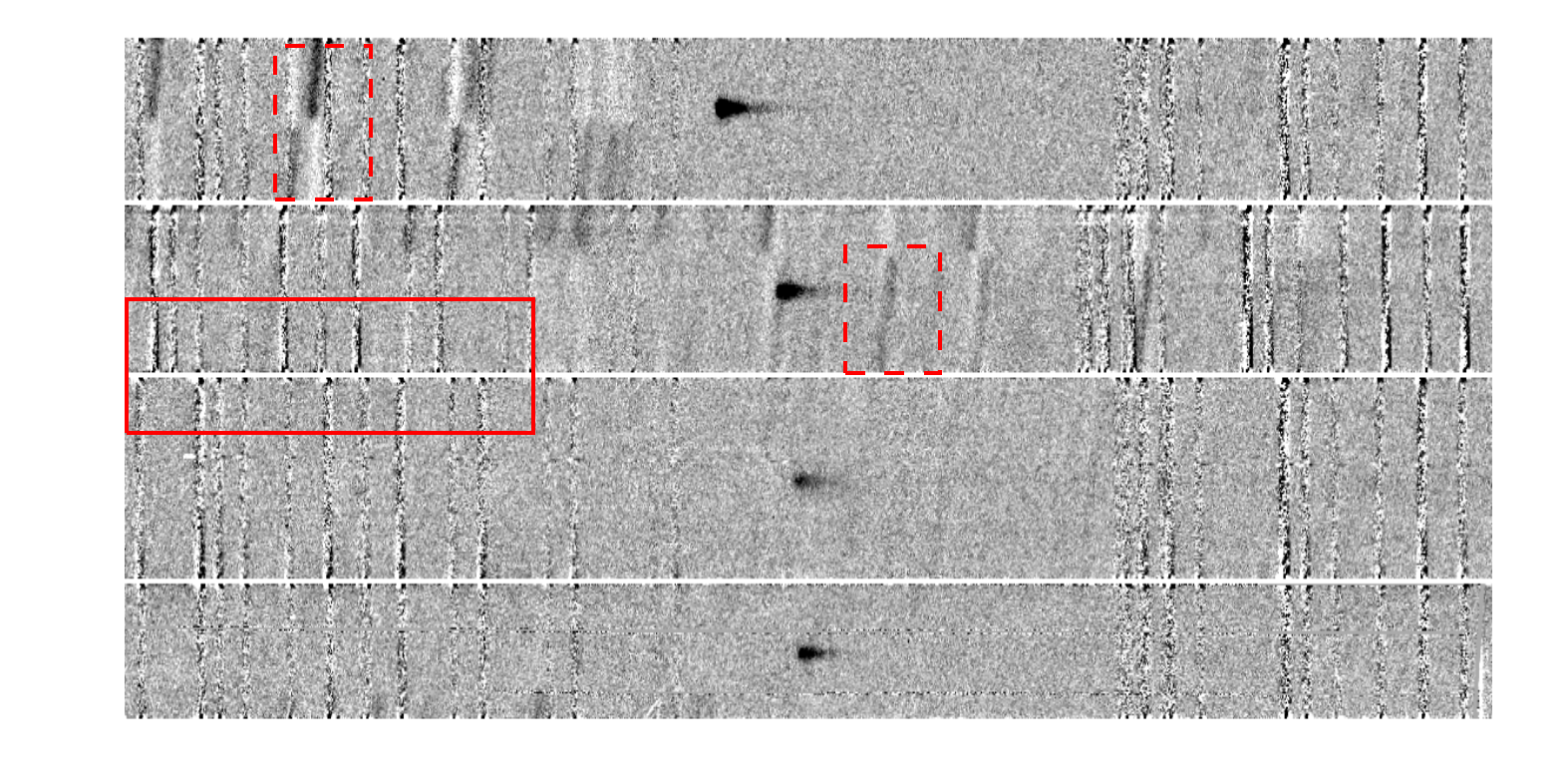} 
\caption{\small Example of a control image from \textsc{spec2d} (`Allslits0.xxxx.fits')
showing four consecutive slits. 
The image is created to check the reduction process
before the 1D spectrum extraction.
Two major problems can be identified. 
First, red dashed line rectangles show 
sky emission probably from other slits.
The reflected light is not removed by the model 
of the background and produces undesired residuals 
in all the slits. 
Second, the red solid line rectangle shows that
the skylines residuals of consecutive
slits are not aligned, which indicates
that the automatic wavelength calibration does
a poor job with our data.
These effects can be avoided with an additional 5$^{\circ}$ tilt 
to the position angle of the slits.}
\label{f:allslit}
\end{figure}

The \textsc{spec2d} package does a good job, in most cases, 
in the spatial direction along the slit which allows for a correct 
curvature rectification. However, in the dispersion direction, 
the 2D output spectra shows a wavelength shift in different 
directions respect with each other. 
This effect is obvious in Figure \ref{f:allslit} that shows a section 
of an image created by the \textsc{spec2d} pipeline for quality inspection 
(`Allslits0.[{\footnotesize name\_of\_the\_mask}].fits'), just before the 
1D spectra are extracted. The  solid line rectangle highlights the sky 
residuals of two consecutive slits which are not aligned. 
In order to solve this problem, a new wavelength dispersion 
solution for each 1D object spectrum was obtained, using the 
spectrum of the sky as comparison spectrum.

Standard \textsc{iraf} tasks (e.g. \textsc{identify}, 
\textsc{refspectra} and \textsc{dispcor}) and the high-resolution 
spectral atlas of night-sky emission lines from \citet{osterbrock1996} 
as reference were used to calculate the correct wavelength 
dispersion solution and to apply it to the objects spectra.
This procedure was applied to the main data set only since
the wavelength dispersion of the additional lower resolution data 
is properly calculated by \textsc{spec2d}.
First, the 1D spectrum of each object was boxcar extracted from 
the 2D spectrum output of \textsc{spec2d} by direct sum of 
the counts in pixels with constant wavelength.
The variance was obtained as the direct sum of the variances 
(i.e. errors added in quadrature). The extraction box position 
was defined by visual inspection to include all the pixels that 
showed flux in the 2D spectrum at the position of the object. 
Secondly, the spectrum of the sky was extracted from the same 
slit and position as the object before the b-spline model of 
the background sky was removed from the data spectrum. 
Thirdly, we obtained a wavelength dispersion solution for the 
sky emission lines measured in the same CCD columns as 
the object (pixel columns in \textsc{deimos} data correspond 
to the dispersion direction). 
The typical RMS of the fit is $\sim 0.03$--0.05\,\AA.
Finally, the same wavelength solution was assigned to the object 
spectrum. 
The output of this process is the reduced 1D spectra 
of the 33 line emitters presented in the following section.

\section{Spectroscopic catalogue of \zlae\ LAEs} \label{s:spectroscopic-LAEs}

This section presents the spectroscopic sample and summarises 
the possible sources of contamination. 
First, we present the data. Second, we comment on the contamination 
level and the approach adopted to remove contaminant sources.
Third, we search for evidence of Active Galactic Nuclei (AGN) among 
the objects in the sample.

The complete sample contains 33 spectra of which 28 are photometric 
LAE candidates and five are extra detections that were not photometrically 
selected due to insufficient signal-to-noise in the NB\civ\ band 
(S/N$_{\text {NBC\,{\sc iv}}}{\simlt} 5$). Spectroscopic quantities 
and other estimates based on the spectroscopic redshift of each 
LAE are presented in Tables \ref{t:spec-LAE-J1030}, \ref{t:phot-LAE-J1030}, 
\ref{t:spec-LAE-J1137} and \ref{t:phot-LAE-J1137} in Appendix \ref{app:spec-tables}.
Contamination was determined based on the asymmetry of 
the emission lines and visual inspection of the spectra.
The contamination level in the photometric sample is $\sim10$--20 per cent, 
based on 25 LAE candidates with reliable spectroscopy (i.e. excluding 
the three low resolution spectra in J1030+0524).
Finally, no evidence of AGN (e.g. emission from 
\Nv(1241\,\AA), \siivoiv(1400\,\AA) and \Niv(1483,1487\,\AA)) 
was found in the sample. However, the search is limited by the lack 
of information on other wavelengths (e.g.: X-rays, sub-mm, etc.) 
and the high contamination of skylines in the observed wavelength range.

\subsection{The sample}\label{s:spectroscopic-LAEs-sample} 

\begin{figure}
\centering 
\includegraphics[width=80mm]{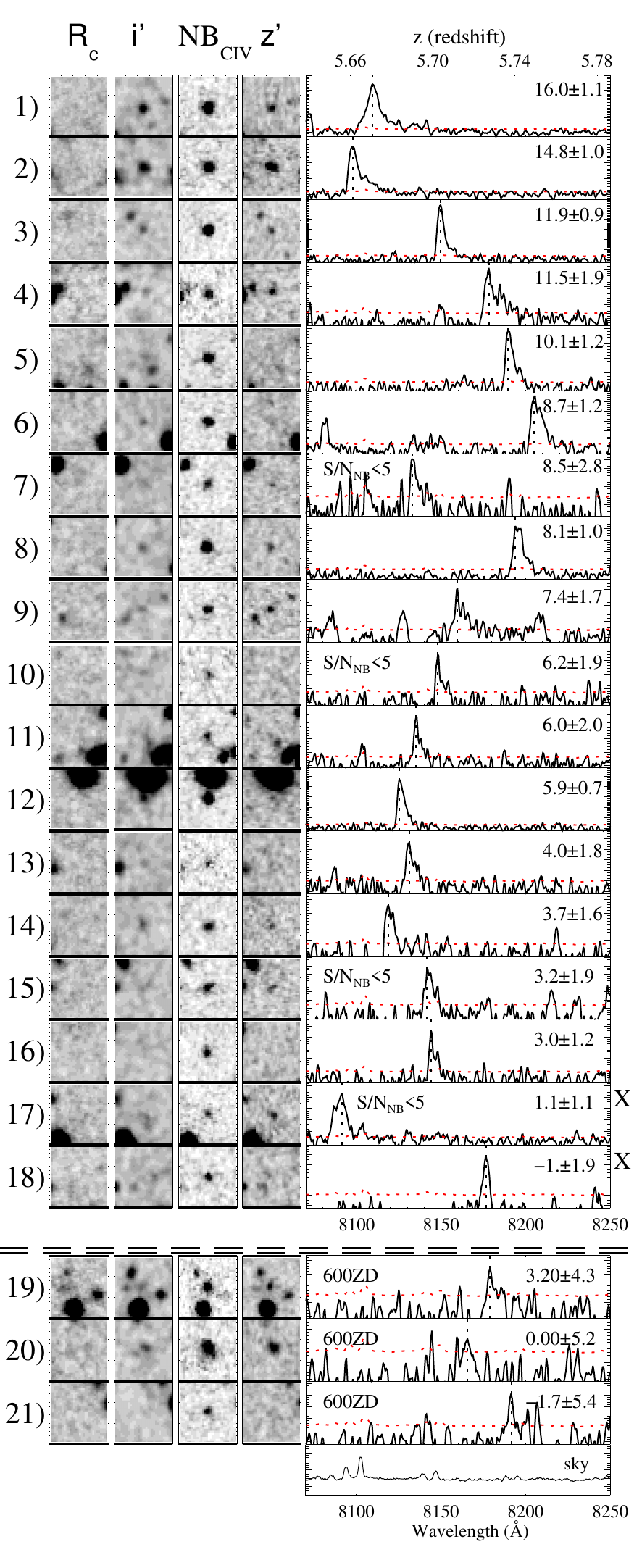} 
\caption{\small Snapshots and spectra of the \zlae\ LAEs in the field 
J1030+0524. Images on the left correspond to R$_c$, $i'$, NB\civ\ and 
$z'$ bands, the snapshot boxes are 6$\times$6 arcsec.
In the spectra, the vertical scale is arbitrary, the peak of 
the emission lines are indicated with vertical dotted lines,
and the 1$\sigma$ error spectra are plotted with red dotted lines.
The objects are arranged by decreasing weighted skewness 
S$_{w,10\%}$ (defined in Section \ref{app:spectroscopic-LAEs-skewness}) 
which is given in the top-right corner of each spectrum. 
The `X' indicates objects with S$_{w,10\%}< 3.0$ and
the bottom panel shows the sky spectrum.
The 600ZD spectra are below the double-dashed line
and do not have the sensitivity required for our analysis.
Nevertheless, emission lines are detected in the three cases,
hence we present the data for completeness.}
\label{f:phot-spec-LAE-1030}
\end{figure}

In the J1030+0524 field, spectra of 17 photometric LAE candidates
have been obtained. Three of the LAEs were observed using the 
600ZD grating and short exposure times, 
thus the asymmetry of their emission line 
cannot be measured with sufficient precision. 
We opted not to include these three objects 
in our discussion but, for completeness, 
we do measure the position of the emission line 
and comment on them when we consider it relevant.
This leaves us with 14 photometric candidates 
with reliable spectroscopy. 
In addition, four objects with  S/N$_{\text{NBC\,{\sc iv}}}{\simlt} 5$ 
were spectroscopically confirmed to have 
an emission line and were included in the analysis. 
Thus, the total sample contains 21 spectra 
of which 18 ($14{+}4$) have sufficient S/N 
for a fair assessment of the asymmetry of the line, 
while the three low resolution spectra 
are only used to measure the position of the line. 
The sample is presented in Figure \ref{f:phot-spec-LAE-1030}, 
which shows the spectra and photometry snapshots 
on the four bands R$_c$, $i'$, NB\civ\ and $z'$, used 
in the photometric selection of Paper I.

\begin{figure}
 \centering 
\includegraphics[width=80mm]{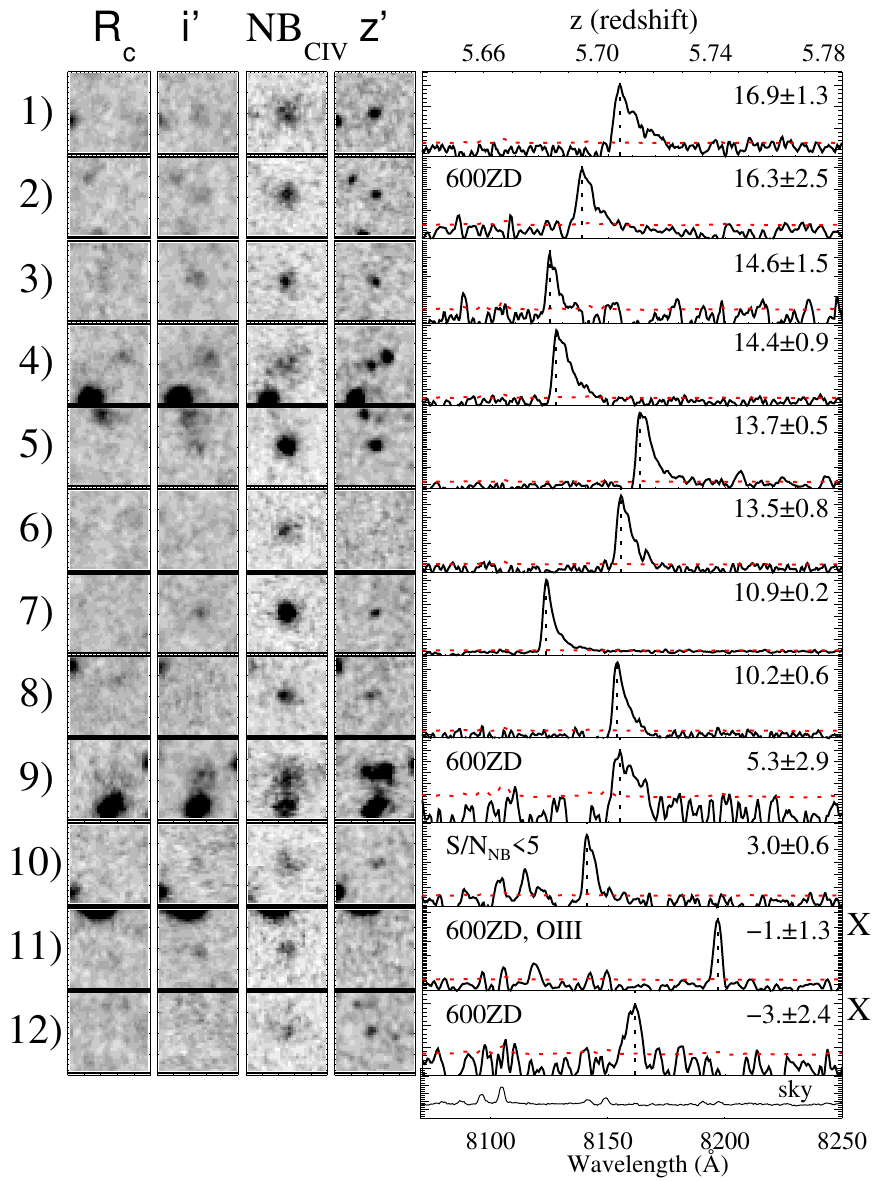} 
\caption{\small Snapshots and spectra of the \zlae\ LAEs in the field J1137+3549.
Similar to Figure \ref{f:phot-spec-LAE-1030}, images in the left hand
correspond to 6$\times$6 arcsec snapshot boxes of R$_c$, $i'$, NB\civ\ 
and $z'$ bands, and the objects are arranged by decreasing weighted 
skewness S$_{w,10\%}$. Spectra obtained with the 600ZD grating 
are labeled. The confirmed \oiii\ emitter is ranked \#11.}
\label{f:phot-spec-LAE-1137}
\end{figure}

In the field J1137+3549, we acquired spectra 
of 11 photometric LAE candidates.
Four of them were observed using the 600ZD grating 
with exposure times long enough 
to provide good quality spectroscopy 
to determine the asymmetry of the emission line.
Thus, we include the low-resolution data 
of this field in our discussion.
One additional object with S/N$_{\text{NBC\,{\sc iv}}}{\simlt} 5$ 
was included in the analysis after 
the spectroscopic confirmation of 
an emission line. 
Hence, the sample in this field 
contains 12 spectra (11+1) which 
are presented in Figure \ref{f:phot-spec-LAE-1137}.

The objects in both Figures \ref{f:phot-spec-LAE-1030} 
and \ref{f:phot-spec-LAE-1137} are sorted according to 
the asymmetry of the emission line, decreasing from top to bottom. 
The ID numbers on the left hand side are for reference to Tables 
\ref{t:spec-LAE-J1030}--\ref{t:phot-LAE-J1137} (Appendix \ref{app:spec-tables}),
which present the information on each of the candidates.
Spectra obtained with the 600ZD grating are labeled as such. 
All the other spectra were obtained with the 830G grating.
Additional objects with S/N$_{\text{NBC\,{\sc iv}}}{\simlt} 5$ 
(i.e. not in the sample of photometric LAE candidates) 
are also indicated with the corresponding label.
The `X' symbol indicates if the object does not meet 
the asymmetry condition for \zlae\ LAE (see Section 
\ref{app:spectroscopic-LAEs-skewness})
The next section presents the search for contamination 
from foreground sources in the sample of LAEs.

\clearpage

 \subsection{Contamination in the spectroscopic sample}\label{s:spectroscopic-LAEs-contamination}
 
Photometric LAE candidates are selected based on an
excess of flux in the NB\civ\ filter.
This excess could result not only from \Lya\ at \zlae\ 
but also from \oii\ ($\lambda$3726, 3729\,\AA) at $z\sim1.17$--1.20,
\oiii\ ($\lambda$5007\,\AA) at $z\sim0.61$--0.64
and \Ha\ ($\lambda$6563\,\AA) at $z\sim0.23$--0.25.
Lower redshift contaminants are removed by a broad band 
colour condition (e.g. (R$_c{-}z'\text)>1.3$, Paper I) 
that aims to identify the flux decrement produced at 
$\lambda_{\text{rest}}\,\simlt\,1216$\,\AA\
and $\lambda_{\text{rest}}\,\simlt\,912$\,\AA\
by the absorbing neutral hydrogen in the Universe at \zlae .
However, some contamination remains, in particular among
fainter emitters.

This section presents the criteria adopted to
remove lower redshift contaminants based on
two observational characteristics of \zlae\ LAEs.
First, the \Lya\ line at this redshift is asymmetric due to neutral 
hydrogen in the galaxy, the CGM and IGM close to the source
\citep[][]{zhengzheng2010} whilst the emission line from a lower 
redshift contaminant is typically symmetric (e.g. \oiii\ and \Ha ), 
or a doublet (e.g. \oii). Secondly, if the emission is \Lya, there are 
no other nebular emission lines in the observed wavelength window.
The presence of additional UV emission lines 
in a \zlbg\ LAE is typically associated with an AGN.
Because AGNs produce well identified emission
lines such as \Nv , \siivoiv , \Niv\ and \civ , the sole 
detection of any of these transitions would indicate an AGN.
However, the fraction of AGNs among \zlbg\ LAEs 
is negligible \citep[e.g.][]{ouchi2010}.
Hence, the most common cases present no emission 
lines other than \Lya. As a result, lower redshift contaminants 
can be spectroscopically identified from the presence of 
additional emission lines and/or from the 
shape (asymmetry) of the emission line.

\subsubsection{Weighted skewness}\label{app:spectroscopic-LAEs-skewness} 

The \Lya\ emission in our sample of LAEs is affected 
by the neutral hydrogen content of the post-reionization Universe.
The number of intervening neutral hydrogen (\HI )
systems increases 
to higher redshifts \citep[e.g.][]{becker2013a}.
As a result, the net effect in the \Lya\ emission line 
is a complete absorption of the blue side. 
This asymmetry could be enhanced by \Lya\ photons 
emitted from the galaxy in a direction opposite 
to the observer, which are scattered back towards 
the Earth after interacting with the receding side of a galactic wind 
\citep[e.g.][]{hansen2006, verhamme2006, dijkstra2010, verhamme2014}.
Therefore, these photons are doppler shifted 
according to the speed of the wind and, by the time 
they reach the neutral IGM towards the observer, 
they have been redshifted and cannot be absorbed 
by the intervening neutral hydrogen.
The result is a typical red `tail' that in the 2D spectrum 
can be seen like a `tadpole' as in the four examples of 
Figure \ref{f:allslit}. For this reason, the asymmetry of 
the high redshift \Lya\ emission has become the main 
discriminant for lower redshift interlopers 
\citep[e.g.][]{rhoads2003, kashikawa2006b, shimasaku2006, kashikawa2011}.

In this work, we adopted the weighted skewness 
criteria proposed by \citet{kashikawa2006b} to quantify 
the asymmetry of the line profile. 
We estimate the weighted skewness S$_{w,10\%}$ 
using $\lambda _{10\%,\text R}$ and $\lambda_{10\%,\text B}$ 
which are the wavelengths where the flux from the emission 
drops to 10 per cent of the peak value on the red and blue sides, 
respectively. 

In particular, from the 18 spectroscopic detections 
in the field J1030+0524, we find 16 with a skewness 
S$_{w,10\%}\geq 3.0$ which are considered LAEs and 
two with S$_{w,10\%} < 3.0$ which are considered contaminants. 
In the field J1137+3549, 10 emission lines have S$_{w,10\%} \geq 3.0$ 
and two have S$_{w,10\%} < 3.0$, of which one is confirmed to be 
\oiii$({\lambda 5007})$ by the detection of \oiii$({\lambda 4959})$. 

Regarding the photometric sample 
(i.e. not including the S/N$_{\text{NBC\,{\sc iv}}}\,\simlt\,5$ objects),
we report that 13 out of 14 photometric LAE candidates ($\sim93$ per cent) 
and 9 out of 11 photometric LAE candidates ($\sim82$ per cent)
are spectroscopically confirmed LAEs in the fields J1030+0524
and J1137+3549, respectively. These translates to a contamination
fraction in the photometric sample of $\sim10$--20 per cent. 
Interestingly, the objects in both fields that do not satisfy
the asymmetry criteria are among the fainter emitters. 
Finally, we note that the asymmetry criterion as a discriminant of lower redshift
emitters is supported by the skewness measurement of an \oii\ emitter 
and an \oiii\ emitter in our data.
\subsubsection{The \oii\ doublet} \label{app:contamination-oii} 

\begin{figure} 
\centering 
\includegraphics[width=85mm]{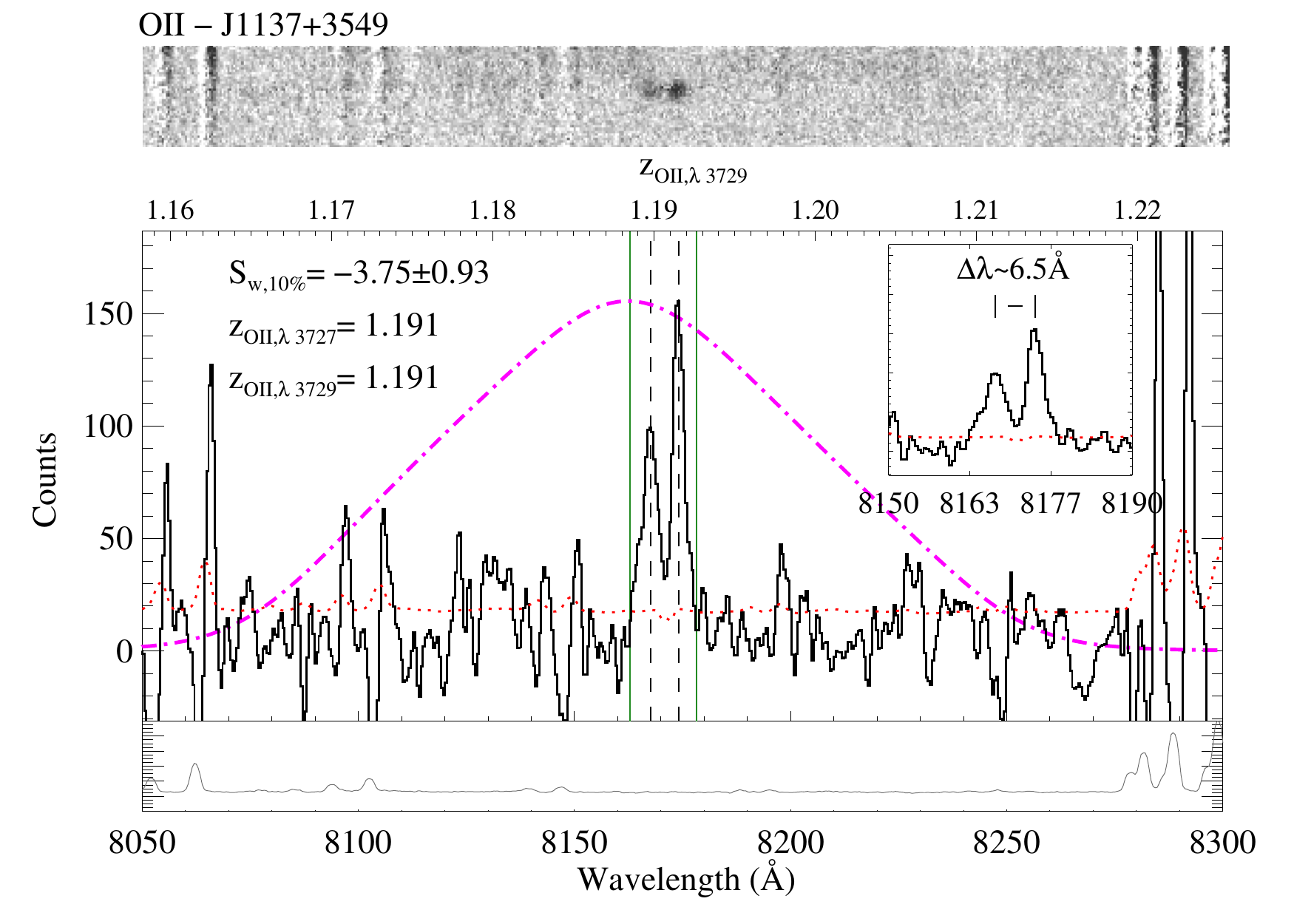}
\caption{\small Example of an \oii\ emitter at $z_{\text{ \oii}}=1.191$
observed with the 830G grating.
The observed wavelength of the \oii\ doublet 
is in the range covered by the NB\civ\ filter
and, therefore, is an example of a possible source of contamination.
{\it Top:} Snapshot of the 2D slit spectrum 
in the wavelength range covered by the NB\civ\ filter.
{\it Middle: } 1D spectrum smoothed over 3 pixels 
and close-up of the emission line. The doublet is clearly resolved
and the continuum is below the 1$\sigma$ error (red dotted line).
The magenta dot-dashed line is the scaled transmission curve of the NB\civ\ filter. 
The green vertical lines enclose the wavelength range
used to determine the skewness. 
{\it Bottom:} Sky spectrum for reference.} 
\label{f:OII}
\end{figure}

Among the possible contamination, the main sources are \oii\ emitters at 
$z\sim1.17$--1.2 because the Balmer break could mimic 
the \Lya\ forest decrement at \zlbg.
Fortunately, the \oii\ doublet has a wavelength separation 
of 2.8\,\AA$\times(1 + z)$, which is $\sim 6.1$\,\AA\ at $z\sim1.18$. 
Thus, the instrument configuration used in our observations
provides the spectral resolution to identify the \oii\ doublet if detected with sufficient S/N.
For example, Figure \ref{f:OII} shows the 2D and 1D spectrum of a \oii\ emitter 
detected in the field J1137+3549 but not a photometric LAE candidate.
The 1D spectrum has been smoothed with a three pixel size boxcar
and the \oii\ doublet is clearly distinguishable.
The maximum of each line was measured at 
$\lambda_{3727.092}=8167.56$\,\AA\ and $\lambda_{3729.875}=8174.05$\,\AA\
which correspond to $z=1.1914{\pm}0.0001$.
The skewness of the doublet is $S_{w,10\%}=-3.7{\pm}0.9$
which indicates that the asymmetry is opposite to that of
the \Lya\ line: the red emission line $\lambda{3729.875}$ is brighter than the blue
emission line $\lambda{3727.092}$.
Therefore, the asymmetry of the doublet is easily detected with
the resolution of our observations and
we are able to discriminate \oii\ from \Lya .

\subsubsection{Additional emission lines}\label{app:contamination-additional}

\begin{figure}
\centering
\includegraphics[width=85mm]{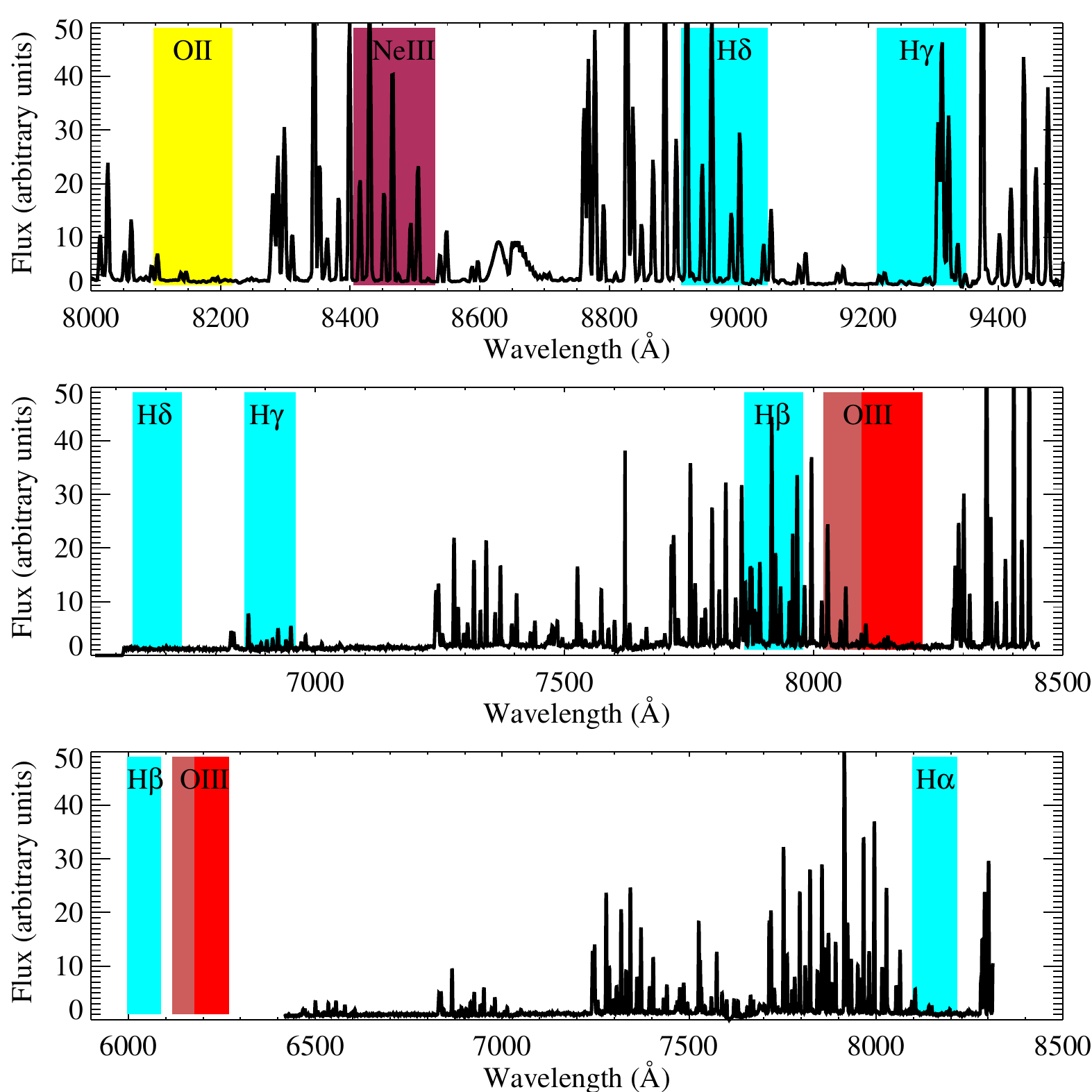} 
\caption{\small Expected wavelength of emission lines
typically present in lower redshift star-forming galaxies
that could be observed if the emission in the NB\civ\ is either \oii , \oiii\ or \Ha .
The sky spectrum is shown for reference. Many of the additional lines
are buried in sky emission lines which reduces our ability to
identify them.
{\it Top:} \oii\ emitters at $z=1.17$--1.20.
The wavelength intervals of emission lines that, within the observed
spectral range, are commonly associated with \oii\ ($\lambda$3726, 3729\,\AA )
are labeled in the plot: \Neiii , \Hd\ and \Hg .
{\it Middle:} \oiii\ emitters at $z=0.61$--0.64.
In this case, the associated emission lines that could
be observed are \oiii ($\lambda$4959\,\AA), \Hb , \Hg\ and \Hd .
{\it Bottom:} \Ha\ at $z=0.23$--0.25.
The spectral range of the data does not sample 
the next brighter lines after \Ha\ (i.e.: \oiii\ and \Hb ).}
\label{f:interloper-search}
\end{figure}

Detection of nebular emission lines provides
a robust identification of lower redshift galaxies.
The spectral range covered by the data allowed us to search for
other emission lines associated with \oii\ emitters and \oiii ($\lambda$5007\,\AA) emitters.
In the first case, we looked for \Neiii, \Hd\ and \Hg\ at the corresponding 
wavelengths, as shown in Figure \ref{f:interloper-search} (top panel).
In the second case, 
we searched for the presence of \oiii ($\lambda$4959\,\AA), \Hb , \Hg\ and \Hd,
in the wavelength windows highlighted in the middle panel of Figure \ref{f:interloper-search}.
A careful inspection of the 2D and 1D spectra resulted in 
the detection of an \oiii\ emitter in the spectroscopic sample of 
LAE candidates in the field J1137+3549.

\begin{figure}
\centering 
\includegraphics[width=85mm]{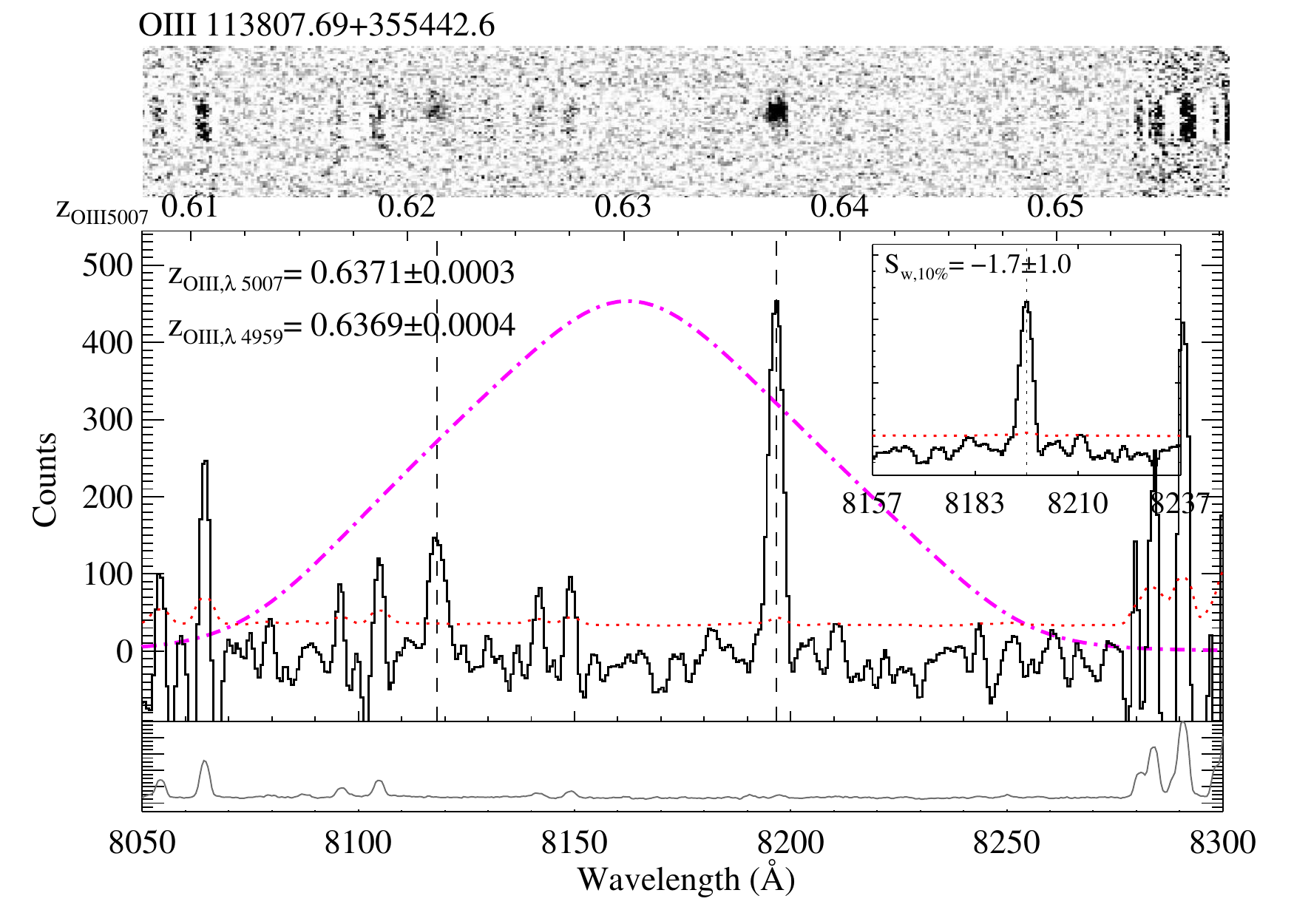} 
\caption{\small \oiii\ emitter in the photometric sample of LAE candidates
observed with the 600ZD grating.
Two emission lines are clearly detected and correspond to
\oiii($\lambda$4959\,\AA) and \oiii($\lambda$5007\,\AA)
at $z_{\text{\oiii}}\sim0.637$.
{\it Top:} Snapshot of the 2D slit spectrum 
in the wavelength range covered by the NB\civ\ filter.
{\it Middle:} 1D spectrum smoothed over 3 pixels 
and close-up of the emission line at $\lambda_{\text{rest}}=5007$\,\AA ,
which has a skewness $<3$.
The red dotted line is the 1$\sigma$ error and the magenta dot-dashed line 
is the scaled transmission curve of the NB\civ\ filter. 
{\it Bottom:} Example of sky spectrum for reference.} 
\label{f:OIII}
\end{figure}

The object is labeled `\oiii ' in Figure \ref{f:phot-spec-LAE-1137},
and is shown in better detail in Figure \ref{f:OIII}.
The \oiii ($\lambda$4959\,\AA) emission line is seen at $\lambda=8117.57$\,\AA .
No other emission lines were detected.
Moreover, the weighted skewness of the brightest emission line
($\lambda_{\text{rest}}=5007$\,\AA ) is $S_{w,10\%}=-1.7{\pm}1.0$
which validates the asymmetry criterion.

Although the 830 line mm$^{-1}$ grating provides
wavelength coverage from 6500 to 10000\,\AA ,
the effective wavelength range observed in each slit depends
on its position on the mask. Therefore, not all the spectra
cover the wavelength range necessary to explore the presence
of all the possible additional emission lines that would reveal the low-redshift
nature of an object.
Moreover, many of the possible additional emission lines
that would confirm a low redshift interloper
are buried in a forest of sky-lines residuals, which makes them 
more difficult to be detected (Figure \ref{f:interloper-search}). 
This problem is more severe for faint objects detected 
with low S/N.

Finally, if the emission in the NB\civ\ filter was \Ha ,
the wavelength coverage of the data does not reach the wavelength of
the nearest strong emission lines like \Hb\ and \oiii\ 
(Figure \ref{f:interloper-search}, bottom panel).
Therefore, \Ha\ cannot be 
identified by the presence of other emission lines. 
As a result, because \Ha\ is symmetric, identification
of \Ha\ emitters at $z\sim0.23$--0.25 
relies mainly on the symmetry of the line.
However, the spectral energy distribution is also a discriminant
if continuum is detected because, for example,
an \Ha\ emitter would show continuum flux at both 
sides of the emission line.

In summary, the abundant emission from the night sky
reduces our ability to detect additional emission lines
in the galaxies' spectra.
The \oii\ emitter in Figure \ref{f:OII} and the LAE candidate 
that shows clear evidence for additional 
emission lines (Figure \ref{f:OIII}) also have negative asymmetry line profiles.
Therefore, these objects support the asymmetry criterion adopted
in this work to discriminate lower redshift contaminants in the LAE sample.
We conclude that additional emission would confirm the low redshift 
nature of an object but cannot be the only
criterion used to detect contamination in the sample.

\subsubsection{Magnitude dependant contamination fraction}\label{s:contamination-fraction}

Among fainter emitters, low S/N reduces
the ability to resolve a doublet from 
a single line and to accurately 
measure the asymmetry of the emission.
Therefore, the level of contamination 
is expected to increase towards fainter magnitudes.
The data set used in this work seems 
to agree with this expectation since, first, 
sources brighter than NB\civ$\,=24.7$ mag 
are confirmed LAEs (S$_{w,10\%} \geq 3$)
and, second, LAE candidates with S$_{w,10\%}<3$
are fainter than NB\civ$\,=24.7$ mag.

\begin{figure}
\centering
\includegraphics[width=85mm]{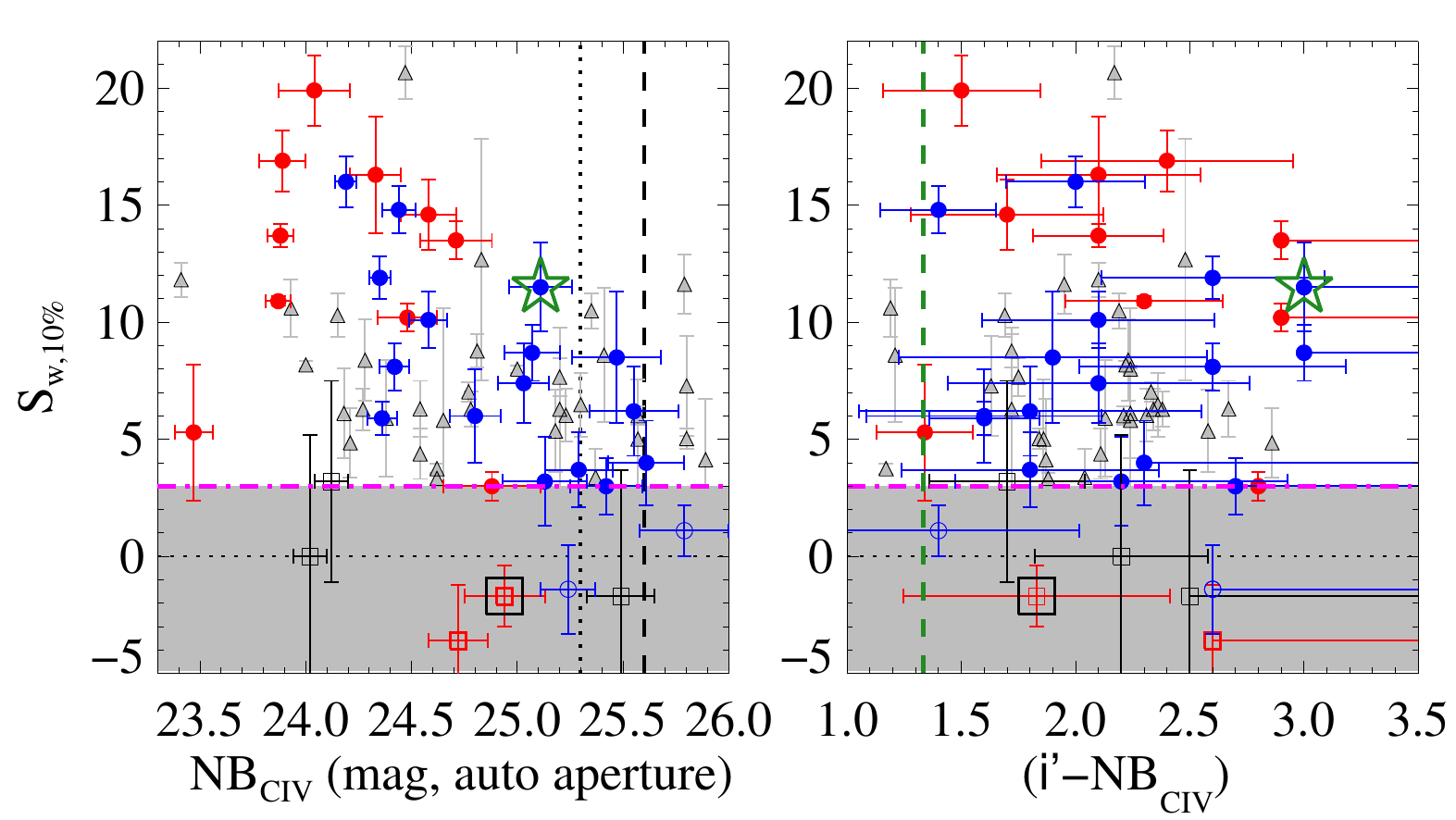}
\caption{\small Weighted skewness as a function of NB\civ\ magnitude
and ($i'-$NB\civ) colour.
Blue and red circles correspond to the fields
J1030+0524 and J1137+3549, respectively.
The star symbol highlights LAE 103027+052419
(see Section \ref{s:lae-civ-pair}).
Triangles are SDF LAEs \citep{shimasaku2006}.
Open squares correspond to low resolution data
obtained with the 600ZD grating in the field J1030+0524,
which are presented only for completeness.
The dot-dashed horizontal line is the boundary S$_{w,1\sigma}=3$
above which objects are considered LAEs (solid symbol).
{\it Left:} NB\civ\ magnitude versus S$_{w,10\%}$.
Brighter objects tend to have higher S$_{w,10\%}$. 
In particular, spectroscopic LAEs with 
NB\civ$\,<24.7$ have S$_{w,10\%} \geq 3.0$.
The two vertical lines indicate the 5$\sigma$ 
limit magnitude in the field J1030+0524 (dashed line) 
and the field J1137+3549 (dotted line).
{\it Right:} ($i'-$NB\civ) versus S$_{w,10\%}$.
In this case, there is not clear correlation 
between the two quantities.
The vertical dashed line shows the condition 
($i'-$NB\civ)$\,>1.335$ used in the photometric 
selection criteria for LAEs in Paper I.}
\label{f:Swb-color-mag}
\end{figure}

Figure \ref{f:Swb-color-mag} shows the relation between
S$_{w,10\%}$ and two photometric quantities: NB\civ\ magnitude
and the flux excess in NB\civ\ measured by the colour ($i'-$NB\civ).
Blue and red circles are LAEs in the fields J1030+0524 
and J1137+3549, respectively, observed with the 830G grating.
Objects observed with the 600ZD grating in the later field
are represented by red squares. We do not consider
observations of J1030+0524 obtained with this grating 
because the lower S/N results in skewness measurements with large errors,
but for completeness we show them in the plot with black open squares.
The star symbol indicates LAE 103027+052419 at $z=5.724$,
which is the closest LAE to the 
\civ\ absorption system at $z_{\text{\civ,b}}=5.7242$ 
in the field J1030+0524 (see Section \ref{s:lae-civ-pair}).
Points above and below S$_{w,10\%}=3$ (grey area) are 
represented by solid and open symbols, respectively. 
In the left panel, we note a trend for fainter objects to have lower skewness, 
with objects in both fields following the same tendency.
In comparison, however, the spectroscopic sample of $z\sim5.7$ LAEs 
in the Subaru Deep Field \citep[SDF,][]{shimasaku2006}, over-plotted with
grey triangles in Figure \ref{f:Swb-color-mag}, 
seems more uniformly scattered across the range of magnitudes.
The spectroscopic sample of SDF LAEs includes
data with different resolution (FWHM$=$9.5, 7.1 and 3.97\,\AA).
Moreover, the spectral resolution of the data affects the ability to
measure the line skewness, and is therefore a source of systematic error 
when comparing data with different resolutions, like in Figure \ref{f:Swb-color-mag}.
As a result, we need larger and homogeneous samples to explore
the morphology of the high-redshift \Lya\ emission and the information
behind it in a consistent way.

Finally, no obvious correlation is seen between 
S$_{w,10\%}$ and ($i'-$NB\civ), which are plotted 
in the right panel of Figure \ref{f:Swb-color-mag}.
Objects above the grey area (S$_{w,10\%} \geq 3$)
show significant scatter in ($i'-$NB\civ). 
This is observed in our
sample and SDF LAEs, and means that robust and bright 
spectroscopically confirmed LAEs
can show small or large narrow-band flux excess, 
regardless of the asymmetry of the line.
Below S$_{w,10\%}=3$, photometric candidates can also 
drop-out of the $i'$-band and reach ($i'-$NB\civ)$\,>2.5$, 
in which case the colour ($i'-$NB\civ) 
is only a lower limit and is more than 
a magnitude greater than the condition 
($i'-$NB\civ)$\,>1.335$ in the photometric 
selection criteria for LAEs of Paper I.
We conclude that the current data cannot robustly confirm nor rule out
the existence of a correlation between the measured asymmetry
of the emission line and the narrow-band photometric properties of the source.
Nevertheless, we recall that contaminants are fainter than NB\civ$\,=24.7$ mag
which is not enough to claim a correlation, but is useful
to keep in mind in further analysis of the sample. 

\subsection{Active Galactic Nuclei in the \zlae\ LAE sample}\label{s:spectroscopic-LAEs-AGN} 

\begin{figure}
\centering 
\includegraphics[width=85mm]{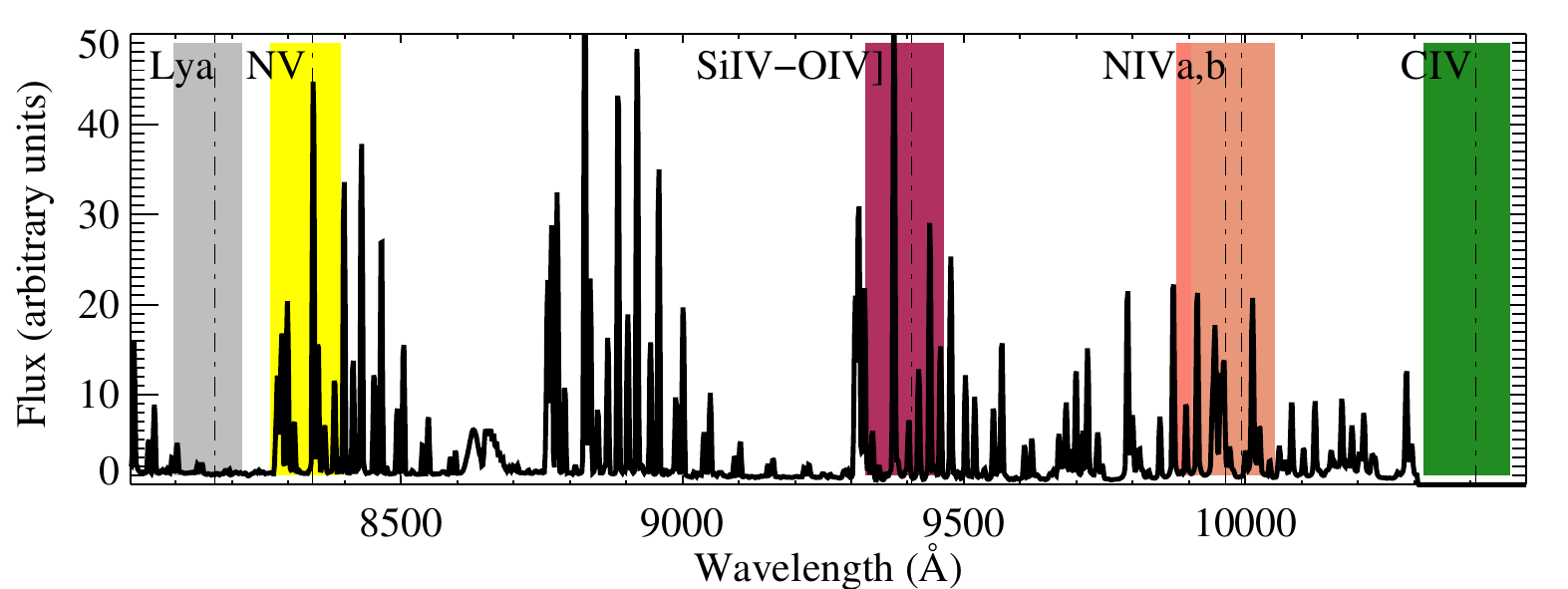} 
\caption{\small Sky spectrum and observed wavelength of typical 
emission lines produced in AGNs: \Nv , \siivoiv , \Niv\ and \civ ;
with \Lya\ in the wavelength range of the NB\civ\ filter.
The vertical dot-dashed lines correspond to the particular case of a galaxy at \zlae .
No evidence of potential emission other than \Lya\ was found in the
spectra (2D and 1D) of confirmed LAEs, 
although the spectra do not cover the wavelength of the \civ\ emission
and most of the emission lines are in regions severely affected by sky lines.}
\label{f:agn-search}
\end{figure}

The \Lya\ emission can also be powered by an AGN.
However, AGNs are rarely found among high redshift LAEs
\citep[e.g.][]{wang2004, ouchi2008, zheng2010}.
Unfortunately, we do not have information from other wavelengths 
(e.g. radio, X-ray, infrared, submillimeter) 
nor we have information on the rest-frame optical emission lines 
to search for AGN activity in the \zlae\ LAE sample.
Therefore, we can only explore the presence of AGNs
by searching for high ionization UV emission lines produced by AGN activity.
However, the wavelength range of our data does not cover the typically
strong \civ\ emission, thus we only searched for \Nv , \siivoiv\ and \Niv.
Moreover, \Nv\ and \siivoiv , which are the brighter emission lines
that could have been detected in the spectral range of the data, would be in 
regions of the spectrum that are heavily affected by sky emission lines. 
This is clearly seen in Figure \ref{f:agn-search}.
As a result, our ability to identify AGNs is very limited.
Nevertheless, after careful visual inspection of each 2D and 1D spectra,
we report no evidence of AGN activity in the spectroscopic sample.

\section{Colours and magnitudes of \zlae\ LAEs and contamination in the photometric sample}\label{s:colmag-spec}
\subsection{Broad-band colours}\label{s:bbcol}
The \Lya\ emission line can affect the broad-band colours of high redshift galaxies.
In Paper I, we find that the $z\sim6$ $i'$-dropout
colour criterion is not a good tracer 
of galaxies at $z\sim5.7$. The effect is more significant for objects 
with strong \Lya\ emission. 
The reason is that the $i'$-dropout selection is based on a condition 
of red ($i'-z'$) colour (e.g. ($i'-z'$)$\,> 1.3$) and the \Lya\ emission 
of a galaxy at $z\sim5.7$ would increase the flux in the $i'$-band 
therefore reducing the colour ($i'-z'$). 

\begin{figure}
\centering
\includegraphics[width=85mm]{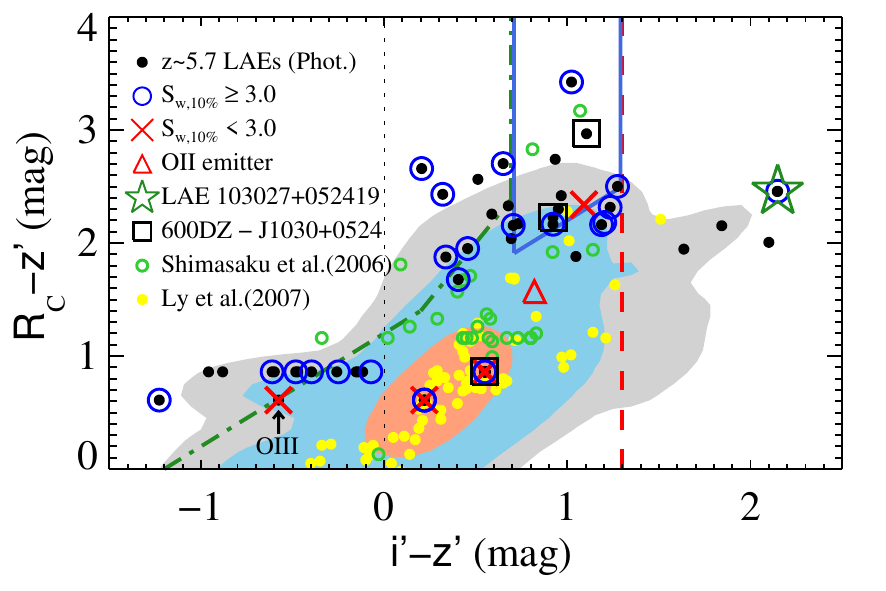}
\caption{\small Broad-band (R$_c-z'$) versus ($i'-z'$) colours of 
the complete photometric sample of \zlae\ LAEs (black points).
The contours correspond to the photometric catalogue of detections 
and contain 50, 90 and 97 per cent of the sources in the catalogue.
The spectroscopically confirmed LAEs (S$_{w,10\%} \geq 3.0$)
are indicated with blue circles, and the contaminants  (S$_{w,10\%}<3.0$)
are indicated with red crosses. 
The open star symbol highlights LAE 103027+052419 
(see Section \ref{s:lae-civ-pair}).
The red dashed vertical line is the ($i'-z'$)$\,= 1.3$ boundary
condition for $i'$-dropouts, the blue solid lines
indicate the colour criteria for \zlbg\ LBGs of Paper I. 
and the green dot-dashed line shows the colour selection 
for $z\sim5.0$ LBGs of \citet{ouchi2005a}.
Green open circles are SDF LAEs \citep{shimasaku2006}
and solid yellow circles are lower redshift emitters 
in SDF from \citet{ly2007}.}
\label{f:BBcol-LAE}
\end{figure}

The spectroscopic data clearly confirms this expectation.
Figure \ref{f:BBcol-LAE} shows the position of the photometric 
LAE candidates (black points) in the  (R$_c-z'$) versus ($i'-z'$) colour diagram.
Spectroscopically confirmed LAEs are indicated with blue circles 
and contaminants are indicated with red crosses.
Open squares are the three LAE candidates in the field J1030+0524
observed with the 600ZD grating. 
The red dashed line is the ($i'-z'$)$\,=1.3$ colour boundary 
of the $i'$-dropout selection and only 4/33 LAEs reach this condition.
Interestingly, the only spectroscopically confirmed LAE 
that satisfies the $i'$-dropout condition is LAE 103027+052419 
(green star symbol), which is the closest LAE to the 
\civ\ absorption system at $z_{\text{\civ,b}}=5.7242$ 
in the field J1030+0524 (see Section \ref{s:lae-civ-pair}).
The (R$_c-z'$) colours are only lower limits in 
the majority of the objects because they are non-detected in R$_c$. 
Nevertheless, most of them are detected in $i'$, 
which means that their ($i'-z'$) colours are more robust
than their (R$_c-z'$) colours.
For those objects that are non-detected in $z'$, the ($i'-z'$) 
colours are upper limits.
In particular, the rows at (R$_c-z'$)$\,=0.86$ and 0.61
are made of objects only detected in $i'$ and NB\civ .

\begin{figure*}
\begin{minipage}{150mm}
\centering 
\includegraphics[width=160mm]{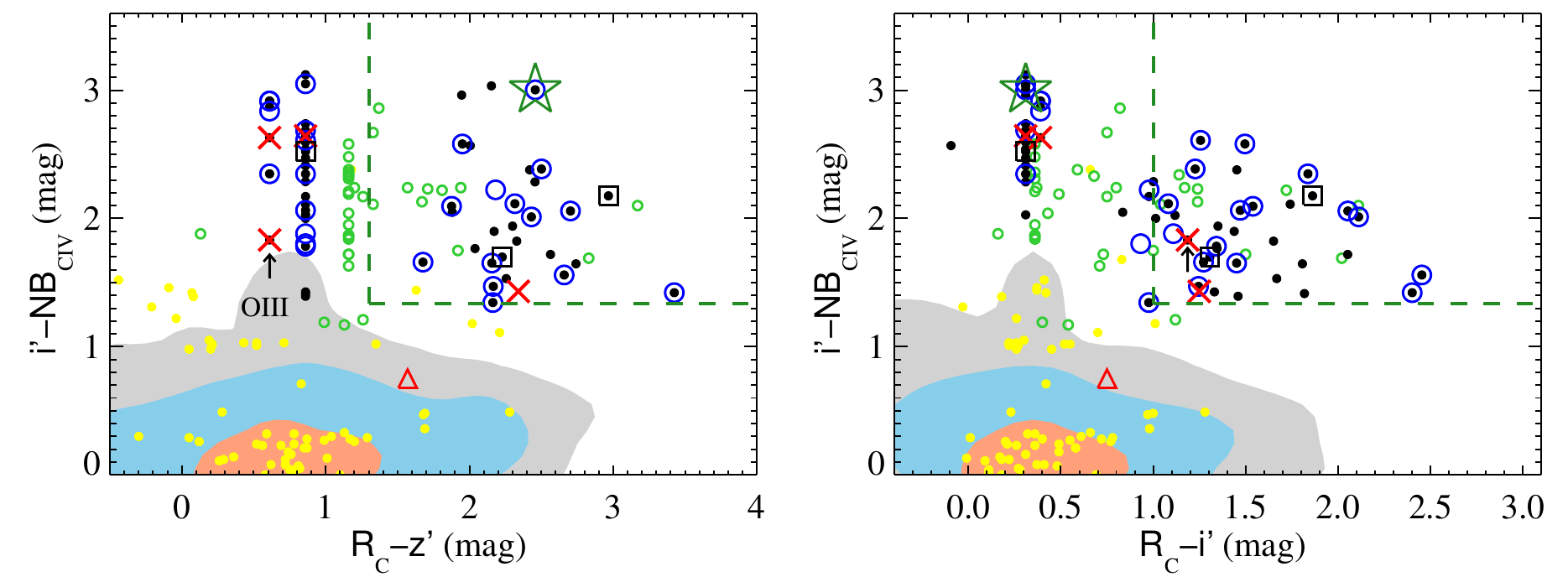}
\caption{\small  Broad-band colours and the excess in NB\civ .
Symbol key is as per Figure \ref{f:BBcol-LAE}.
The green dashed lines show the boundaries of the LAE colour selection criteria.
Objects non-detected in R$_c$ and $z'$ accumulate at (R$_c-z'$)$=0.86$ 
and 0.61 ({\it left}), and objects non-detected in R$_c$ and $i' $
accumulate at (R$_c-i'$)$=0.31$ and 0.39 ({\it right}).}
\label{f:NBcol-LAE}
\end{minipage}
\end{figure*}

Moreover, good agreement is found between our LAE sample and 
the confirmed LAEs at \zlbg\ from \citet{shimasaku2006} (green open circles).
This sample also shows bluer ($i'-z'$) than expected for $i'$-dropout,
since none of them have ($i'-z'$)$\,>1.3$.
The concentration of green circles around (R$_c-z'$)$\,\sim1.6$
and ($i'-z'$)$\,=0.4$--0.9 corresponds to non-detections in the R$_c$ and $z'$
bands, for which these colours are lower and upper limits, respectively.
Keeping in mind that the \zlae\ LAE selection is not defined in the colour--colour
diagram of Figure \ref{f:BBcol-LAE},
it is encouraging that the broad-band colours of the LAE sample 
agree with the expectations from the spectrophotometry of composite spectra of LBGs
used to define the \zlbg\ LBG selection of Paper I.

The red open triangle is the \oii\ emitter presented in Figure \ref{f:OII} 
and is found below the colour selection window for \zlbg\ LBGs (blue solid lines),
in agreement with the contamination from low redshift galaxies 
analysed in Paper I.
The colours of low redshift ($z\sim0.07$--1.47) narrow-band
emitters from \citet{ly2007} are represented by yellow solid circles. 
This sample contains \oii , \oiii\ and \Ha\ emitters, which are 
the dominant contamination source in LAE searches. 
Moreover, the sample covers a redshift range larger than 
that covered by the NB\civ\ transmission curve.
Figure \ref{f:BBcol-LAE} shows that lower redshift emitters do not
occupy the broad-band colour--colour region of higher redshift LAEs.
As a result, the discrepancy in the colour--colour distribution of interlopers
and spectroscopically confirmed LAEs provides a good reason
to expect a low fraction of contamination among the LAE photometric candidates.

\subsection{Narrow-band colours and the LAE photometric selection criteria}\label{s:bbcol}

Figure \ref{f:NBcol-LAE} presents the colour--colour 
diagrams where the LAE colour selection criteria 
are defined (green dashed lines).
The photometric sample of LAEs (black points) 
was selected from a catalogue of robust detections 
(S/N$_{\text{NBC\,{\sc iv}}} \geq  5$). They
can be seen to the right and to the left of the vertical dashed lines
because the condition on the broad-band colours of LAEs (Paper I)
requires that either colour is `red', or
there is no detection in all three broad-bands:
 [(R$_c-i'$)$\,>$1.0]$\cup$[(R$_c-z'$)$\,>$1.3]$ \cup$[(R$_c<1\sigma$)$\cap$($i'<1\sigma$) 
 $\cap$ ($z'<1\sigma$)].

We have spectroscopic confirmation (blue circles) 
of photometric candidates. Those that show 
significant NB\civ\ excess but are not detected in more than one 
broad band are found at (R$_c-z'$)$=0.86$ and 0.61 (left hand panel), 
and at (R$_c-i'$)$=0.31$ and 0.39 (right hand panel).
We recall that the majority of the objects are non-detected in R$_c$ 
(see Tables \ref{t:phot-LAE-J1030} and \ref{t:phot-LAE-J1137}) and therefore
both (R$_c-i'$) and (R$_c-z'$) are lower limits.

The LAE discussed in Section  \ref{s:lae-civ-pair} 
(LAE 103027+052419, open star symbol)
is one of the objects at (R$_c-i'$)$\,=0.31$ which 
would have been missed if only one condition (R$_c-i'$)$\,>1.0$ 
(vertical green dashed line in the right hand panel) was applied.
This means that the inclusion of the colour condition 
$\cup$ (R$_c-z'$)$\,>1.3$ (vertical green dashed line in the left hand panel) 
in the selection of \zlae\ LAEs is successful among objects 
like LAE 103027+052419 which are non-detected in the $i'$-band.
Furthermore, we confirm five LAEs from the seven photometric 
candidates selected by the NB\civ\ excess but not detected 
in any broad-band, for which we have spectra with sufficient S/N 
(i.e. excluding the 600ZD data in the field J1030+0524).
Figure \ref{f:NBcol-LAE} also shows \zlbg\ LAEs 
from SDF \citep{shimasaku2006} in green open circles.
The narrow--band magnitude corresponds to the filter
NB816, which is similar to NB\civ .
Thus, when the measured magnitudes are compared, 
we consider them equivalent.
We note that the selection criteria of these objects 
were based on their (R$_c-z'$) colour and, therefore,
many of them do not reach (R$_c-i'$)$\,>1.0$.
Since \zlbg\ LAEs are faint broad-band sources 
typically non-detected in R$_c$,
we find objects that are only detected in one 
broad-band ($i'$ or $z'$), or non-detected in all 
three broad-bands, by using a flexible broad-band colour condition
(based on an `OR' operator) in the selection criteria. 

Objects with S$_{w,10\%}<3.0$ (red crosses)
do not seem to occupy a defined region. 
The red open triangle that indicates the colours of the \oii\ emitter 
discussed in Section \ref{s:spectroscopic-LAEs-contamination}
is below the colour selection condition ($i'-$NB\civ)$\,=1.335$ 
and bluer than (R$_c-i'$)$\,>1.0$.
This is in agreement with the distribution of lower redshift emitter 
from \citet{ly2007} (yellow circles),
whose narrow-band magnitude corresponds to NB816.
As a result, we find that lower redshift emitters have less 
significant narrow-band ($i'-$NB\civ) excess.

\begin{figure}
\includegraphics[width=85mm]{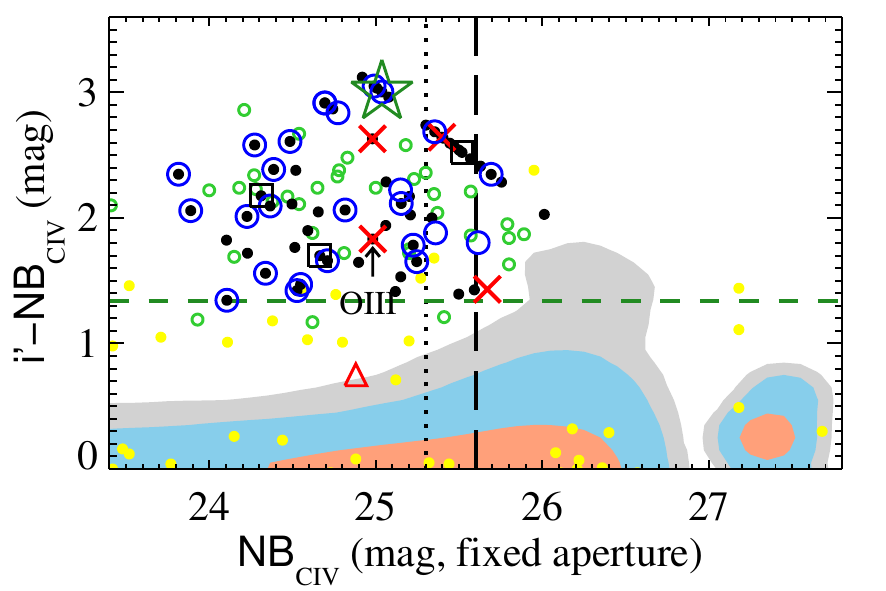}
\caption{\small  Colour--magnitude diagram ($i'-$NB\civ) versus NB\civ .
Vertical lines indicate the NB\civ\ $5\sigma$
limiting magnitudes in the field J1030+0524 (dashed) and the field J1137+3549 (dotted).}
\label{f:NBmag-LAE}
\end{figure}

The colour--magnitude diagram ($i'-$NB\civ) versus NB\civ\
in Figure \ref{f:NBmag-LAE} shows that the \oii\ emitter 
is among other lower redshift emitters from \citet{ly2007}
which can have brighter narrow-band magnitudes than
the LAE sample. 
Moreover, it is clear that spectroscopically confirmed LAEs (including SDF LAEs)
have larger ($i'-$NB\civ) colours than the contaminant population.
In particular, we note that SDF LAEs and our LAE sample
are evenly distributed and mixed in the colour-magnitude diagram ($i'-$NB\civ) versus NB\civ ,
meaning that both samples have similar brightness and colours.

In summary, our results validate the colour selection criteria for 
\zlae\ LAEs of Paper I and suggest low contamination among 
bright (NB\civ$\,<25$ mag) photometric LAE candidates.

\section{Spectroscopic redshift distribution}\label{s:redshift-dist}
\begin{figure}
\centering 
\includegraphics[width=85mm]{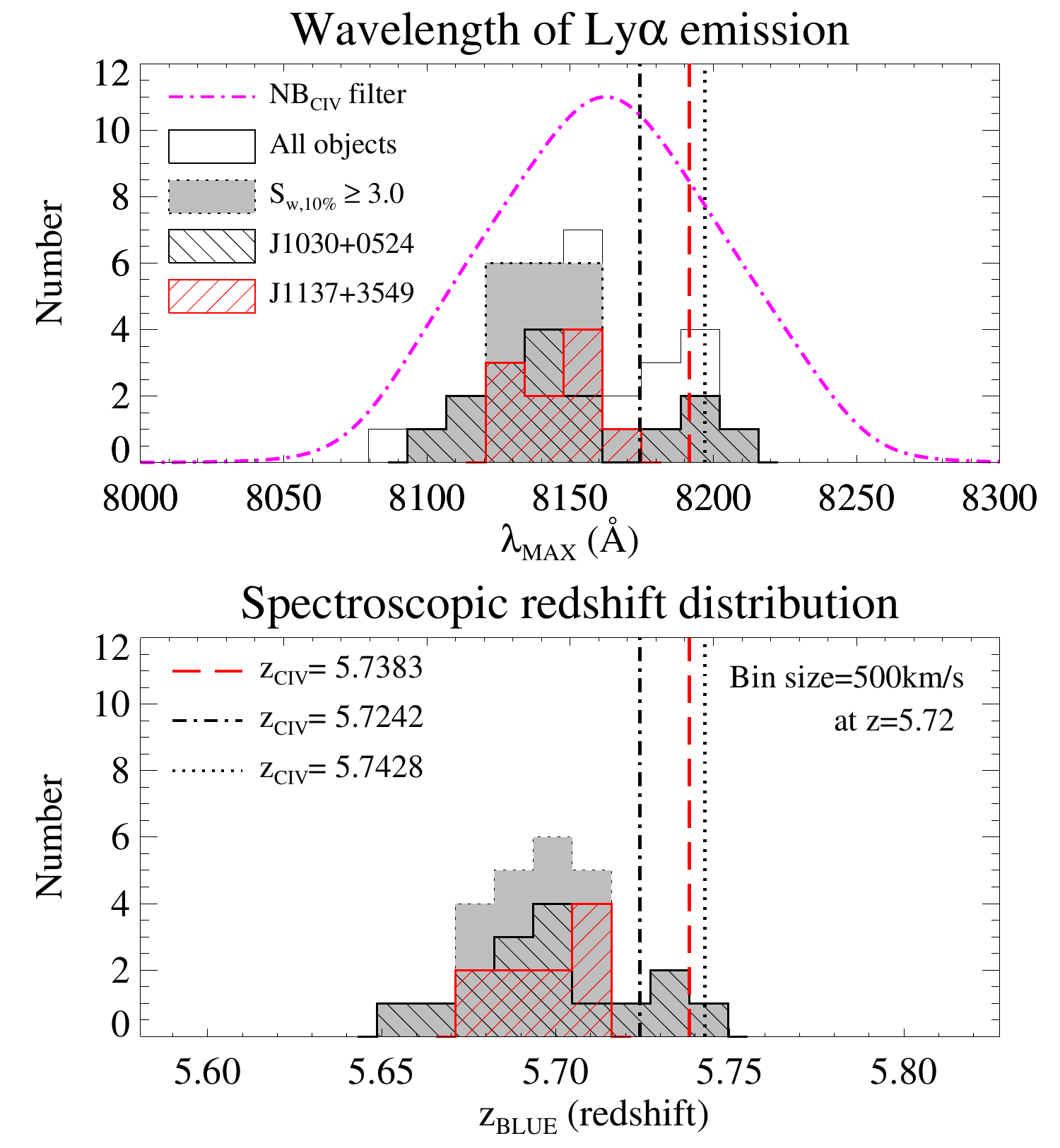} 
\caption{\small {\it Top:} Wavelength distribution of the maximum 
of the emission line of all the LAE candidates in both fields 
(black solid histogram). 
The grey filed histogram corresponds to candidates
with S$_{w,10\%} \geq 3$ in both fields.
Individual fields are shown by the black-line filled histogram (J1030+0524)
and red-line filled histogram (J1137+3549). 
The vertical lines indicate the wavelength of \Lya\ 
at the redshift of the \civ\ absorption systems in each field, 
black dot-dashed and black dotted lines for the field
J1030+0524 and red dashed for the field J1137+3549.
{\it Bottom:} Spectroscopic redshift distribution of confirmed LAEs
(S$_{w,10\%} \geq 3$), as measured from the bluest pixel of the emission line 
with flux above the $1\sigma$ level.
Histogram key is the same as the top panel.}
\label{f:spec-redshift-LAE}
\end{figure}

Figure \ref{f:spec-redshift-LAE} shows
the wavelength distribution of the
\Lya\ peak emission of all the detections
and the NB\civ\ filter transmission curve for comparison.
The solid line includes all the objects in both fields and
the grey histogram correspond to objects with S$_{w,10\%} \geq 3.0$.
We find that the wavelength distribution is biased towards the blue half of the 
filter curve.
This effect as been found in all previous narrow-band selected samples of 
high-redshift LAEs \citep[e.g.][]{shimasaku2006, ouchi2008, kashikawa2011}.
The reason is that the flux contribution from the red side of the \Lya\ emission
to the NB\civ\ magnitude decreases towards higher redshift.
If the emission is bluer than the middle point of the NB\civ\ 
transmission curve (maximum throughput), the red wing of \Lya\
is covered by a more sensitive section of the filter and there is more 
rest-frame UV flux under the filter transmission curve.
If the emission is redder than the middle point of the NB\civ\ transmission curve, 
there is little rest-frame UV flux included in the NB\civ\ magnitude, 
and the red wing of the \Lya\ is covered by a progressively less sensitive 
section of the filter. 

Considering that \Lya\ is a resonant transition,
radiative transfer plays a very important role 
in the escape of \Lya\ photons.
Cosmological hydrodynamical simulations 
combined with radiative transfer have shown the effect of
the dynamics of the surrounding \HI\ gas (CGM and/or IGM) 
in the emission line profile.
For example, \citet{zhengzheng2010} find that 
radiative transfer is needed to properly model
frequency diffusion in which \Lya\ photons close to the central
wavelength diffuse out to the wings of the emission line.
The implication of this effect in the observed \Lya\ profile
is clearly seen in Figure 5 of \citet{zhengzheng2010}.
They show simulated observations of \Lya\ profiles 
at \zlae\ in which the emission peak
is at a longer wavelength than the peak of the intrinsic profile.
Moreover, the flux at the centre of the line is strongly 
suppressed, which results in an observed emission profile 
fully contained red-wards of the true line centre
\citep[e.g.][]{jensen2013}.

Because we are aware that redshift estimates from the
\Lya\ line are affected by the \HI\ content of the near environment
of the source and its kinematics, to measure the spectroscopic redshift
of a \zlae\ LAE, we prefer to use the bluest pixel of the line profile 
with flux over the $1\sigma$ level instead of the pixel with maximum
flux, which is usually adopted in the literature \citep[e.g.][]{ouchi2005a, shimasaku2006, hu2010}.
Our choice should bring our measurement in closer
agreement with the true redshift of the galaxy.

The bottom panel of Figure \ref{f:spec-redshift-LAE} 
shows the redshift distribution of the LAEs in each field with line-filled 
histograms: black lines for J1030+0524 and red lines for J1137+3549.
We notice that the redshift distribution is roughly centred at $z\sim 5.7$, 
which is lower than the redshift of the \civ\ absorption systems reported 
in the two lines of sight. 
The redshift of the \civ\ systems are indicated by the
vertical lines: dot-dashed black line and dotted black line for the two \civ\ systems 
in the field J1030+0524 and dashed red line for the \civ\ system in the field J1137+3549.
 
\begin{figure}
\centering
\includegraphics[width=85mm]{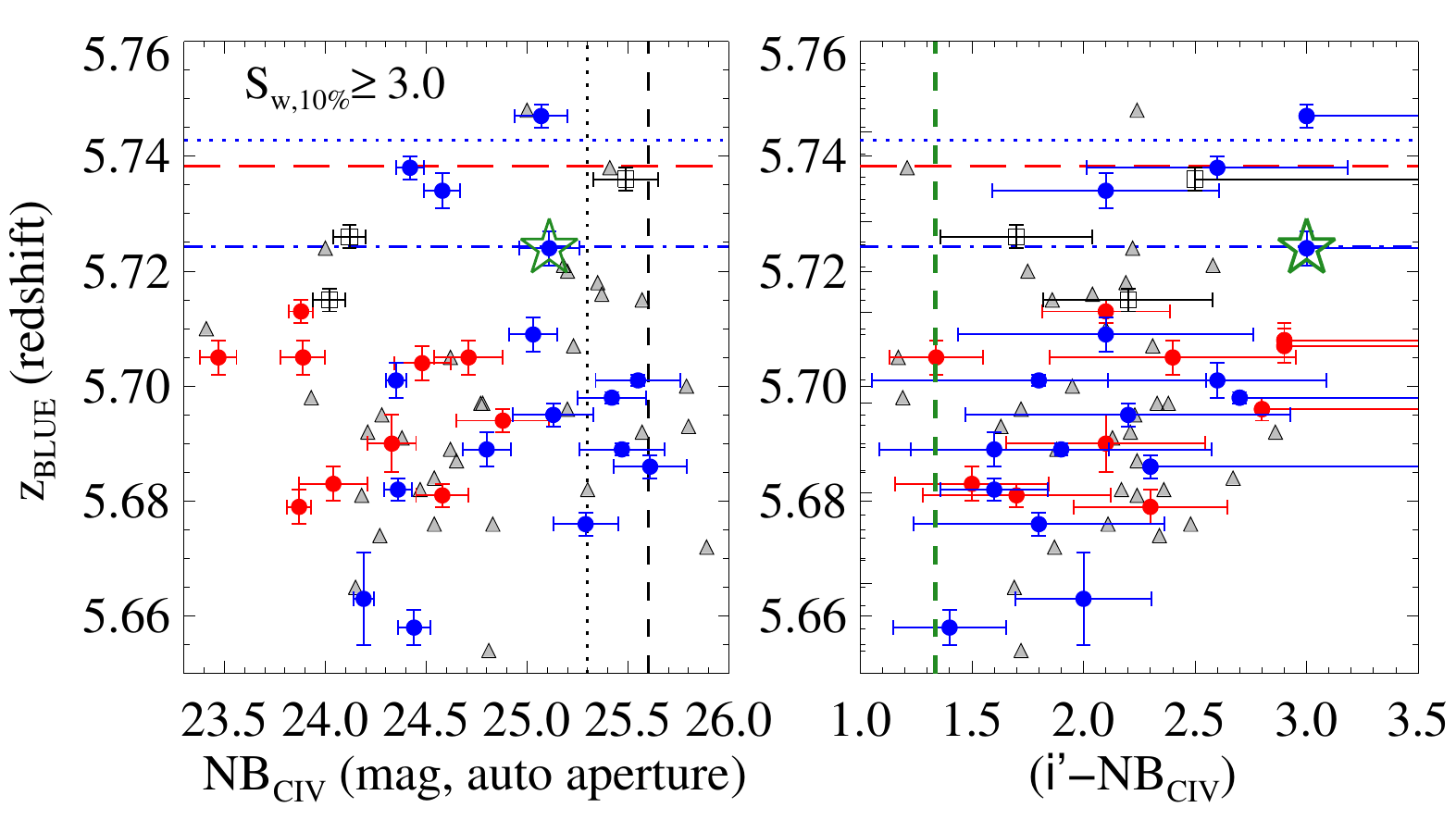}
\caption{\small Narrow-band photometry versus spectroscopic redshift
of emission lines with S$_{w,10\%} \geq  3.0$.
Blue and red circles correspond to the two fields
J1030+0524 and J1137+3549, respectively.
LAE 103027+052419 is highlighted with a star symbol.
Grey triangles are SDF LAEs \citep{shimasaku2006}.
The open squares represent the low resolution data (i.e. 600ZD grating) 
in the field J1030+0524, and although the sensitivity is not enough, if the
emission is \Lya, these objects will fill the top half of the plots 
which are less populated.}
\label{f:z-color-mag}
\end{figure}

The position of the emission under the narrow-band filter
transmission curve depends on the redshift of an object.
We searched for potential correlation between redshift and
the narrow-band magnitude and colour 
and find no obvious trend.
Figure \ref{f:z-color-mag} presents the redshift 
of the LAEs as a function of the NB\civ\ magnitude (MAG\_AUTO) 
and the ($i'-$NB\civ) colour (MAG\_APER).
Red circles correspond to LAEs in the field J1137+3549 
and are contained in the range $z\sim5.68$--5.72, 
as in Figure \ref{f:spec-redshift-LAE}, 
showing no correlation with NB\civ.
Blue circles correspond to the field J1030+0524
and cover a wider redshift range $z\sim5.66$--5.75.
Also in this case, the apparent magnitudes of the objects in the sample 
do not show a dependance with redshift.

The horizontal lines indicate the redshift of the \civ\
systems using the same line style as per Figure \ref{f:spec-redshift-LAE}. 
The LAE indicated with the star symbol is LAE 103027+052419 
at $z=5.724 \pm 0.001$ (Section \ref{s:lae-civ-pair}).
The right hand panel of Figure \ref{f:z-color-mag} shows 
the spectroscopic redshift of the LAEs versus the NB\civ\
excess ($i'-$NB\civ). We note there seems to be a weak trend for higher 
redshift objects to have larger narrow-band excess.
However, large photometric errors 
and several non-detections in the $i'$-band 
prevent us to reach a meaningful conclusion.
Finally, we include \zlbg\ LAEs from SDF
and find very good agreement with the distribution
of our sample.

In summary, the distribution in redshift of the 
spectroscopic sample of LAEs is centred at $z\sim5.7$ and
no strong relation is found between 
the redshift and the narrow-band photometry of the
LAE sample. Moreover, SDF LAEs at \zlbg\ show a distribution 
similar to our sample in the $z$ versus NB and $z$ versus ($i'-$NB) planes
(Figure \ref{f:z-color-mag}).

\section{The large-scale environment of \civ\ absorption systems}\label{s:discussion-large-scale}

This section presents the spatial distribution of the complete LAE photometric sample
based on spectroscopic information.

\begin{figure*}
\begin{minipage}{150mm} 
\centering 
\includegraphics[width=145mm]{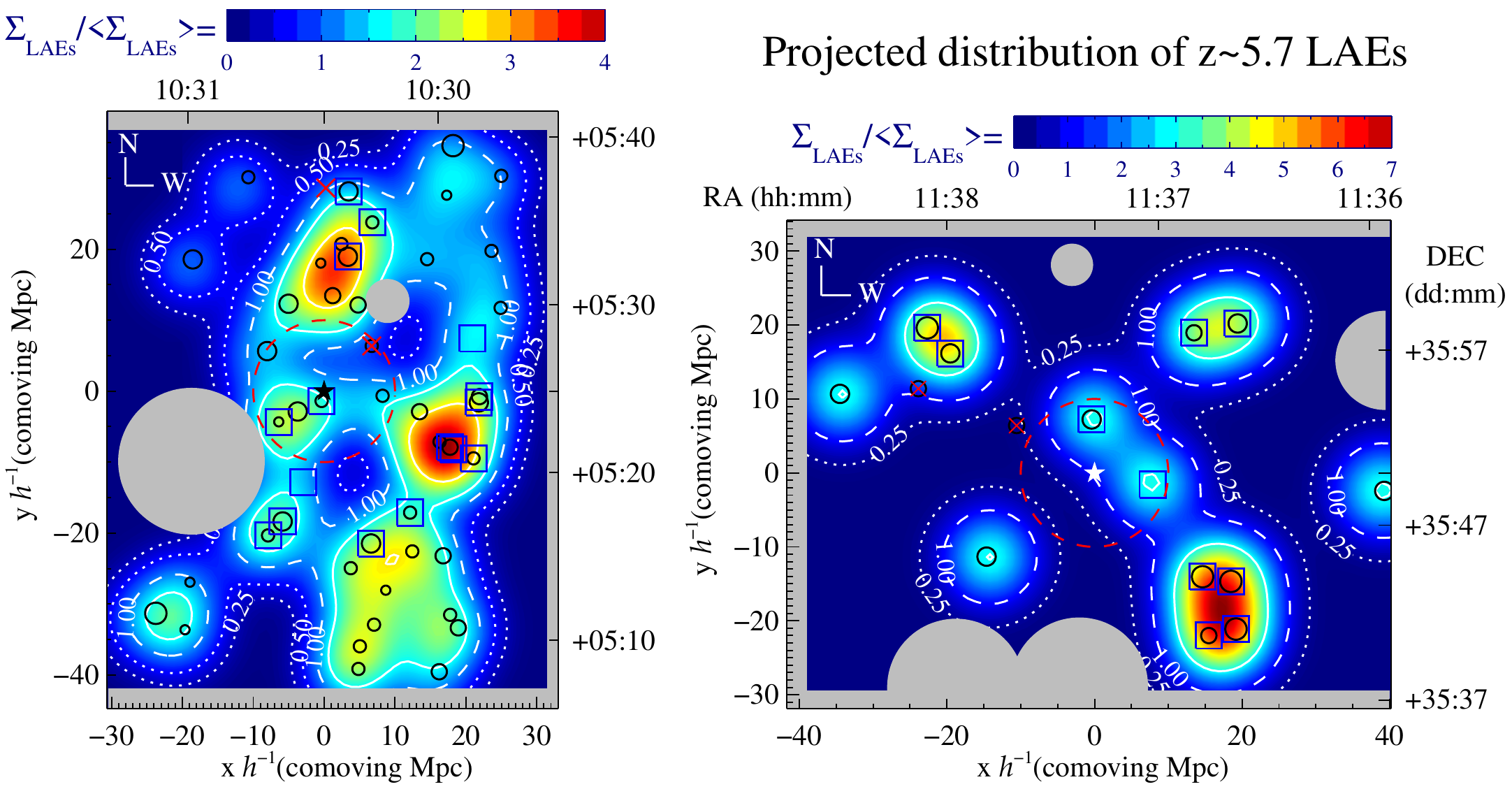} 
\caption{\small 
Projected distribution of LAEs in the field J1030+0524 ({\it left}) 
and the field J1137+3549 ({\it right}).
Open circles are photometric candidates and
the size of each circle indicates the apparent magnitude bin
with larger circles for brighter magnitudes. 
Open squares are confirmed by spectroscopy (S$_{w,10\%} \geq 3.0$)
and red crosses are non-LAEs (S$_{w,10\%}<3.0$).
The star symbol indicates the line of sight to the \civ\ systems
and the the red dashed circle centred on the star symbol
has a radius of $10{\it h}^{-1}$ comoving Mpc.
Masked areas of the field (bright Galactic stars and edges of 
the CCD) are shaded grey.
The colour-coded contours correspond to constant levels of 
surface density contrast $\Sigma/ \langle \Sigma \rangle$,
obtained using 1${\it h}^{-1}$ comoving Mpc bin size 
and a Gaussian smoothing kernel with FWHM of $10{\it h}^{-1}$ 
comoving Mpc. 
Dotted contours correspond to under-dense 
regions ($\Sigma / \langle \Sigma \rangle<1$), 
solid contours correspond to over-dense regions 
($\Sigma / \langle \Sigma \rangle>1$), 
and dashed contours correspond to mean 
density regions ($\Sigma / \langle \Sigma \rangle=1$).} 
\label{f:maps_specLAE}
\end{minipage}
\end{figure*}

Figure \ref{f:maps_specLAE} shows the projected distribution of LAEs in the 
fields J1030+0524 (left hand panel) and J1137+3549 (right hand panel)
in comoving Mpc with respect to the \civ\ lines of sight, indicated by the
black star and white star, respectively.
Photometric LAE candidates are represented by open circles 
whose sizes indicate the NB\civ\ magnitude in the range 23.0--26.0 mag,
in bins of 0.5 mag, with smaller circles for fainter objects.
Squares indicate spectroscopic confirmation of LAEs (S$_{w,10\%} \geq 3.0$),
and although most of them are photometric candidates (circles),
the few squares without a circle correspond to object that are
not included in the photometric sample because the S/N in the NB\civ\
band is slightly below 5. These objects are included in the analysis of the density contrasts.
Finally, the red crosses are contaminants (S$_{w,10\%}<3.0$),
which are not included in the density contrast.

Following the analysis in Paper I, 
the contours in Figure \ref{f:maps_specLAE}
correspond to constant levels of density contrast 
$\Sigma_{\text{LAE}}/\langle \Sigma \rangle_{\text{LAE}}$,
where $\langle \Sigma \rangle_{\text{LAE}}$ is the mean 
surface density averaged over the size of the field and
$\Sigma_{\text{LAE}}$ is the surface density of LAEs obtained using 1${\it h}^{-1}$ 
comoving Mpc bin size and a Gaussian smoothing kernel with FWHM$\,=10{\it h}^{-1}$ 
comoving Mpc.
Solid contours correspond to over-dense regions,
dashed contours correspond to mean density regions
and dotted contours correspond to under-dense regions.

In the field J1030+0524 (left hand panel), the contours include all the photometric 
LAE candidates plus the spectroscopy-only LAEs.
The spectroscopic campaign confirms the projected structure 
reported in Paper I. The line of sight to the \civ\ system is found in
a projected over-density of LAEs of 
$\Sigma_{\text{LAE}}(10)/\langle \Sigma \rangle_{\text{LAE}}\sim1.5$,
where the surface density within 10${\it h}^{-1}$ 
comoving Mpc radius is $\Sigma_{\text{LAE}}(10)=89{\times}10^{-3}$ gal./arcmin$^2$,
and the mean surface density of the field is 
$\langle \Sigma \rangle_{\text{LAE}}=58{\times}10^{-3}$ gal./arcmin$^2$.
In the field J1137+3549 (right hand panel), 
the distribution reported in Paper I is also confirmed.
After removing
contaminants, the surface density of the field 
is $\langle \Sigma \rangle_{\text{LAE}}=16{\times}10^{-3}$ gal./arcmin$^2$.
Within 10${\it h}^{-1}$ comoving Mpc 
the number of LAEs is
so $\Sigma_{\text{LAE}}(10)=36{\times}10^{-3}$ gal./arcmin$^2$.
As a result, the revised density contrast increased to
$\Sigma_{\text{LAE}}(10)/\langle \Sigma \rangle_{\text{LAE}}\sim2.2$.

\subsection{Distribution of \zlae\ LAEs in the line of sight}

\begin{figure}
\centering
\includegraphics[width=85mm]{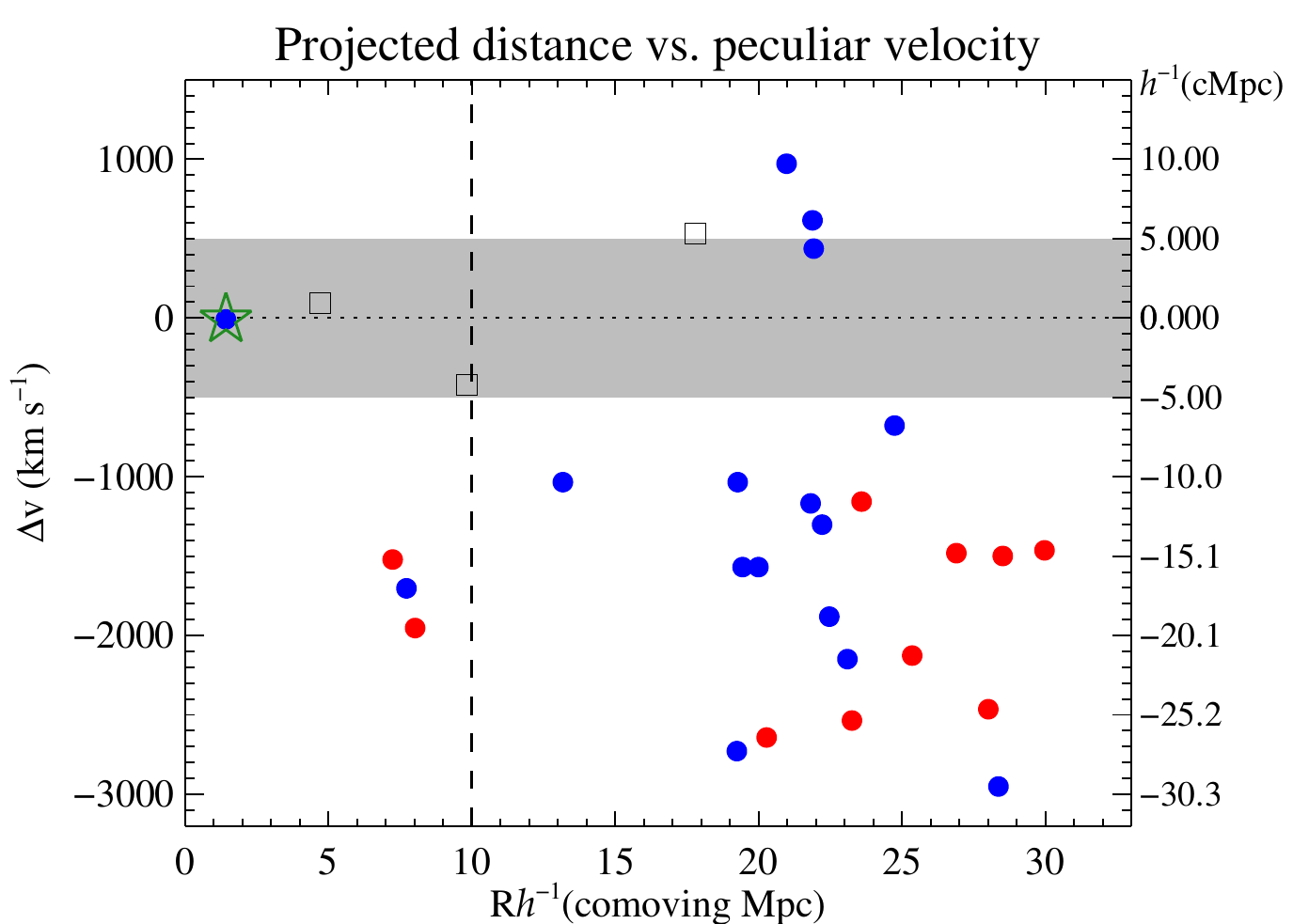} 
\caption{\small Projected distance to the \civ\ system
measured from the position of the corresponding \civ\ line of sight,
versus radial velocity measured respect to the \civ\ system at 
$z_{\text{\civ}}=5.7242$ for J1030+0524 LAEs
and the \civ\ system at $z_{\text{\civ}}=5.7383$ for J1137+3549 LAEs.
Blue and red circles correspond to confirmed 
LAEs in the field J1030+0524 and J1137+3549, respectively.
The only confirmed LAE within ${\pm}500$\,\kms\ (star symbol) is  
the closest object at $212.8h^{{-1}}$ physical kpc.
Open squares are LAEs observed with the 600ZD grating in J1030+0524.
Although the data prevent a solid confirmation/rejection of these LAE candidates,
we note that the position of the emission line put two of them within 
${\pm}500$\,\kms\ from the \civ\ at $z_{\text{\civ}}=5.7242$.} 
\label{f:dist-vel}
\end{figure}

The projected distribution of LAEs 
shows an over-density towards each \civ\ line of sight.
However, spectroscopic follow-up suggests that
the photometric LAE sample is at a slightly lower redshift
than the \civ\ system in the J1137+3549 field (Figure \ref{f:spec-redshift-LAE}).
This result is clear in Figure \ref{f:dist-vel}
showing the projected distance (R) versus 
the radial velocity ($\Delta v$)
to the corresponding \civ\ system 
in each field.
Red solid circles and blue solid circles 
are spectroscopic LAEs in the field J1137+3549
and the field J1030+0524, respectively.
Open squares are LAE candidates in the field J1030+0524 
observed with the 600ZD grating (assuming that the emission is \Lya).
Projected distances are measured from the position 
of the corresponding \civ\ line of sight, and radial velocities 
are measured respect to the \civ\ system at 
$z_{\text{\civ}}=5.7242$ for J1030+0524 LAEs
and the \civ\ system at $z_{\text{\civ}}=5.7383$ for J1137+3549 LAEs.

The LAEs in the field J1137+3549 are at 
17.9${\pm}6h^{-1}$ comoving Mpc in the line of sight direction
($\langle z\rangle_{\text{LAEs}}=5.698{\pm}0.013$,
$\langle \Delta v\rangle = {-}1784.6{\pm}585.9$\,\kms)
from the \civ\ system.
Figure \ref{f:dist-vel} shows that all the red circles
are at $\,>1000$\,\kms\ from the \civ\ system in the field.
Thus, the photometric LAE sample 
is not in the immediate environment of the \civ\ system at $z=5.7383$.
Consequently, we cannot test the idea that the \civ\ system 
is associated with LAEs with our current data. Instead it
should be tested with NB photometry better centred at 
the wavelength of the \Lya\ emission at $z=5.7383$.
Nevertheless, our spectroscopic findings presented here 
do not negate the results based on photometry presented in Paper I
because the narrow redshift range is very useful
for the interpretation of the comparison of projected 
distributions of galaxies.
In addition, no \civ\ system is found in the QSO's
spectrum at the redshift of the LAE sample.
Since the closest LAE is at $\sim7$ transverse comoving 
Mpc to the line of sight region without \civ ,
this observation does not contradict the idea that
\civ\ systems are preferentially detected 
near LAEs, as supported by the result in J1030+0524 where 
an LAE-\civ\ system pair has been confirmed.

In the field J1030+0524, we find that, except for one, all spectroscopically confirmed
LAEs are at ${\simgt}5h^{-1}$ comoving Mpc from the \civ\ system
in the line of sight direction ($|\Delta v | \,\simgt\,500$\,\kms).
This LAE is represented by the green open star symbol
and is at the same redshift as the high-column density 
\civ\ absorption system in the field. This galaxy-\civ\ 
system pair is presented with more detail in the next section.
The mean redshift of the sample is $\langle z\rangle_{\text{LAEs}}=5.702{\pm}0.025$
($\langle \Delta v\rangle = {-}989{\pm}1131$\,\kms), which
shows that the sample is slightly biased towards lower redshifts.
However, this value does not include the 600ZD grating data
plotted as open squares. 
Interestingly, two of these objects are within $10h^{-1}$ comoving Mpc
of projected distance and $\pm500$\,\kms\ of radial velocity.
Although the main spectroscopic data set does not show an over-density in the
line of sight direction at the position of the \civ\ system,
if the tentative detections on the low
resolution data are confirmed to be \Lya , then three out of the five
LAEs within $10h^{-1}$ projected comoving Mpc are in the near
environment of the \civ\ absorption ($|\Delta v |\,\simlt\,500$\,\kms).
Considering the depth sampled by the LAE selection ($\sim40h^{-1}$ comoving Mpc),
the confirmation of the low resolution data will imply
an over-density at $10h^{-1}$ comoving Mpc scale 
of $(3/10)/(5/40)\sim2.4$, where five LAEs are detected in  
$\sim40h^{-1}$ comoving Mpc depth
of which three are within $\sim10h^{-1}$ comoving Mpc depth.
In conclusion, if the emission lines in the low resolution data 
of the field J1030+0524 are found to be \Lya , it will confirm
that the \civ\ system is associated with an excess of LAEs.

\section{Spectroscopic confirmation of an LAE-\civ\ absorption system pair at \zlae}\label{s:lae-civ-pair}

One of the motivations to study the fields J1030+0524 and J1137+3549 
is the search for galaxies physically connected to the \civ\ absorption 
systems.
In this work, we report the redshift of 
LAE 103027+052419 (\#4 in Figure \ref{f:phot-spec-LAE-1030})
to be $z_{\text \Lya}=5.724\pm0.001$ measured at 
the position of the bluest pixel of the emission line
with S/N$\,=1$, as a proxy for the centre of the intrinsic \Lya\ line.
This confirms that the LAE is at
212.8$h^{-1}$ physical kpc 
from a high column density \civ\ absorption system at 
$z_{\text{\civ,b}}= 5.7242\pm0.0001$.

\begin{figure}
\centering 
\includegraphics[width=85mm]{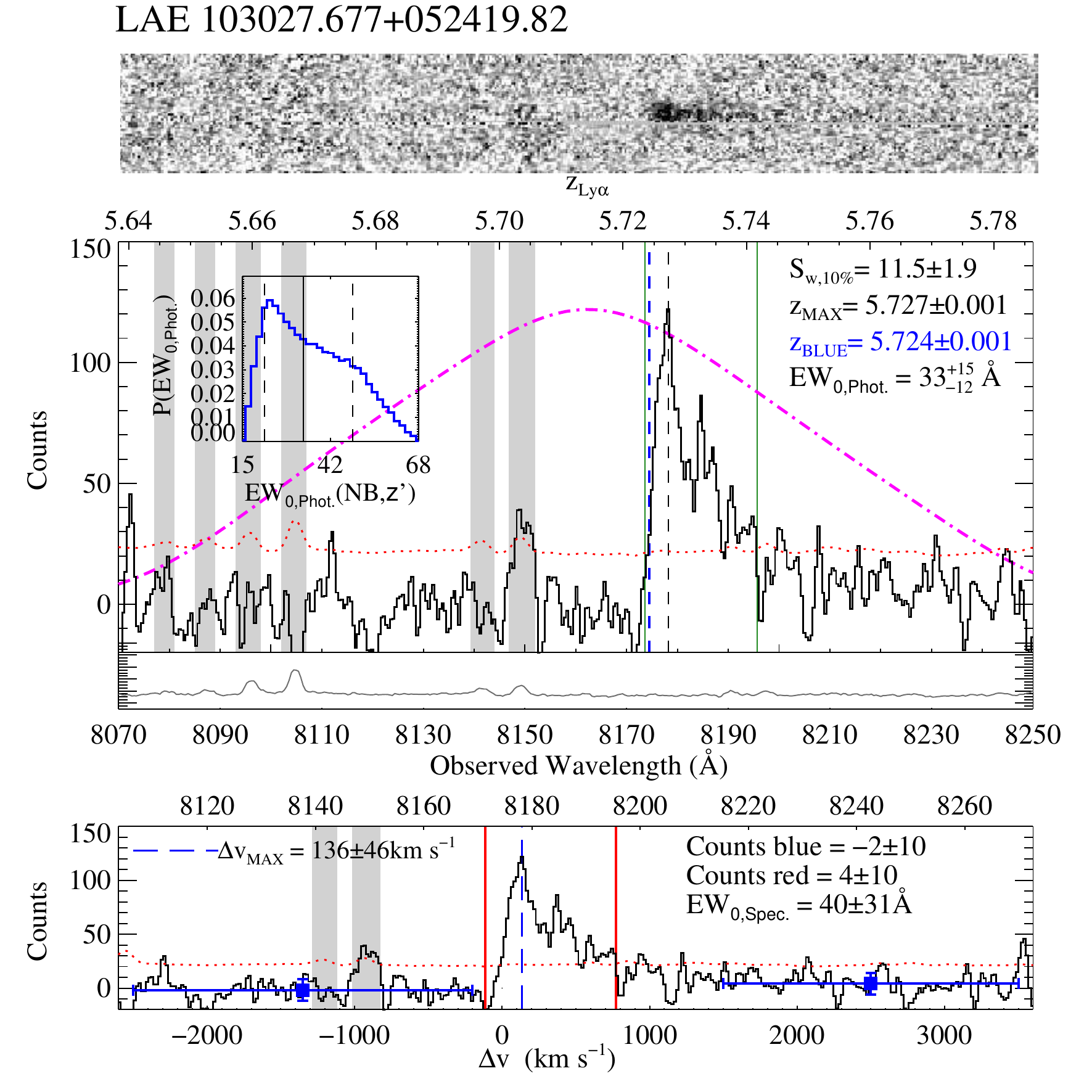} 
\caption{\small {\it Top:} Snapshot of the 2D slit spectrum 
in the wavelength range covered by the NB\civ\ filter.
{\it Middle:} 1D spectrum of the emission line. 
The red dotted line indicates one standard deviation
of the counts in the extraction box (1$\sigma$ error), 
the magenta dot-dashed line is the scaled transmission curve of the NB\civ\ filter, 
the vertical dashed lines indicate the position of the maximum (black line)
and the bluest pixel (blue line), and the grey shaded areas show
the wavelength of the skylines where residuals can be large.
The sky spectrum in arbitrary scale is given for reference in the bottom plot.
The weighted skewness is measured in the range contained
by the vertical green solid lines that indicate $\lambda_{10\%,\text B}$ and 
$\lambda_{10\%,\text R}$.
{\it Inset:} Equivalent width probability distribution obtained from
the photometry and the redshift of the object (see Section \ref{s:eqw}).
The solid line and the dashed lines indicate the central value 
of the distribution (33\,\AA) and the range enclosing a probability of 0.68,
respectively.
{\it Bottom:} Velocity with respect to the bluest pixel of the emission.
The mean count level on the red side of the emission line
is slightly higher than the mean value on the blue side, 
which is consistent with the presence of a flux decrement 
due to the \Lya\ forest.
However, large errors make the values consistent with each other 
and with zero.
The vertical solid lines indicate the wavelength range where
the spectroscopic equivalent width was measured.} 
\label{f:LAE04a}
\end{figure}

Figure \ref{f:LAE04a} presents a closer view of the spectrum of 
LAE 103027+052419. 
The emission line shows a clear asymmetry that is quantified 
by a weighted skewness of S$_{w,10\%}= 11.5{\pm}1.9$. 
The vertical green solid lines (middle panel) indicate the wavelength 
range $\lambda_{10\%,\text B}$--$\lambda_{10\%,\text R}$
used to measure S$_{w,10\%}$.
The maximum of the emission is indicated by the black vertical 
dashed line at 8178\,\AA , which corresponds 
to $z_{\text{MAX}}=5.727{\pm}0.001$.
However, the blue edge of the \Lya\ emission is very sharp 
whereas the red wing extends for at least 500\,\kms\ 
suggesting that the line centre is bluer than the observed
maximum of the emission.

\begin{figure}
\centering 
\includegraphics[width=85mm]{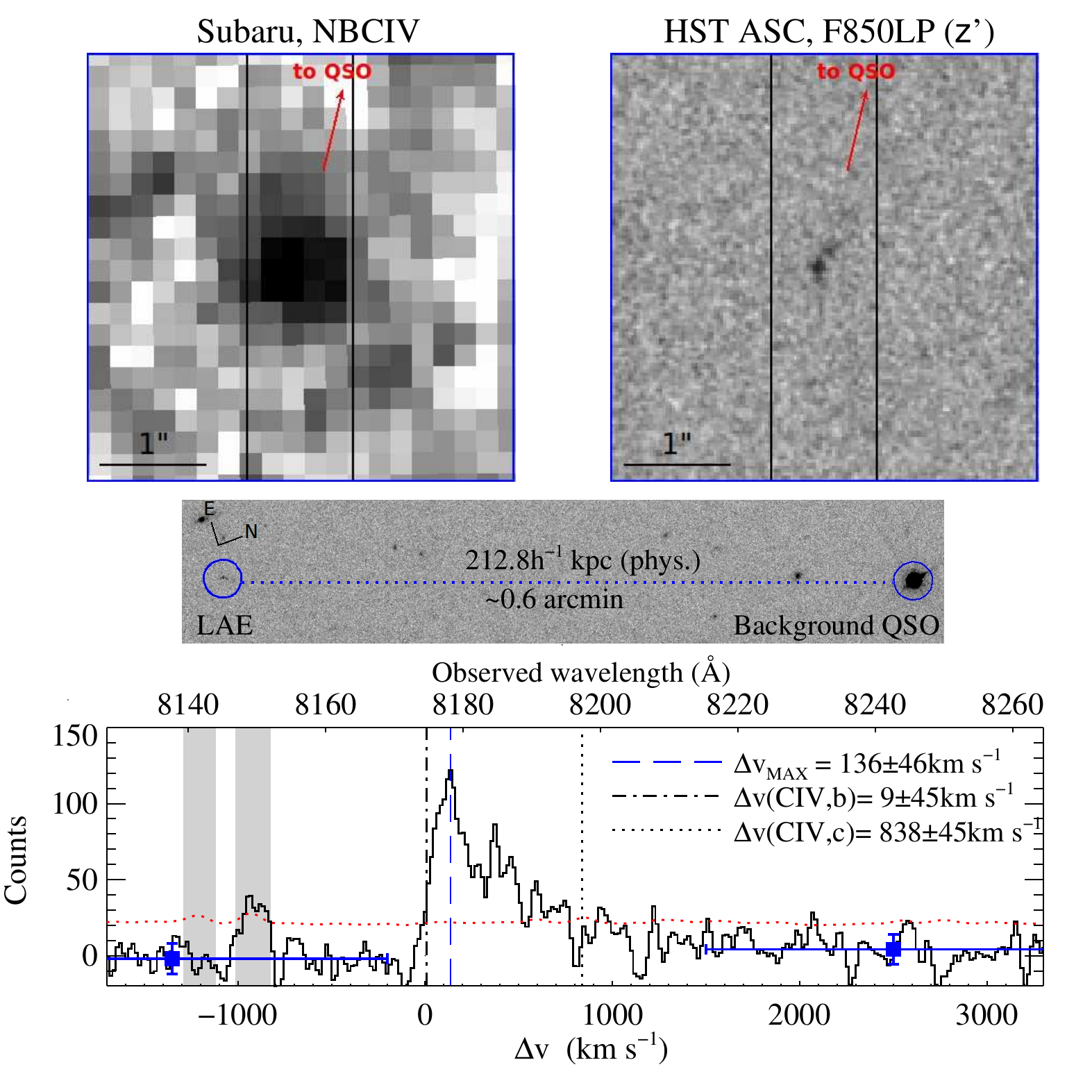}
\caption{\small {\it Top:} Thumbnails (4$\times$4 arcsec) of LAE 
103027+052419 in NB\civ\ with Suprime-Cam on the Subaru 
Telescope ({\it left}) and F850LP ($z'$) with ACS 
onboard HST ({\it right}). 
The black lines are the edges of the slit projected 
on the sky. The direction towards the background QSO is indicated 
with red arrows. The galaxy is distorted and elongated in the direction
towards the background QSO.
{\it Middle:} ACS image in the $z'$-band showing the projected distance 
from LAE 103027+052419 to the QSO line of sight. 
{\it Bottom:} Spectrum in \kms\ with respect to the bluest pixel
above 1$\sigma$. The vertical dashed line indicates the position of 
the emission peak, and the vertical dotted line and dot-dashed line
show the velocity of the two \civ\ systems 
at $z_{\text{\civ,b}}=5.7242$ and $z_{\text{\civ,c}}=5.7428$ \citep{ryan-weber2009,
simcoe2011b, dodorico2013}.} 
\label{f:LAE04b}
\end{figure}

The rest-frame equivalent width of \Lya\ 
obtained from the photometry is EW$_0=33^{+15}_{-12}$\,\AA , 
which is the lowest EW$_0$ of the LAE sample in the field 
J1030+0524 (see Section \ref{s:eqw-distribution}). 
This value is consistent with the spectroscopic measurement of EW$_0$.
In the bottom panel of Figure \ref{f:LAE04a}, the blue squares 
with error bars at each side of the emission line show the mean 
count level of the continuum, where the vertical error bars 
indicate one standard deviation of the flux in counts and
the horizontal error bars enclose the wavelength range used to 
estimate the mean. At both sides of the emission, we avoided 
regions of significant skyline residuals (grey areas).
We measure F$_{\lambda 1210}={-}2{\pm}10$ counts
for the blue side of the emission,  
which is consistent with zero.
For the red side of the emission, 
the flux level is F$_{\lambda 1225}=4{\pm}10$ counts, which is 
also consistent with zero.
Nevertheless, we estimate EW$_{0,\text{Spec}}$ of the observed emission 
within the vertical solid lines using the continuum level 
F$_{\lambda 1225}$ 
and we find EW$_{0,\text{Spec}}=40{\pm}31$\,\AA , where the dominant 
source of error is the uncertainty in the continuum.

The \Lya\ emission peak has an offset of
$\Delta v_{\text{MAX}}=136{\pm}45$\,\kms\ from 
the redshift defined by the observed blue edge of the line.
This velocity shift is smaller than the typically observed 
velocity offset of \Lya\ with respect to the systemic redshift 
of $z\sim2$--3 LAEs:
$\langle \Delta v (\Lya)\rangle=217{\pm}29$\,\kms\
\citep{hashimoto2013}. 
However, these two velocities are not directly comparable.
On one hand, the velocity of \Lya\ with respect to the systemic 
redshift measured from nebular emission
is interpreted as evidence for gas motion in the galaxy 
\citep[e.g.][]{mclinden2011, hashimoto2013}.
On the other hand, our measurement of $\Delta v_{\text{MAX}}$ 
is used as an indicator of the `sharpness' of the blue edge 
of the emission, meaning that smaller $\Delta v_{\text{MAX}}$ 
corresponds to line profiles with a steeper blue side.
This is affected by a combination of several factors
likely dominated by 
(a) the neutral hydrogen content in the IGM
in the line of sight to the LAE, 
(b) \HI\ intrinsic to the LAE,
and (c) outflowing gas.
Although we cannot differentiate between these effects
without knowing the systemic redshift, they are probably 
related since galactic scale outflows can affect the distribution 
of the neutral gas which regulates the escape fraction 
of \Lya\ photons. Thus, although $\Delta v_{\text{MAX}}$ 
does not directly probe the speed of the outflowing gas, 
it is interesting to see how it compares in relation to the 
rest of the sample. 
This is discussed in Section \ref{s:dv}.

Figure \ref{f:LAE04b} presents 4$\times$4 arcsec thumbnails of 
LAE 103027+052419 (top panels), showing the slit used for 
spectroscopy and the direction to QSO J103027+052455. 
The image in the F850LP band (which equates to $z'$) 
obtained with the Advanced Camera for Surveys (ACS)
onboard the {\it Hubble Space Telescope} 
(right hand panel) shows a distorted shape slightly elongated towards 
the background QSO.
The close proximity of this LAE to the line of sight of the \civ\ absorption
systems is shown in the middle panel of the figure, and the velocity of 
the two \civ\ systems near the galaxy are shown in the bottom panel. 
The dot-dashed line corresponds to the second \civ\ absorption system 
at $z\geq5.5$ in this line of sight at $z_{\text{\civ,b}}= 5.7242{\pm}0.0001$ 
($\log$\nciv$\,= 14.52{\pm}0.08$, \citealt{dodorico2013}),
with $\Delta v (\text{\civ}_b)=9$\,\kms .
The dotted line corresponds to \civ$_c$, at 
$z_{\text{\civ,c}}= 5.7428{\pm}0.0001$ 
 \citep[$\log$\nciv$\,= 13.08{\pm}0.1$,][]{dodorico2013}, 
with $\Delta v (\text{\civ}_c)=838$\,\kms .

In summary, the spectroscopic redshift of LAE 103027+052419 
confirms that this galaxy is the closest neighbour to the 
highest column density \civ\ absorption system known to date
at $z \geq 5.5$.
In addition, the red side of the emission line is very extended
and reaches the wavelength corresponding to \Lya\ at the 
redshift of the \civ$_c$ absorption system.
This finding demonstrates that star-forming galaxies 
can be detected in close proximity to highly ionized 
metal absorption systems immediately after the EoR,
leading the path to exploring in better detail the origin 
and the physical state of the absorbing gas.

\section{Equivalent width}\label{s:eqw}

\subsection{Measures of the \Lya\ emission line at \zlbg }\label{s:eqw-measure}

The faint continuum of \zlae\ LAEs is the main obstacle 
to obtaining precise spectroscopic measurements of EW$_0$.
Therefore, EW$_0$ is usually measured using narrow-band 
and broad-band photometry \citep[e.g:][]{malhotra2002, hu2004, shimasaku2006, zheng2014}.
On one hand, if the broad-band overlaps with the narrow-band 
(e.g. $i'$ and NB\civ), an estimate of the flux of the continuum
requires the subtraction of the contribution of the \Lya\
emission and a correction for the IGM transmission, 
which are usually both unknown.
On the other hand, if the broad-band is redder than 
the narrow-band containing \Lya\ (e.g. $z'$ and NB\civ), 
then no IGM correction and no \Lya\ correction are needed
to estimate the continuum flux.
In this work, we have tested both methods and the approaches
by several authors.
We find that using a non-overlapping broad-band tends to give
stable results that are independent of the method, whereas
using an overlapping broad-band is more sensitive to the corrections 
applied to the continuum flux density.
Therefore, we adopt the EW$_0$ obtained from the z' and 
NB\civ\ using the following procedure.

The emission line flux was estimated from the NB\civ\ band and the
continuum flux density was measured with the $z'$-band at $\lambda_{\text{rest}}\sim1350$\,\AA .
We follow the approach suggested by \citet{zheng2014} 
and use the equations:
\begin{equation}
\label{eq:ew-N}
\frac{N}{W_N}=a_N(\lambda)\times \frac{F_{\text \Lya}}{W_N}+b_N(\lambda)\times  \frac{F_{\text \Lya}}{\text{EW}_{0}\times(1+z)}
\end{equation}
\begin{equation}
\label{eq:ew-Z}
\frac{Z}{W_Z}=b_Z\times  \frac{F_{\text \Lya}}{\text{EW}_{0}\times(1+z)},
\end{equation}
where $Z$ and $N$ are the integrated fluxes in the $z'$-band 
and the NB\civ\ band, respectively; and $W_N$ and $W_Z$ 
are the corresponding filter width 
$W_N=\int T_N\delta \lambda / max(T_N)$
and  $W_Z=\int T_Z\delta \lambda / max(T_Z)$, 
with $T_Z$ and $T_N$ representing the transmission curve 
of each filter. The observed \Lya\ flux is $F_{\text \Lya}$ and 
the continuum flux density
was replaced by $F_{\text \Lya}/(\text{EW}_{0}\times(1+z))$.
Because the narrow-band filter is not a top-hat, the contribution of 
the \Lya\ line to the NB\civ\ flux depends on the filter transmission 
at the wavelength of the emission. 
The coefficient $a_N(\lambda)$ is a correction for this effect
and is estimated as the ratio of the filter transmission at the 
wavelength of the emission line over the maximum transmission 
at the centre of the filter $a_N=T_N(\lambda_{\text{MAX}})/max(T_N)$.
The \Lya\ emission is not covered by the $z'$-band, thus the first
term of equation \ref{eq:ew-N} is not present in equation \ref{eq:ew-Z}.
The correction for the IGM absorption to the continuum emission 
in the NB\civ\ and $z'$-bands is applied through $b_N$ and $b_Z$.
Since, the broad-band is redder than \Lya , no correction is 
necessary in $z'$, thus $b_Z=1$.
Considering that most of our LAEs are non-detected in R$_c$,
we assume that the contribution from the continuum bluer than 
\Lya\ ($\lambda<\lambda_{\text{BLUE}}$) is negligible to flux measured by the narrow-band.
Under this assumption $b_N$ is calculated as the fractional 
area of the NB\civ\ transmission curve for 
$\lambda \geq \lambda_{\text{BLUE}}$:
$b_N=\int^{\infty}_{\lambda_{\text{BLUE}}} T_N \delta\lambda / \int T_N\delta \lambda$,
where $\lambda_{\text{BLUE}}$ is the wavelength of the bluest pixel
of the emission line.
Finally, our measurements of the EW$_0$ only account for 
the observed \Lya\ flux since no correction for IGM absorption 
to the \Lya\ emission was applied. In summary, 
our approach uses the wavelength of the emission line
to estimate the corrections for individual objects.

The expression for EW$_{0}$ from equations \ref{eq:ew-N} and \ref{eq:ew-Z} is
\begin{equation}
\label{eq:ew}
\text{EW}_{0,{\rm Phot.}}=\frac{N\times W_Z - Z \times b_N \times W_N}{Z\times a_N} \times \frac{1}{(1+z)}.
\end{equation}

The equivalent width probability function P(EW$_0$) 
of each LAE was obtained by Monte Carlo simulation.
Values for the $z'$-band magnitude were pulled from
a Gaussian distribution centred at the measured magnitude 
with a standard deviation equal to the magnitude error $\delta z'$.
The same process was used to sample the NB\civ\ magnitude, 
and for each pair of values we calculate EW$_0$. 
After 10$^5$ iterations, P(EW$_0$) is obtained and from it 
we calculate the mean and the errors in EW$_0$.
Following \citet{shimasaku2006}, the central value EW$_c$ 
is defined as
$\int^{{\rm EW}_c}_{0} P(\text{EW}_0) \delta \text{EW}_0 = 0.5$,
and the $1\sigma$ upper and lower limits that contain 
68 per cent of the distribution are estimated from 
$\int^{{\rm EW}_c}_{{\rm EW}_-} P(\text{EW}_0) \delta \text{EW}_0 = 0.34$,
 and $\int^{{\rm EW}_+}_{{\rm EW}_c} P(\text{EW}_0) \delta \text{EW}_0 = 0.34$, respectively.
For objects non-detected in $z'$, lower limits in the the EW$_{0}$ 
are calculated by adopting the 1$\sigma$ limiting magnitude.
These measurements are reported in Tables \ref{t:spec-LAE-J1030} 
and \ref{t:spec-LAE-J1137} in Appendix \ref{app:spec-tables}. 
The equivalent width probability function of each LAE 
is plotted in Appendix \ref{app:spec-thum}.

\begin{figure}
\centering 
\includegraphics[width=85mm]{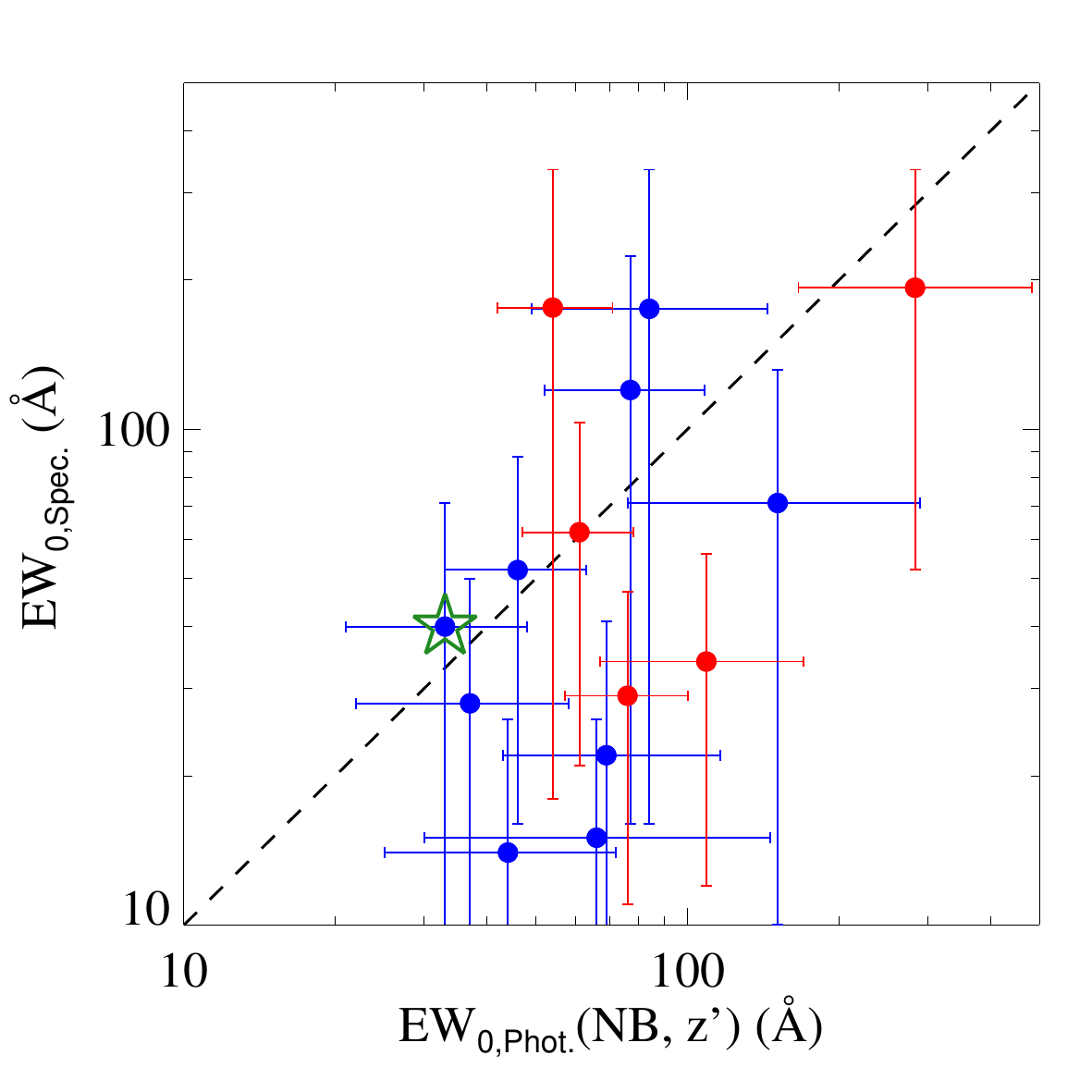} 
\caption{\small Comparison of the spectroscopic and
photometric estimates of the rest-frame equivalent width (EW$_0$)
of nine LAEs in the field J1030+0524 (blue circles) and five
LAEs in the field J1137+3549 (red circles). Spectroscopy
only provides upper limits for 
the remainder LAEs in the sample, which are
not included in this figure. The dashed line is the identity relation 1:1.
Points that depart from the identity towards lower spectroscopic values
can result from the subtraction of sky emission lines affecting \Lya .
The green star is LAE 103027+052419 (Figure \ref{f:LAE04a}) and 
is one of the examples showing agreement in
the EW$_0$ measured with both methods.}
\label{f:comparison-EW}
\end{figure}

In the few examples where it was possible to measure 
EW$_0$ from the spectrum, 
we find that the agreement with 
the photometric estimates using the $z'$ magnitude 
for the continuum is better than using the $i'$ magnitude. 
Figure \ref{f:comparison-EW} compares the photometric 
and spectroscopic measurements of EW$_0$ of 14 LAEs. 
The vertical error bars are dominated by the uncertainty in 
the continuum flux density. In particular, some spectra are 
contaminated by light from 
other slits (shown in Figure \ref{f:allslit}), which is a source 
of error for the continuum determination. Moreover,
the subtraction of sky emission lines can 
have a significant impact in the properties of a \Lya\ emission 
line by modifying the skewness, changing the position of the 
emission peak and diminishing the measured line flux. 
Thus, it is possible that the lower values measured 
from the spectra of the objects below the $x=y$ line (dashed line)
are biased by these effects.
Having said that, although EW$_{{\rm 0, Spec}}$ 
have large errors, we interpret our results in Figure \ref{f:comparison-EW} 
as evidence that the photometric determination of EW$_0$ using a 
non-overlapping broad-band filter returns values that are supported 
by spectroscopy. Nevertheless, we recognise the need for deeper 
spectroscopy to consistently measure the properties of \zlbg\ LAEs.

\subsection{Distribution of EW$_0$}\label{s:eqw-distribution}

\begin{figure}
\centering 
\includegraphics[width=85mm]{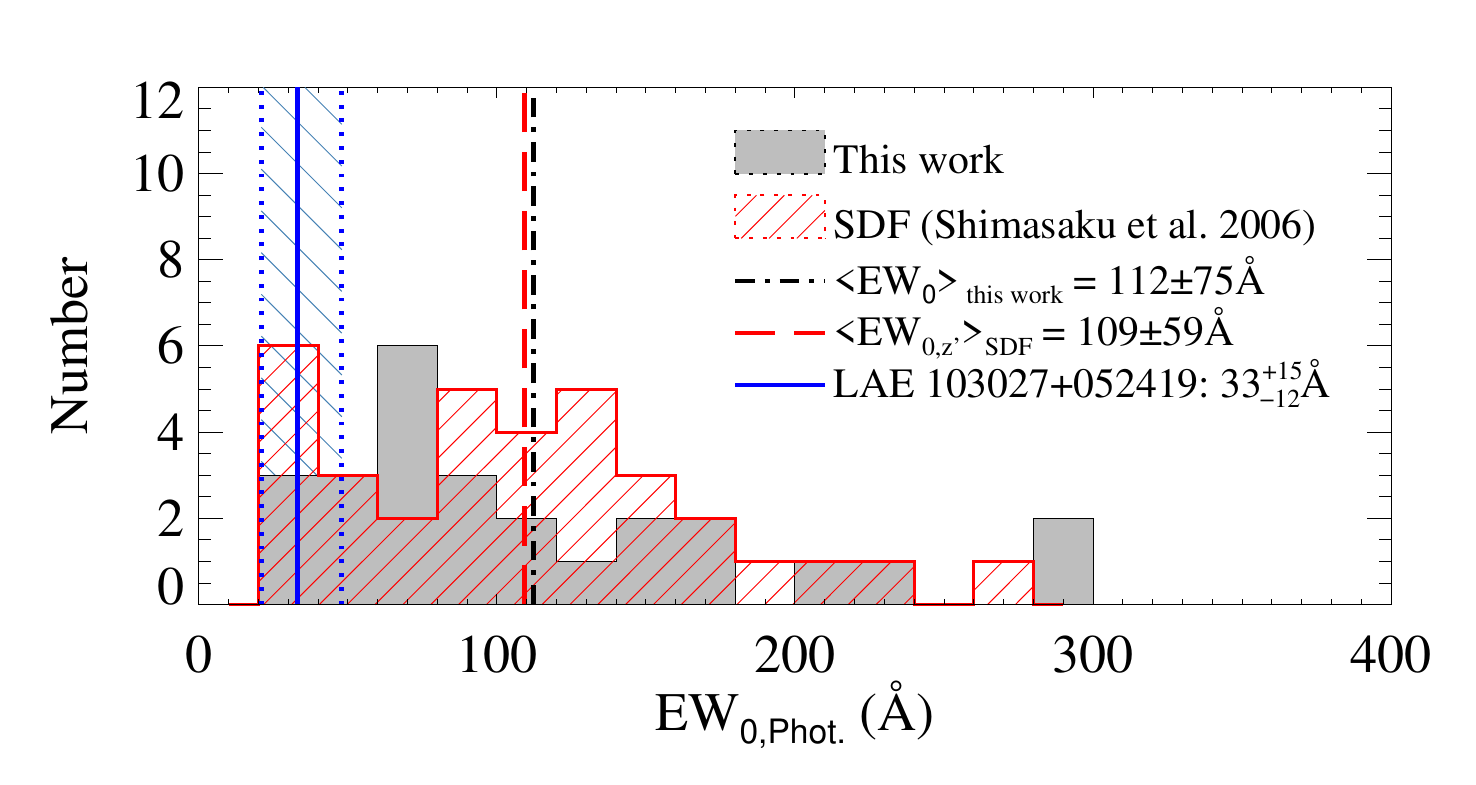} 
\caption{\small Rest-frame equivalent width distribution 
of confirmed LAEs in both fields (grey filled histogram).
The red line hatched histogram corresponds to SDF LAEs
\citep{shimasaku2006}.
The vertical lines indicate the mean value of each distribution,
long-dashed for SDF LAEs and dot-dashed for our sample.
The blue vertical line and line-filled region indicate EW$_0$
and 1$\sigma$ uncertainty
of LAE 103027+052419, which is clearly at the low end of the distribution.}
\label{f:EW-distribution}
\end{figure}

Figure \ref{f:EW-distribution} presents the EW$_0$ distribution 
of the LAE sample of this work (grey histogram) and
the LAE sample from SDF \citep[][red-dashed histogram]{shimasaku2006}.
We find agreement in the mean and the standard deviation 
of both samples. The equivalent width of LAE 103027+052419 is
represented by a solid blue line and a blue-hashed region 
enclosing 68 per cent of the P(EW$_0$).
It is clear from this figure that LAE 103027+052419 
has a lower EW$_0$ than the rest of the sample.
This is interesting considering that lower redshift LAEs ($z=2$--3)
with lower EW$_0$ values tend to display larger velocity shifts
of the \Lya\ emission peak with respect to the systemic redshift 
\citep[e.g.][]{hashimoto2013}.
Moreover, the same effect is observed among LBGs 
\citep[e.g.][]{adelberger2003, shapley2003}.
The implications of this result are discussed in Section 
\ref{s:discussion-origin-CIV}.

\begin{figure}
\centering 
\includegraphics[width=85mm]{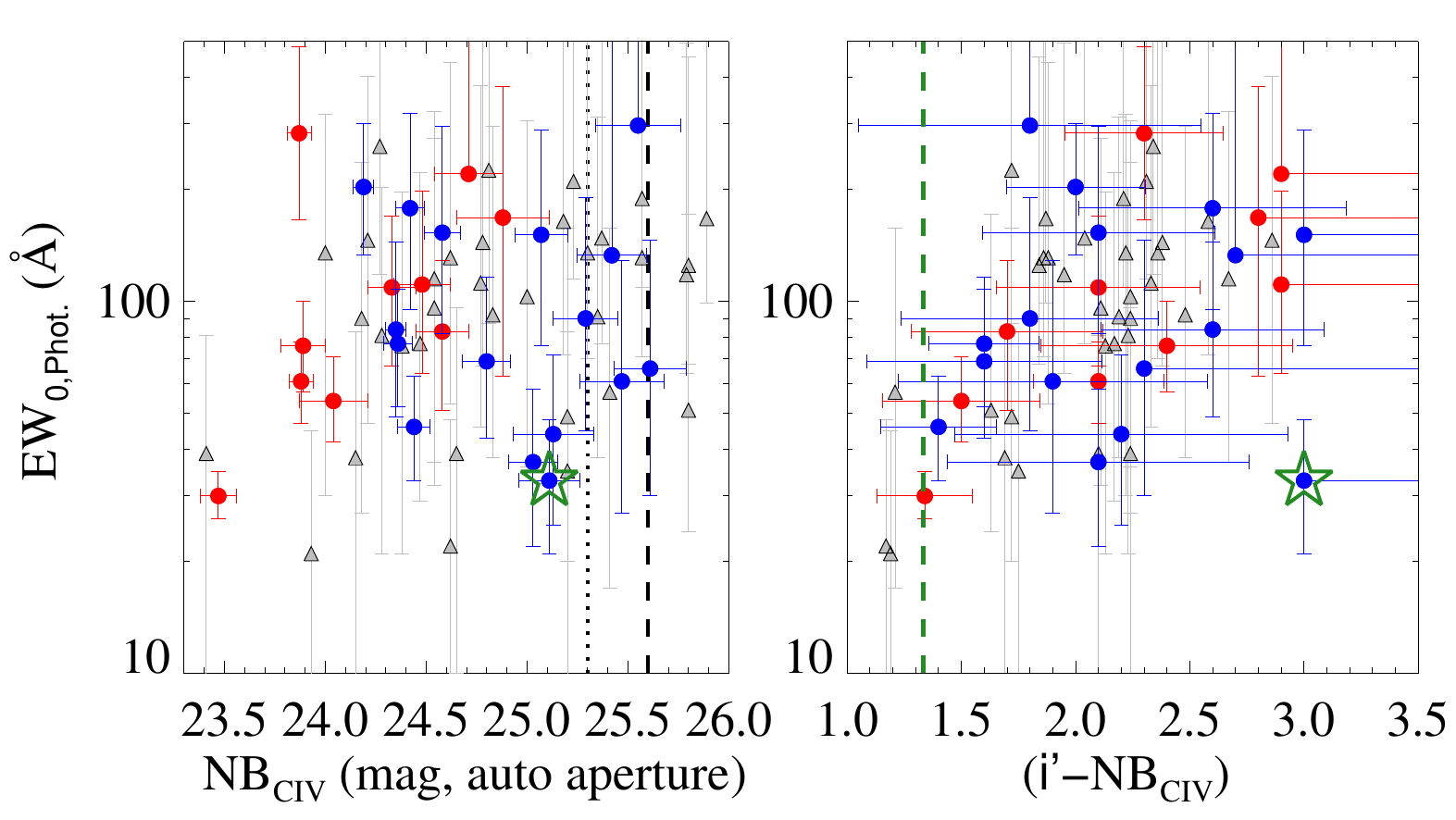} 
\caption{\small Narrow-band photometry versus EW$_0$ of 
confirmed LAEs. Blue and red circles correspond to the two 
fields J1030+0524 and J1137+3549, respectively.
Triangles are SDF LAEs. 
LAE 103027+052419 is highlighted with a star symbol.
{\it Left:} The magnitude in NB\civ\ does not correlate with EW$_0$.
{\it Right:} The excess ($i'-$NB\civ) shows a small trend with EW$_0$
in which LAEs with higher EW$_0$ have redder ($i'-$NB\civ) colour.
This is also seen in the SDF sample.
Considering that both quantities are independent probes of the 
flux excess with respect to the continuum, the existence of this relation 
is reasonable and encouraging. 
However, large uncertainties limit the detection of a
significant correlation.}
\label{f:EW-magcol}
\end{figure}

We compare the EW$_0$ with the NB\civ\ magnitude and 
($i'-$NB\civ) colour of \zlae\ LAEs in Figure \ref{f:EW-magcol}.
Blue and red circles correspond to the samples of the present 
work and grey triangles correspond to SDF LAEs.
In the left panel, we find no trend between EW$_0$
and NB\civ\ magnitude since both LAE samples are 
well distributed across the plot. This is not a surprise 
because the NB\civ\ magnitude measures the flux of 
the emission line and part of the continuum while EW$_0$ 
is a relative measure of the flux in the emission line 
with respect to the continuum. 
A rough estimate of this relation is the ($i'-$NB\civ) colour which is slightly larger 
for LAEs with larger EW$_0$ (right panel of Figure \ref{f:EW-magcol}).
The open star symbol is LAE 103027+052419, and 
it seems to occupy a curious corner in the plot. 
This LAE is not detected in $i'$, so the colour is a 
lower limit. Usually, other non-detections in $i'$
have larger EW$_0$ because they are also very 
faint in $z'$, what brings them into the relation.
However,  LAE 103027+052419 is brighter in $z'$ than the
other LAEs with lower-limit ($i'-$NB\civ) colours,
resulting in a lower EW$_0$ which is closer 
to the expectation for LBGs.

\begin{figure}
\centering 
\includegraphics[width=85mm]{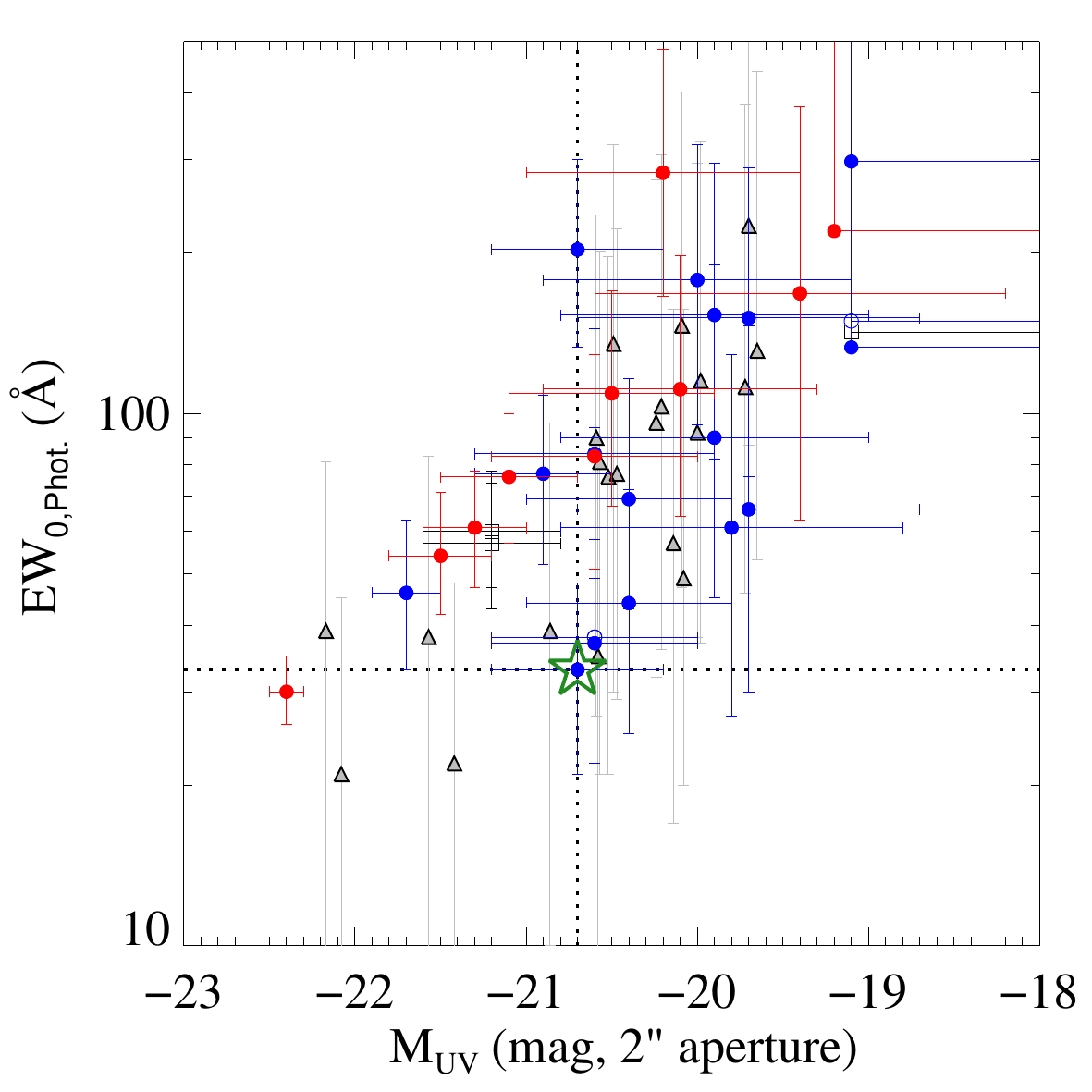} 
\caption{\small Absolute magnitude at $\lambda_{\text{rest}}\sim1350$\,\AA\ 
and EW$_0$ of confirmed LAEs.
As in previous figures, blue and red circles correspond to the fields
J1030+0524 and J1137+3549, respectively, 
and triangles are SDF LAEs.
Both samples seem to follow the same relation
where more luminous objects tend to have smaller EW$_0$.
LAE 103027+052419 is highlighted with a star symbol
and the dotted lines (vertical and horizontal) 
show that fainter LAEs (i.e. to the right of LAE 103027+052419)
have larger EW$_0$ (i.e. above LAE 103027+052419).}
\label{f:EW-Muv}
\end{figure}

This is clearly seen in Figure \ref{f:EW-Muv} which presents
M$_{\text{UV}}$ versus EW$_0$. Symbol key is the same as per
previous figures. The open star representing LAE 103027+052419
is in the middle-bottom of the distribution and seems to agree with the
trend previously noted by many authors that fainter UV galaxies
at $z=5$--6 have larger EW$_0$ \citep{ando2006, shimasaku2006,
stanway2007, ouchi2008}.
The remaining LAEs with lower-limit ($i'-$NB\civ) colours are at the 
top-right corner of the Figure, that corresponds to large EW$_0$ and 
faint M$_{\text{UV}}$.
The horizontal dotted line shows that only one LAE in our sample 
has a lower EW$_0$ than LAE 103027+052419.

In summary, the distribution of EW$_0$ is comparable with
the sample of \zlae\ LAEs from SDF, 
and the closest neighbour to \civ$_b$ (LAE 103027+052419)
has one of the lowest EW$_0$, which is well determined by 
photometry and spectroscopy. 

\section{Velocity of the \Lya\ emission peak}\label{s:dv}

Many authors have found that the wavelength of the \Lya\ emission
peak tends to be shifted from the systemic redshift of a galaxy,
which is measured from nebular emission lines 
\citep{adelberger2003, steidel2010, finkelstein2011, mclinden2011, hashimoto2013, chonis2013, shibuya2014b}.
The velocity shift of the \Lya\ emission is interpreted as the
result of the dynamics of the gas. 
In this study, we have no measurements of the systemic redshift 
from rest frame optical nebular lines, thus we have adopted 
the redshift of the bluest pixel of the \Lya\ emission as 
the best approximation to the real redshift of the galaxy. 
This section presents \vmax , which is the velocity 
of the pixel with the maximum flux 
with respect to the redshift of the galaxy,
and shows how it compares with other observables of LAEs
such as the asymmetry of the emission line profile, 
the equivalent width and the UV luminosity.

\begin{figure}
\centering 
\includegraphics[width=85mm]{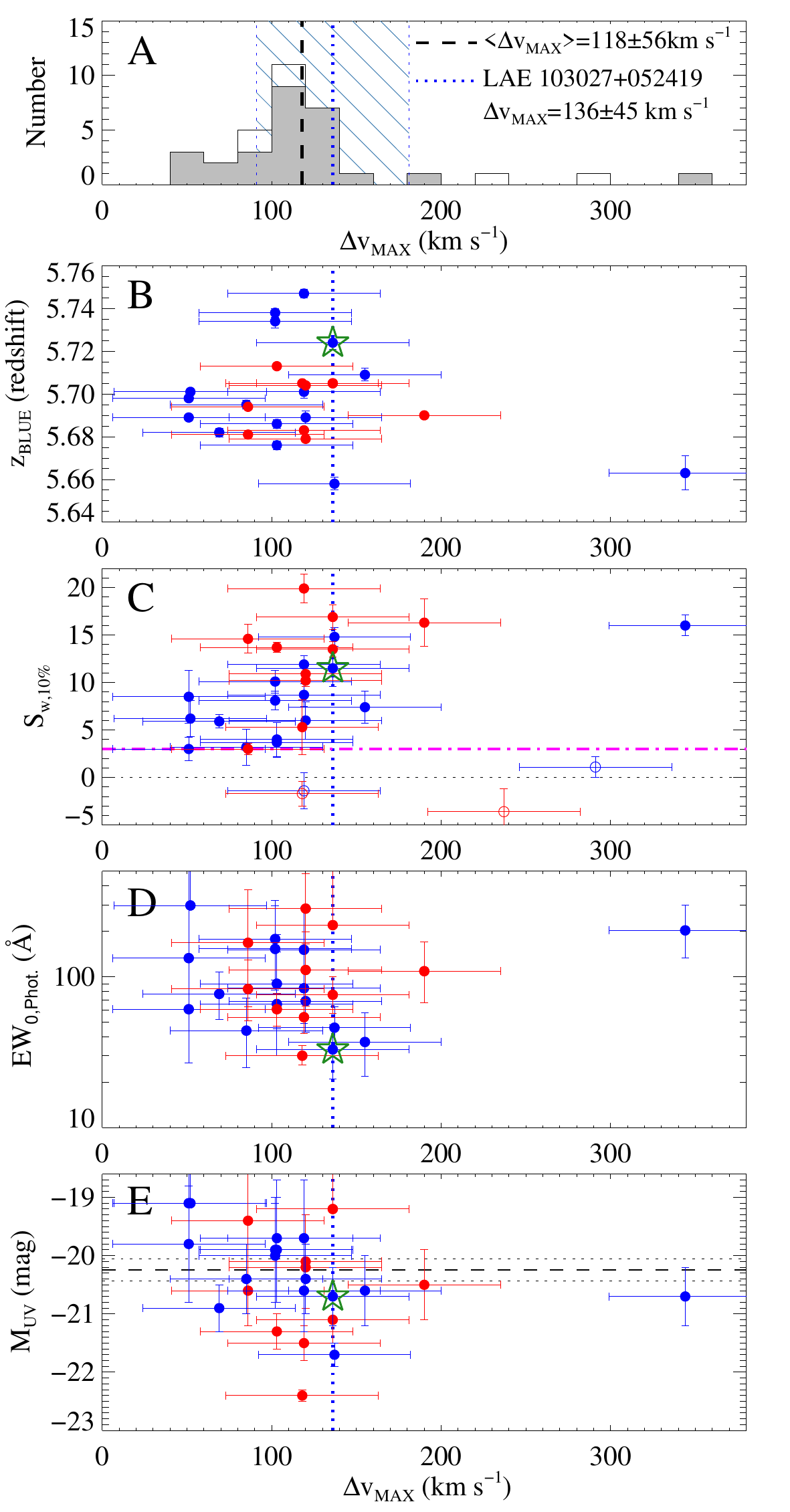} 
\caption{\small 
The velocity of the \Lya\ emission maximum 
is plotted against other properties.
The errors in \vmax\ are conservatively large as they
were estimated from the FWHM of the data in the 
dispersion direction (2.5--3.5\,\AA).
Blue and red circles correspond to the fields
J1030+0524 and J1137+3549, respectively, and 
LAE 103027+052419 is highlighted with a star symbol.
{\bf (A)} Distribution of \vmax\ including both fields. The complete 
sample of candidates is plotted with the solid line histogram and
the sample of confirmed LAEs is plotted with a grey filled histogram.
The dotted vertical line is LAE 103027+052419 and is found
in the high velocity half of the distribution. 
{\bf (B)} \vmax\ versus redshift shows no correlation.
{\bf (C)} \vmax\ versus S$_{w,10\%}$.  A small trend might be present 
in which high skewness LAEs have larger \vmax\
than low skewness LAEs. 
{\bf (D)} \vmax\ versus EW$_0$ shows that LAE 103027+052419 
has one of the lowest EW$_0$ and is among the highest \vmax .
{\bf (E)} \vmax\ versus M$_{\text{UV}}$ shows that 
LAE 103027+052419 is close to M$^{\star}_{\text{UV}}(z=6.0)$ 
(horizontal dashed line)
and galaxies with higher \vmax\ have similar M$_{\text{UV}}$.}
\label{f:dv-LAE}
\end{figure}

Starting with the distribution of \vmax , the top panel of 
Figure \ref{f:dv-LAE} shows 
that the mean velocity of the LAE sample is 
$\langle$\vmax$\rangle = 118{\pm}56$\,\kms .
The velocity shift measured in LAE 103027+052419 is 
\vmax$ = 136{\pm}45$\,\kms\ 
which is higher than the mean of the sample but within 
one standard deviation.
The rest of the section is dedicated to the possible dependancies 
of \vmax\ with other observables plotted in Figure \ref{f:dv-LAE}.
The symbol key is the same we have used before, with the
green star symbol representing LAE 103027+052419.
The errors in \vmax\ are
very conservative since they are
determined from the resolution of the 
observations (FWHM$=2.5$\,\AA\ for the 830G data and 3.5\,\AA\ 
for the 600ZD data, Section \ref{s:observation-specLAE}).

The second panel from the top presents \vmax\ versus 
redshift and no correlation is seen. This is not surprising 
since the factors that regulate the \Lya\ profile 
(e.g. speed of the gas, column density of the neutral gas, etc.)
are not expected to evolve dramatically 
in the small redshift range of the LAE sample ($\sim20$\,Myr).
Moreover, if there was a trend with redshift it would 
suggest that \vmax\ is not dominated by gas dynamics but by some
observational systematic effect, which is not the case.

The resonant nature of \Lya\ implies 
that radiative transfer effects coupled with the dynamics of the gas
result in spatial and wavelength diffusion of \Lya\ photons 
\citep{hansen2006, zhengzheng2010, dijkstra2012, jeesondaniel2012}.
In general, red-shifted \Lya\ photons find it easier to 
reach the observer than blue-shifted \Lya\ photons
because they are not absorbed by the \HI\ 
in the direction to the observer
whereas the latter do.
Furthermore, if the gas has an additional velocity component
caused by some type of outflow mechanism,
the \Lya\ photons reprocessed in the receding side of the 
outflow will also experience an additional redshift
that will take them out of resonance allowing them to reach 
the observer \citep{ahn2003, verhamme2006, dijkstra2010}.
In this case, the observed line profile presents a red `tail'
--that may also show some structure or `bumps'--
which increases its asymmetry.
However, at fixed outflow velocities, larger \HI\ column densities
produce broader \Lya\ profiles and larger
velocity shifts of the \Lya\ peak \citep[e.g. Figure 1 of ][]{verhamme2014}
which would reduce the asymmetry of the line profile.

In the middle panel of Figure \ref{f:dv-LAE} (labeled C), 
there is a weak suggestion that the weighted skewness defined 
in Section \ref{s:spectroscopic-LAEs-contamination} 
increases with \vmax . 
Considering that S$_{w,10\%}$ measures the asymmetry 
of the line profile, this observable increases for a more
extended `red wing' and sharper `blue edge'.
Because \vmax\ is a measurement of the extent of the blue edge
of the emission, a smaller \vmax\ correspond to a
sharper blue edge.
Therefore, a smaller \vmax\ should result in a more 
asymmetric profile, i.e. larger S$_{w,10\%}$.
However, the opposite relation is observed
since S$_{w,10\%}$ increases with \vmax .
This could indicate that the extension of the red wing may correlate, although
slightly, with the shift of the \Lya\ peak to longer wavelengths.

As mentioned before, since outflowing gas would enhance the 
extension of the red wing of the emission, then a galactic wind
would increase S$_{w,10\%}$ with respect to the case without it.
In addition, larger \HI\ column densities produce greater redward
shifts of the \Lya\ peak, although reducing the asymmetry of the profile.
The tentative trend in panel (C) indicates that larger 
\HI\ column densities (larger \vmax) might also host stronger
galactic outflows (larger S$_{w,10\%}$).
Under this picture, since LAE 103027+052419
has a large \vmax\ and one of the
a largest weighted skewness of the sample,
it is possible that this galaxy contains a stronger outflow
and a larger \HI\ column density
compared to the rest of the LAE sample.

Panel D shows \vmax\ versus EW$_0$ where LAE 
103027+052419 is among the lowest EW$_0$ and the highest
\vmax . This is in agreement with lower redshift observations 
that LAEs and LBGs with higher EW$_0$ tend to have lower
\Lya\ velocity shifts \citep[e.g.][]{shapley2003, shibuya2014b}.
However, the scatter shown by our data
prevents us from identifying any trend.

Finally, the bottom plot (panel E) presents M$_{\text{UV}}$ 
versus \vmax\ and shows that 
the galaxies with the largest \vmax\ values
have similar M$_{\text{UV}}$ to LAE 103027+052419 
and close to (or larger than) 
M$^{\star}_{\text{UV}}$ 
\citep[$-20.24\pm0.19$ mag,][horizontal dashed line]{bouwens2007}.
Once again, this result is in agreement with statistics of outflows
in lower redshift galaxies which have found
larger wind speed in more massive galaxies 
\citep[e.g.][]{martin2005, weiner2009, bradshaw2013}.
Furthermore, LAEs at $z\sim2$--3 show smaller velocity
offset than bright LBGs 
\citep{mclinden2011,mclinden2014,chonis2013,hashimoto2013,shibuya2014b}.
Interestingly, if more massive star-forming galaxies 
at \zlbg\ contain larger \HI\ column densities, then
it is logical to expect that, statistically, UV brighter 
galaxies (i.e. more massive) have larger \vmax , lower
EW$_0$ and stronger outflows which result in more
asymmetric profiles (larger S$_{w,10\%}$).

In summary, LAE 103027+052419 presents one of the largest
values of \vmax\ in the spectroscopic sample of LAEs.
Although \vmax\ is not a direct measure of the outflow speed and
the errors in \vmax\ are rather large, the tentative detection of 
trends with S$_{w,10\%}$, EW$_0$ and M$_{\text{UV}}$
could reflect a dependence with the gas dynamics
and/or the mass of the galaxy.
As a result, the position of LAE 103027+052419 
in the plots of Figure \ref{f:dv-LAE} suggest that this 
LAE is a good candidate to host galactic-scale outflows 
among the LAEs in the sample.

\section{The origin of \civ\ absorption systems after the epoch of reionization}\label{s:discussion-origin-CIV}

In this section we discuss the possibility that $\civ_{b}$
is physically associated with LAE 103027+052419.
The proximity in redshift confirms that these two object are close
neighbours, thus the final question explored in this work is
the possibility that the galactic scale wind of LAE 103027+052419
enriched its environment out to ${\simgt}212h^{-1}$ physical kpc.

\subsection{Evidence from the \Lya\ emission}
The distribution and kinematics of the gas in a galaxy 
affects the escape of \Lya\ photons from the galaxy and
the \Lya\ line profile \citep{hansen2006, verhamme2006, verhamme2008, 
dijkstra2010, zhengzheng2010, orsi2012, garel2012, behrens2014}. 
In Figure \ref{f:LAE04b}, 
the two \civ\ absorption systems at $\sim$212$h^{-1}$ projected kpc
show small velocities with respect to the \Lya\ emission
line of LAE 103027+052419.
The strongest \civ\ system known at $z \geq 5.5$, \civ$_b$, 
has a column density $\log$\nciv$= 14.52{\pm0.08}$ and is 
found at 9\,\kms\ from the LAE.
The other \civ\ absorption system at a similar redshift, \civ$_c$,
is weaker ($\log$\nciv$= 13.08{\pm}0.1$) and is found at 
838\,\kms\ from the LAE, which is not too far from 
the red end of the \Lya\ emission. 

The \Lya\ peak of LAE 103027+052419
is at  $\Delta v_{\text{MAX}}=136{\pm}45$\,\kms .
Moreover, it has one of the lowest EW$_0$ values (Section \ref{s:eqw}) and
one of the highest $\Delta v_{\text{MAX}}$ values (Section \ref{s:dv})
of the sample of LAEs in this study, which is in agreement 
with the observed anti-correlation between 
EW$_0$ and $\Delta v_{\text \Lya}$ 
in star-forming galaxies like LBGs and LAEs 
at $z\sim2$--3 \citep[e.g.][]{shapley2003, hashimoto2013, shibuya2014b}.
In other words: larger velocity shifts are found among lower EW$_0$.
Radiative transfer studies \citep[e.g.][]{verhamme2008, verhamme2014}
show that a higher neutral hydrogen column density
results in a larger velocity shift of the \Lya\ peak.
This is valid for outflow velocities ${\simlt}300$\,\kms\
since larger velocities will allow \Lya\ photons to escape more easily
closer to the line centre, thereby reducing $\Delta v_{\text \Lya}$.
Therefore, a low $\Delta v_{\text \Lya}$ could result from a
low column density of \HI\ or a large outflow speed,
whereas a large $\Delta v_{\text \Lya}$ implies not only
larger \HI\ column densities but also lower outflow speeds.

This result is in agreement with \citet{zhengzheng2010} who find
the radiative transfer of \Lya\ photons to be dominated by 
the density and velocity structures
of the gas in the CGM and IGM. As a result, 
the wavelength of the \Lya\ peak tends to shift
to longer wavelengths in more massive dark matter haloes.
Given that more luminous galaxies at \zlbg\ are more massive
\citep{gonzalez2011,mclure2011},
the trend for UV brighter LAEs
to have larger $\Delta v_{\text{MAX}}$
shown in the bottom panel of Figure \ref{f:dv-LAE}
is consistent with this picture.
Therefore, according to the UV luminosity, \Lya\ profile, 
\vmax\ and EW$_0$, it is possible
that LAE 103027+052419 is at the high mass end of the LAEs,
has a large \HI\ column density, and is a good candidate
for a large scale outflow.

\subsection{The outflow scenario}

\begin{figure}
\centering 
\includegraphics[width=85mm]{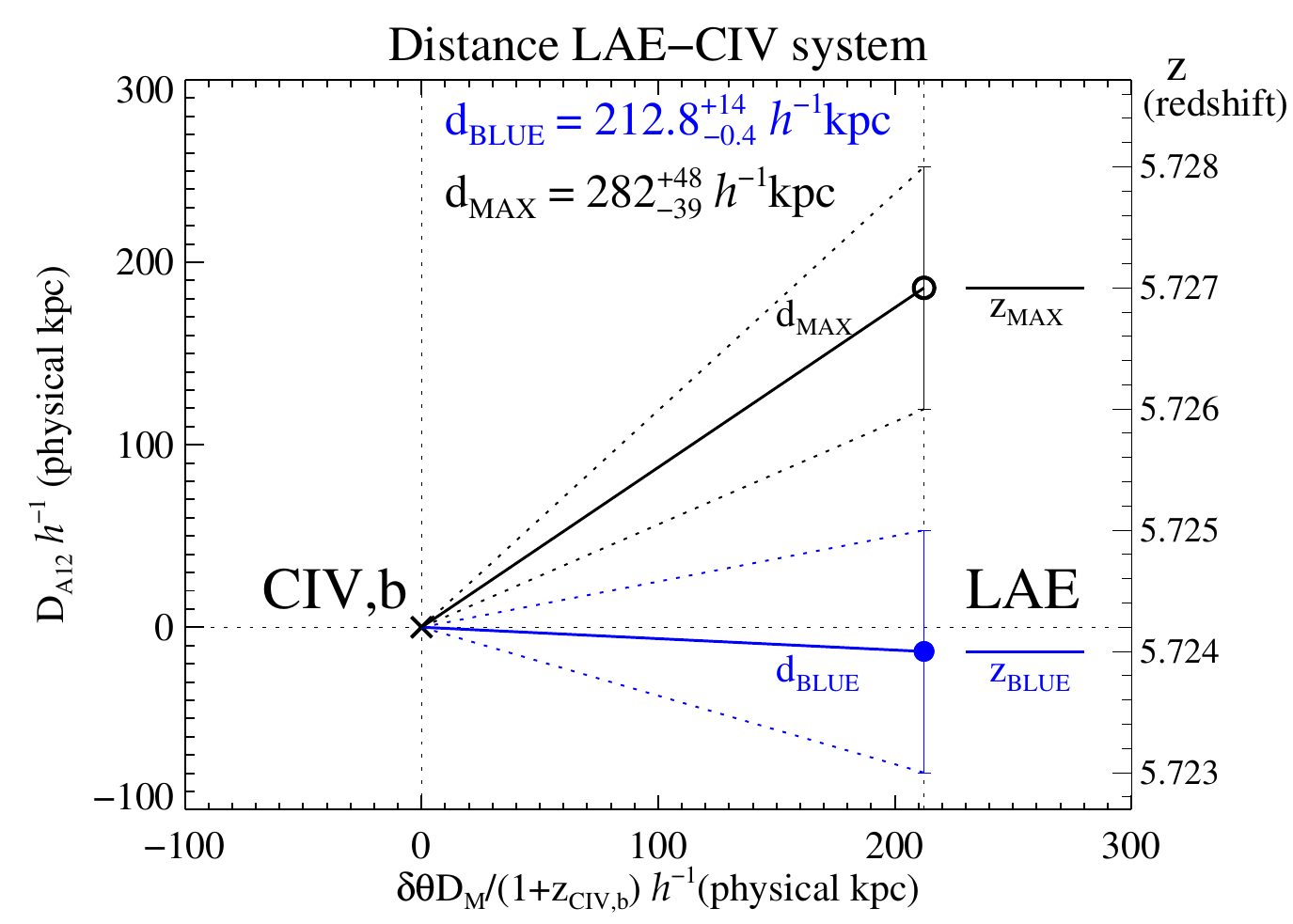} 
\caption{\small 
Distance from \civ ,b to LAE 103027+052419.
The $x$ axis is the transverse distance 
$\delta\theta$D$_{\text{M}}/(1+z_{\text{\civ,b}})$
where $\delta\theta=36.2412$ arcsec is the angular separation 
on the sky.
The $y$ axis is the angular diameter distance D$_{\text{A}}$ between 
$z_{\text{\Lya}}$ and $z_{\text{\civ,b}}$.
The black open circle corresponds to $z_{\text{\Lya}}=z_{\text{MAX}}$ 
and the blue solid circle corresponds to $z_{\text{\Lya}}=z_{\text{BLUE}}$.
The uncertainty on $z_{\text{\civ,b}}$ is an order of magnitude
better than that of the LAE.}
\label{f:trans-pos}
\end{figure}

Although the speed of the outflowing gas cannot be measured 
without the detection of nebular emission lines and 
ISM absorption lines in the continuum of the galaxy,
the distance to the absorption system and the limit imposed by the 
age of the Universe can help us to understand the origin of the metals.
The separation of the LAE-\civ\ system is estimated from
the angular separation on the sky ($\delta\theta=36.2412$ arcsec)
and the corresponding redshifts as
$\text{d}=\sqrt[]{D^2_{\text{A}}(z_{\text{\Lya}},z_{\text{\civ,b}})+
(\delta\theta D_{\text{M}}/(1+z_{\text{\civ,b}}))^2}$,
where D$_{A}(z_{\text{\Lya}},z_{\text{\civ,b}})$ is the angular diameter 
distance between the redshifts of the LAE and the \civ\ system
and $\delta\theta$D$_{\text{M}}$ is the transverse comoving distance
at the redshift of the \civ\ system. Figure \ref{f:trans-pos} shows
a diagram of the distance in the LAE-\civ\ system pair
assuming the redshift $z_{\text{BLUE}}$ and $z_{\text{MAX}}$ 
and their uncertainties. 
The Figure clearly shows that the distance estimated from $z_{\text{BLUE}}$
is close to the minimum distance, which is important 
when considering the time needed for an outflow to reach
the position of the \civ\ system.

\begin{figure}
\centering 
\includegraphics[width=85mm]{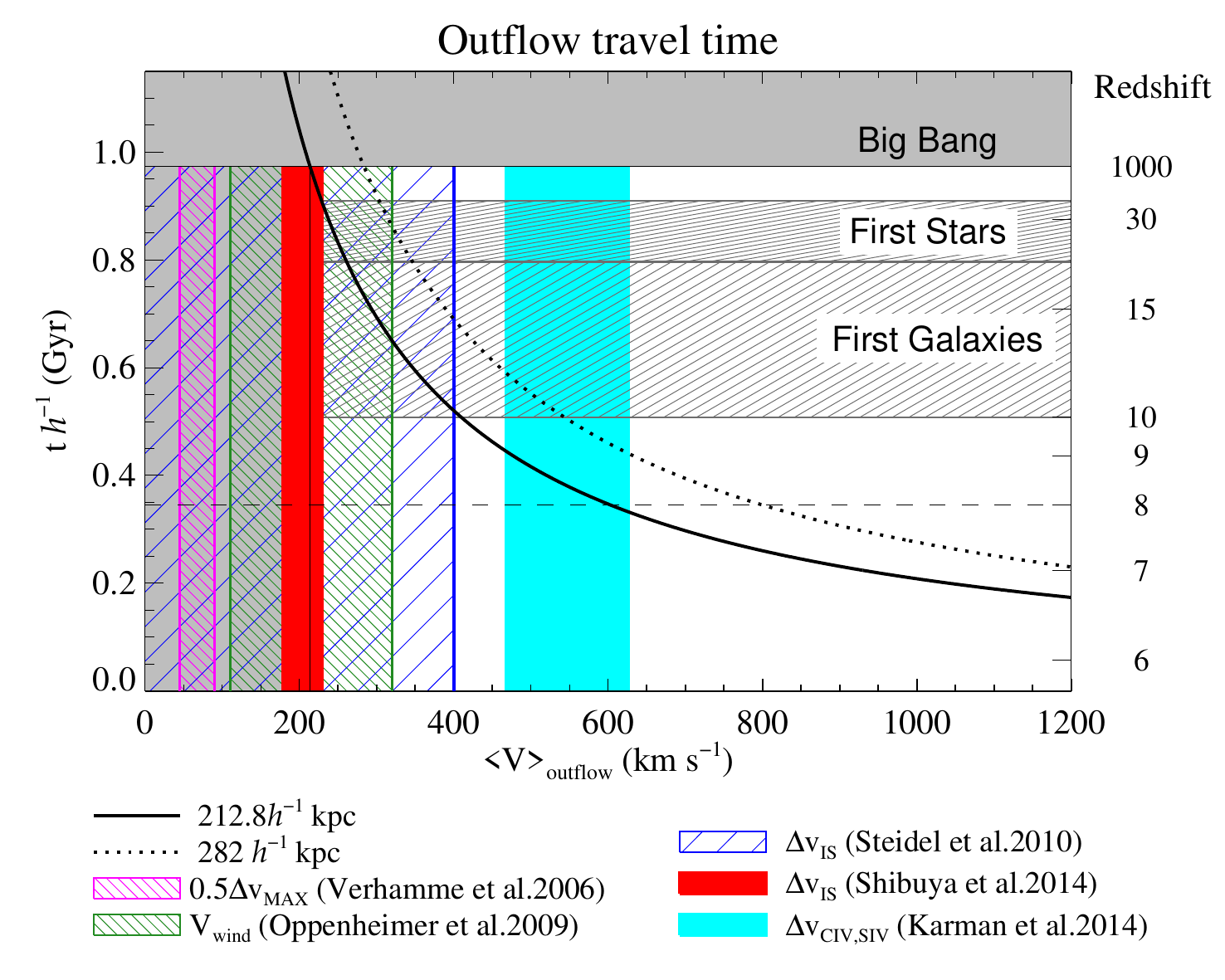} 
\caption{\small Travel time versus mean outflow velocity.
The solid line and the dotted line indicate the time required 
to travel 212.8$h^{-1}$ and 282$h^{-1}$ physical kpc, respectively.
The grey horizontal areas indicate the Big Bang, the time
of formation of the first mini-halos to form stars, and
the time of formation of the first galaxies.
Vertical colour-coded regions represent observed and predicted
outflow speeds from the literature.}
\label{f:v-outflow}
\end{figure}

The time required for outflowing gas to reach a distance of 
212.8$h^{-1}$ kpc depends on the speed of the gas:
a typical outflow speed of $\sim200$\,\kms\ would require a time
longer than the age of the Universe at $z=5.724$. 
Therefore, if the carbon was produced in stars of LAE 103027+052419,
the outflowing material had to have a larger mean velocity
during its journey. For example, at $\sim$400\,\kms\ 
it would have taken 0.518 Gyr to travel 212$h^{-1}$ kpc, implying
that the outflow mechanisms were in place by $z\sim10.1$.
Figure \ref{f:v-outflow} presents the time required
for an outflow travelling at mean velocity $\langle V_{outflow}\rangle$
to reach 212.8$h^{-1}$ physical kpc (solid line) and 
282$h^{-1}$ physical kpc (dotted line)
corresponding to the separation determined using $z_{\text{\Lya}}=z_{\text{BLUE}}$
and $z_{\text{\Lya}}=z_{\text{MAX}}$, respectively.
The age of the Universe at  $z=5.724$ ($\sim0.97$ Gyr)
provides a hard limit for the mean outflow speed.
However, theoretical studies predict that the formation of the first
mini-haloes capable of hosting PopIII star-formation
started around $20<z<30$ \citep[e.g.][]{barkana2001, yoshida2003, gao2007, bromm2011}.
Moreover, the first systems that can be considered `galaxies'
are predicted to have formed around $10<z<20$ 
\citep[e.g.][]{loeb2010, greif2010, bromm2011}.
These limits imply that it is practically impossible
for a low speed outflow ($\langle V_{outflow}\rangle < 250$\,\kms)
from LAE 103027+052419 to reach the position of the \civ\ 
absorption system.
Mean outflow velocities in the range 250--400\,\kms\
should have departed from the LAE during the assembly of the first stars
and galaxies. 
Finally, a strong outflow ($\langle V_{outflow}\rangle>400$\,\kms)
would have had enough time to enrich the region if the outflow
was active since $z\sim10$.
If the redshift of the galaxy is $z_{\text{MAX}}$, then 
the distance to the \civ\ system is 282$h^{-1}$ physical kpc
and stronger winds (additional $\sim150$\,\kms\ in the given examples) 
are required to reach the position of the \civ , which will make the 
`late-outflow' scenario even less plausible.

A comparison with outflow velocities from
the literature is shown in Figure \ref{f:v-outflow}.
\citet{verhamme2006} used 3D radiation transfer 
of \Lya\ photons and report that a model of an expanding shell 
of cold gas would produce a \Lya\ peak with a velocity shift of
$\sim2\times V_{outflow}$, where $V_{outflow}$ is measured
from the blueshift of inter-stellar absorption lines (IS).
Under the assumption that $\Delta v_{\text{MAX}}$ is the velocity
shift of the \Lya\ peak 
with respect to the systemic redshift, the outflow 
speed would be $\frac{\Delta v_{\text{MAX}}}{2}\sim68$\,\kms\ 
(magenta shaded area),
which is too slow to enrich a radius of 212.8$h^{-1}$ kpc.
However, this model is relevant for a single velocity shell and does not
account for the full range of velocities observed in high-resolution
spectra of the absorbing gas on $z\sim2$--3 star-forming galaxies.

Cosmological simulations that reproduce 
observables like the comoving density of \civ\ ions in the IGM
and the luminosity function of galaxies suggest 
that, among intergalactic metals, \civ\ is one of the best tracer 
of diffuse IGM in the redshift range $z=5$--6
and predict that star-forming galaxies in $\simlt 10^{9}$ M$_{\sun}$ 
haloes pollute the IGM for the first time 
and provide the flux to maintain the ionization level
of the systems up to 200 physical kpc \citep{oppenheimer2009}.
Moreover, the work of \citet{oppenheimer2009} 
suggests that \civ\ absorption systems
with column densities in the range $10^{12}-10^{14}$\cm\, 
are probably associated with galaxies of luminosities 
$L \leq L^{\star}$. Their best outflow model predicts mean outflow 
velocities between 200--250\,\kms\
with decreasing trend towards higher redshift.
In particular at $z\sim5.7$, their best-fitting model
predicts outflows with $110<V_{wind}<320$\,\kms\ 
($\pm1\sigma$ range, green hatched area in Figure \ref{f:v-outflow}).
This range is consistent with observations at lower redshift 
\citep[e.g.][]{steidel2010, shibuya2014b} and, considering that 
the impact parameter of the galaxy-\civ\ pair is larger than
similar column density examples at $z \leq 3.5$, it would imply
that enriched gas expelled during the first stages of galaxy
formation was able to reach the diffuse IGM more easily than
the outflowing gas from star-formation episodes at later times,
when galaxies are more massive.

\citet{steidel2010} report that LBGs at $2<z<3$ show 
low-ionization absorption blue-shifted from the galaxy rest-frame,
with the centroid of the IS line profiles (the bulk of the outflowing material) 
reaching 
$| \Delta v_{\text{IS}}| \simlt 400$ (dark blue hashed area).
Figure \ref{f:v-outflow} clearly shows that such a velocity range
implies an early onset of the outflowing mechanism ($z\simlt10$) in the LAE.
In the same work, the authors confirm that the velocity profile of 
the strongest IS lines can reach $|v_{max}|\sim700$-800 \kms, 
which suggest that lower column density gas could be moving at 
these high speeds \citep[Figure 8 of][]{steidel2010}.
Therefore, if the LAE has similar outflows and the velocity
remains constant along the outflows' path, then low column density gas 
could reach the region of the \civ\ system by the time of observation. 
However, it is important to note that their sample of LBGs contains 
massive star-forming galaxies ($M_{\text{DM}}\sim 9 \times 10^{11}$ \msun) 
whose gas kinematics are probably different from LAE 103027+052419.

A better comparison sample is given by \citet{shibuya2014b} who
study LAEs at $z\sim2.2$ and measure the velocity of \Lya\ and IS lines 
with respect to nebular emission lines. From a sample of four LAEs,
they find a mean blue-shift of 
$\langle \Delta v_{\text{IS}}\rangle=204{\pm}27$\,\kms\ (red filled area)
which is comparable to LBGs and not enough to disperse metals 
out to 212.8$h^{-1}$ kpc by $z\sim5.724$.
This analysis suggests that typical outflow velocities
found among lower redshift LAEs and LBGs struggle
to support the scenario of a recent outflow from 
LAE 103027+052419 as the chemical origin of the \civ\ absorbing gas.
Hence, it seems possible that 
outflows are active from very early times in the life of a
star-forming galaxy.
This early onset of galactic outflow supports the pre-galactic
enrichment scenario where a significant fraction of the metals 
in the IGM were produced during the time of formation of 
the first galaxies or before.

So far, the comparison has been made with outflowing cool gas
traced by low-ionization species (e.g. \cii, \siii).
However, highly ionized gas traced by high-ionization absorption,
like \civ\ and \siiv , usually has larger outflow speed.
For example, Figure \ref{f:v-outflow} shows the outflow velocity
measured by \citet{karman2014} from \civ\ and \siiv\ in one of 
the star-forming galaxies in their sample at $z=$2.5--3.5
(cyan filled region), for which
they estimate the stellar mass $M_{\star}=9.29\times10^9M_{\odot}$ 
and age$\,=1.43\times10^8$yr. 
Thus, if LAE 103027+052419 had a similar outflow velocity 
around $z\sim8$, such a highly ionised galactic wind
could have reached 212.8$h^{-1}$ physical kpc by $z=5.724$.

Considering that the details of the distribution of metals around high-$z$ galaxies
are not fully understood, it is not clear yet if the amount of high-velocity 
material would be enough to produce the level of absorption that is observed
($\log$\nciv$= 14.52{\pm0.08}$).
For this reason and because the system has the highest \civ\ column density 
know to-date at $z>5.5$, we have chosen the conservative approach to compare with
the speed of the bulk of the outflowing material (mean speed).
From this comparison, we conclude that low-speed
outflows are less probable to have enriched the distances observed
in our example whereas high-speed low-density gas could transport 
some \civ\ to 212.8$h^{-1}$ kpc in a reasonable cosmological time.
Another alternative is that the carbon was produced and 
distributed by a closer undetected dwarf galaxy.

\subsection{Undetected dwarf galaxies}
Finally, we consider the scenario in which the \civ\ system
is produced by a dwarf satellite galaxy gravitationally bound
to the LAE.
Our photometric sample reaches a luminosity limit of
 $L_{Ly\alpha}\sim 0.5 L_{Ly\alpha}^\star$ 
(star formation rate of 5--10~$M_\odot$~yr$^{-1}$ uncorrected for IGM and dust absorption), 
where $L_{Ly\alpha}^\star = 6.8\times 10^{42}\,h^{-2}$ erg s$^{-1}$ 
at $z=5.7$ \citep{ouchi2008}. 
Assuming the luminosity function of \citet{ouchi2008} 
and a uniform distribution of galaxies, 
the number of LAEs brighter than $0.5 L_{Ly\alpha}^\star$ expected 
inside a projected area of radius $\sim212h^{-1}$ physical kpc
and within $\pm$500\,\kms\ from the \civ\ system is $\sim 0.03$.
Having detected one LAE implies very low chances of
having a second bright galaxy in such a small volume. 
However, under the hierarchical formation 
scenario, galaxies grow by accretion of less massive haloes.
For this reason, it is possible that smaller haloes hosting 
galaxies below $0.5 L_{Ly\alpha}^\star$ are falling into 
LAE 103027+052419.
However, the predicted number of faint LAEs depends on 
the faint end slope of the luminosity function ($\alpha$).
For example, using $\alpha=-1.5$ \citep[e.g.][]{ouchi2008} 
the number of LAEs expected in the range
0.01--$0.5\, L_{Ly\alpha}^\star$ is $\sim 0.79$.
Interestingly, recent measurements of the \Lya\ 
luminosity function at \zlae\ indicate a steeper 
faint end with $-2.35< \alpha <-1.95$ \citep[$1\sigma$, ][]{dressler2014}.
Using $\alpha=-2.1$ and the luminosity function from 
\citet{dressler2014}, the predicted number of LAEs with 
$L_{Ly\alpha}=0.01$--$0.5 L_{Ly\alpha}^\star$ is 0.98 (i.e. $\sim 1$).
In other words: the most recent estimate of the LAE luminosity function
at \zlae\ predicts at least one LAE in the range
$0.01$--$0.5 L_{Ly\alpha}^\star$ that would lie
within a radius $\sim212h^{-1}$ physical kpc
and $\pm$500\,\kms\ from the \civ\ system.
As a result, it is possible that the source of \civ\ is a galaxy fainter 
than our detection limit and closer to the absorption 
than LAE 103027+052419. However, this question will remain open until deeper 
observations are available.

A related possibility is that a faint galaxy is
aligned exactly with the background QSO 
and therefore outshone by the QSO's brightness.
Given the peculiarly high incidence of
\civ\ absorption systems in this line of sight
where four \civ\ systems have been found 
in the redshift range $5.5<z<6$ \citep{ryan-weber2009,
simcoe2011b, dodorico2013},
the probability of having four galaxies and one QSO 
highly aligned with the observer is negligible.
Moreover, low ionization absorptions
like \oi\ and \cii\ are typically found in
lines of sight to QSOs that intercept a galaxy
``disc", indicating cool gas capable of forming stars.
In this case, the absence of low ionization lines
is in conflict with this scenario.

In summary, the LAE-\civ\ system pair with an impact parameter
of 212.8$h^{-1}$ physical kpc at $z=5.724{\pm}0.001$ reported in this work,
is in marginal agreement with the expectations from 
\citet{oppenheimer2009} that \civ\ absorption systems 
at this redshift can be found 
at $\sim200\,h^{-1}$ physical kpc of the galaxy providing the 
ionizing flux (the ionization `bubble' model).
Nevertheless, typical outflow velocities cannot support 
the idea that the LAE is the chemical source of the absorbing gas,
unless highly ionised outflows can reach velocities
twice as fast (or more) as the cool gas.
Hence, the results agree with the idea that \civ\ systems are
tracing lower density (diffuse) highly ionized IGM.
On one hand, the possibility that the carbon was 
processed and distributed by an undetected dwarf satellite galaxy
cannot be ruled out, whereas the late-outflow scenario would
require a particularly high outflow speed (e.g. $\sim800$--1000\,\kms\ from LAE 103027+052419)
not yet seen among lower redshift galaxies in the mass range of LAEs.
On the other hand, if the progenitor of the carbon is indeed 
LAE 103027+052419, 
assuming a typical average speed of lower redshift outflows,
we find a time of production and distribution of metals
which is in agreement with the expectations from 
a cosmic chemical enrichment during a pre-galactic stage
of galaxy formation (i.e. $z{\simgt}8$).
Therefore, regardless of the origin of the carbon, the results 
are more easily explained by the pre-galactic enrichment scenario.

\section{Summary and conclusions}  \label{s:conclusion-spec}

Following our findings of an excess of
LAE candidates in the projected environment of
\zlbg\ \civ\ absorption systems resported in Paper I,
this work presents the results from the spectroscopic 
campaign using the \textsc{deimos} spectrograph 
on the Keck-II telescope for LAEs in the fields J1030+0524 
and J1137+3549. The main results are:
\begin{itemize}
\item The broad-band colours of confirmed LAEs 
are in agreement with the predictions in Paper I from the 
spectrophotometry of LBGs with \Lya\ emission. 
We find that LAEs at \zlbg\ are not recovered by 
the $i'$-dropout criteria since most of them have ($i'-z'$)$\,<1.3$.

\item The contamination level in the LAE sample 
based on the asymmetry of the line is 10--20 per cent.
We find that contaminant sources are more common
at fainter magnitudes (NB\civ$\,>25$ mag).
No evidence for AGN was found in the spectroscopic
sample of LAEs, although the search is heavily limited
by the abundance of sky emission lines.

\item Four out of the five LAEs within 
10$h^{{-}1}$ projected comoving Mpc of 
the \civ\ line of sight in the field 
J1030+0524 have been spectroscopically observed.
If the emission line is \Lya, three are within 
${\pm}500$\,\kms\ from the high-column density \civ\ system.
This would confirm an excess of LAEs, which
would become more significant ($\Sigma / \langle \Sigma \rangle\sim2.4$)
considering the volume sampled by the narrow-band.
However, two LAEs were observed with 
low resolution and short exposure times and, although
we identify the wavelength of the emission line,
we cannot accurately measure the asymmetry of the line.
Therefore, we cannot secure their LAE nature
and the confirmation of the excess of LAEs on large scales
will remain open until better data is acquired. 

\item  We find that the LAE sample of the field 
J1137+3549 is not centred at the redshift 
of the \civ\ system but at lower redshift, as a result of 
the NB\civ\ sensitivity. 
Therefore, we cannot test the idea that LAEs dominate the 
environment of this absorption
because the sample of LAEs is too distant from
the environment of the \civ\ system.
However, the absence of \civ\ at the redshift of 
the LAE sample and the large transverse 
distance to the closest LAE, would easily
be explained if all LAEs are able to produce
\civ\ absorption systems.
This does not imply that all \civ\ systems
are produced by LAEs.

\item The redshift of the closest LAE 
to the \civ\ line of sight in J1030+0524
(LAE 103027+052419) is $z_{\text\Lya}=5.724\pm0.001$.
This galaxy is at 212.8$h^{-1}$ physical kpc transverse 
from the \civ\ absorption system
at $z_{\text{\civ,b}}=5.7242\pm0.0001$.
This galaxy-\civ\ absorption system 
pair is the highest redshift example 
known to date, and involves the highest 
column density absorption system at $z \geq 5.5$.

\end{itemize}

This work has shown that our LAE sample is 
comparable to other \zlbg\ LAE searches,
since in many instances we compare our measurements
with the \zlae\ LAEs from SDF and find very good agreement.
Although the tentative detection of a small excess of LAEs
in the line of sight around the \civ\ system will 
require better data to be confirmed,
the current picture seems to
supports the idea that 
\civ\ systems inhabit the surroundings of LAEs.

In the field J1030+0524, spectroscopy revealed
the closest galaxy to a \civ\ absorption system at $z>5.5$. 
We compared the observed properties
of this LAE with the rest of the sample,
in order to explore possible scenarios for the 
existence of a galaxy-\civ\ absorption system pair
at \zlae . The main results can be summarised as follows:

\begin{itemize}
\item The \Lya\ emission line of LAE 103027+052419
shows a clear asymmetry (S$_{w,10\%}= 11.5{\pm}1.9$),
low rest-frame equivalent width (EW$_0=33^{+15}_{-12}$\,\AA)
and a velocity shift of the emission peak \vmax$ = 136{\pm}45$ \,\kms .
The shape of the galaxy in the $z'$-band image from HST
seems elongated  towards the background QSO.

\item Agreement between our sample and the sample of LAEs from SDF
is found in the EW$_0$ distribution and the trend for UV fainter
galaxies to have larger EW$_0$.

\item Compared with the rest of the sample, 
LAE 103027+052419 has one of the lowest values of
EW$_0$ and the highest values of \vmax .

\item The UV luminosity, \Lya\ profile, $\Delta v_{\text{MAX}}$
and EW$_0$ of LAE 103027+052419 suggest that this
galaxy is at (or near) the high mass end of the LAEs,
might contain a large \HI\ column density 
and is a good candidate to host an outflow.

\item The temporal limits imposed by the Big Bang 
and the formation of the first stars
imply that outflow velocities typically observed at lower redshift
are inconsistent with 
the LAE as the chemical source
of the absorbing gas, unless highly ionised gas is moving
more than twice the speed of the cool gas.
Even in this case, the time of production and dispersion
of metals would agree with predictions from
the pre-galactic enrichment scenario.

\item The luminosity function of LAEs at \zlae\ 
predicts at least one LAE below our detection limit 
and within $\sim212h^{-1}$ 
physical kpc radius (projected) and $\pm$500\,\kms\ (line of sight) 
from the \civ\ system. Therefore, outflows from 
undetected galaxies cannot be ruled out as the chemical 
origin of the absorption system.

\end{itemize}

The confirmation that LAE 103027+052419 is at 
 212.8$^{+14}_{-0.4}h^{-1}$ physical kpc from the \civ\ system
demonstrates that galaxies can be detected 
in association with high column density highly 
ionized metal absorption systems
shortly after the epoch of reionization.
However, our analysis suggests that typical outflow velocities
found among lower redshift LAEs and LBGs do not
support the scenario of an outflow from 
LAE 103027+052419 as the chemical origin of the \civ\ absorbing gas.
The main implication is that the time when an outflow should have departed 
from the LAE agrees with the expectations from the pre-galactic enrichment scenario.
Finally, we cannot rule out the possibility that the carbon was produced and distributed by 
an undetected dwarf galaxy fainter than LAEs. 
However, we note the low likelihood of dwarf galaxies 
aligned with the background QSO being responsible for all the \civ\ systems
reported in this particular line of sight.
Both options point towards an early ($z\,\simgt\,6$) production and distribution
of metals, which supports the idea that a significant fraction of the metals
in the IGM were injected during a pre-galactic stage of galaxy formation.

As shown in this work, the search for galaxies around metal systems at high$-z$ requires deep 
broad-band and narrow-band photometry, which must be followed-up by spectroscopy
resulting in a two-year process. In addition, observations are limited to sources 
identified from the photometry and the fraction of the sky with spectroscopic
coverage is minimal. This procedure is far from ideal but has been the
only option available so far.
The optimal approach to study metal absorption systems and their 
galaxy counterparts would be a deep blind spectroscopic search around
the absorptions.  
As a result, this field of research will be significantly benefited by
new generation integral field units like \textsc{MUSE} on the {\it Very Large Telescope},
which provides the field of view, wavelength coverage and sensitivity
that enable a blind spectroscopic search for galaxies around multiple
absorption systems in the same line of sight.
Instruments like MUSE are crucial to understand the physical state of the gas 
around high-$z$ galaxies and its evolution, which is strongly related to mechanisms 
of chemical enrichment of the CGM and IGM.

\section*{Acknowledgments}

We are extremely grateful to Kym-Vy Tran whose
data contribution obtained in February 2014
has significantly enriched the scientific output of this work.
Many thanks to Neil Crighton, Valentina D'Odorico and Thibault Garel
for their comments and the interesting conversations that we had
over this work. C.G.D. acknowledges the support from the 
Victorian Government through the
Victorian International Research Scholarship program.
E.R.W. acknowledges the support of Australian 
Research Council grant DP1095600.
J.C. acknowledges the support of Australian 
Research Council grant FF130101219.
Y.K. acknowledges the support from the Japan Society for the
Promotion of Science (JSPS) through JSPS research fellowships
for young scientists. Y.K. also acknowledges the support by JSPS KAKENHI
Number 26800107.
M.O. acknowledges the supports from
World Premier International Research Center Initiative (WPI Initiative), 
MEXT, Japan, and KAKENHI (23244025) Grant-in-Aid 
for Scientific Research(A) through Japan Society 
for the Promotion of Science (JSPS).

\appendix
\section{\\Spectroscopic sample of \zlae\ LAEs.} \label{app:spec-tables}

\begin{table*}
\begin{minipage}{170mm} 
\caption{ LAEs in the field J1030+0524. 
The double line separates LAEs observed with the 600ZD grating, which 
are not selected by their weighted skewness
because the emission lines are barely detected, 
but are presented for completeness.}
\label{t:spec-LAE-J1030}
\begin{tabular}{crrrrrrrrr}
\hline
  \multicolumn{1}{c}{ID} &
  \multicolumn{1}{c}{Grating} &
  \multicolumn{1}{c}{RA} &
  \multicolumn{1}{c}{Dec.} &
  \multicolumn{1}{c}{$z_{\text{MAX}}$} &
  \multicolumn{1}{c}{$z_{\text{BLUE}}$} &
  \multicolumn{1}{c}{S$_{w,10\%}$} &
  \multicolumn{1}{c}{EW$_{0,\text{Spec.}}$} &
  \multicolumn{1}{c}{EW$_{0,\text{Phot.}}$} &
  \multicolumn{1}{c}{$\Delta v_{\text{MAX}}$} \\
  \multicolumn{1}{c}{} &
  \multicolumn{1}{c}{} &
  \multicolumn{1}{c}{(J2000)} &
  \multicolumn{1}{c}{(J2000)} &
  \multicolumn{1}{c}{} &
  \multicolumn{1}{c}{} &
  \multicolumn{1}{c}{} &
  \multicolumn{1}{c}{(\AA)} &
  \multicolumn{1}{c}{(\AA)} &
  \multicolumn{1}{c}{(\kms)} \\
\hline
1 & 830G & 10:30:21.535& +5:32:56.33 &5.671$\pm$0.001 & 5.663$\pm$0.001 & 16.0$\pm$1.1 & $>$35 & 203$^{+97}_{-70}$ & 344$\pm$46 \\ 
2 & 830G & 10:30:21.442& +5:36:49.99 &5.661$\pm$0.001 & 5.658$\pm$0.001 & 14.8$\pm$1.0 & 52$\pm$36 & 46$^{+17}_{-13}$ & 137$\pm$46 \\
3 & 830G & 10:30:36.902& +5:17:08.33 & 5.704$\pm$0.001 & 5.701$\pm$0.001 & 11.9$\pm$0.9 & 175$\pm$159 & 84$^{+60}_{-35}$ & 119$\pm$46 \\ 
4 & 830G & 10:30:27.677& +5:24:19.82 & 5.727$\pm$0.001 & 5.724$\pm$0.001 & 11.5$\pm$1.9 & 40$\pm$31 & 33$^{+15}_{-12}$ & 136$\pm$46\\
5 & 830G & 10:29:49.853& +5:24:37.91 & 5.737$\pm$0.001 & 5.734$\pm$0.001 & 10.1$\pm$1.2 & $>$8 & 153$^{+142}_{-71}$ & 102$\pm$46 \\
6 & 830G & 10:30:06.358& +5:17:42.10 & 5.749$\pm$0.001 & 5.747$\pm$0.001 & 8.70$\pm$1.2 & 71$\pm$61 & 151$^{+138}_{-75}$ & 119$\pm$46\\
7 & 830G &10:29:56.062& +5:21:28.64 & 5.690$\pm$0.001 & 5.689$\pm$0.001 & 8.50$\pm$2.8 & $>$5 & 61$^{+68}_{-34}$ & 51$\pm$46\\
8 & 830G & 10:29:49.994& +5:24:17.49 & 5.740$\pm$0.001 & 5.738$\pm$0.001 & 8.10$\pm$1.0 & $>$25 & 178$^{+142}_{-83}$ & 102$\pm$46\\
9 & 830G &  10:30:15.722& +5:34:59.51 & 5.712$\pm$0.001 & 5.709$\pm$0.001 & 7.40$\pm$1.7 & 28$\pm$22 & 38$^{+21}_{-15}$ & 155$\pm$46 \\
10 & 830G &10:30:32.050& +5:19:28.76 & 5.702$\pm$0.001 & 5.701$\pm$0.001 & 6.20$\pm$1.9 & $>$2 & $>$297 & 52$\pm$46 \\
11 & 830G & 10:29:56.844& +5:21:36.72 & 5.692$\pm$0.001 & 5.689$\pm$0.001 & 6.00$\pm$2.0 & 22$\pm$19 & 69$^{+47}_{-26}$ & 120$\pm$46 \\
12 & 830G & 10:30:15.689& +5:15:50.90 &5.684$\pm$0.001 & 5.682$\pm$0.001 & 5.90$\pm$0.7 & 120$\pm$104 & 77$^{+31}_{-25}$ & 69$\pm$46 \\
13 & 830G & 10:30:37.937& +5:23:04.58 &5.688$\pm$0.001 & 5.686$\pm$0.001 & 4.00$\pm$1.8 & 15$\pm$11 & 66$^{+80}_{-36}$ & 103$\pm$46 \\
14 & 830G & 10:29:51.226& +5:20:57.26 &5.678$\pm$0.001 & 5.676$\pm$0.001 & 3.70$\pm$1.6 & $>$8 & 90$^{+100}_{-45}$ & 103.0$\pm$46 \\
15 & 830G & 10:29:51.562& +5:28:05.61 &5.697$\pm$0.001 & 5.695$\pm$0.001 & 3.20$\pm$1.9 & 14$\pm$12 & 44$^{+28}_{-19}$ & 85$\pm$46 \\
16 & 830G & 10:30:40.385& +5:16:18.11 &5.699$\pm$0.001 & 5.698$\pm$0.001 & 3.00$\pm$1.2 & $>$4 & $>$133 & 51$\pm$46 \\
\hdashline
17 & 830G & 10:30:26.911& +5:37:01.61 &5.656$\pm$0.001 & 5.649$\pm$0.001 & 1.10$\pm$1.1 & 27$\pm$20 & 38$^{+56}_{-31}$ & 291$\pm$46 \\
18 & 830G & 10:30:15.727& +5:27:37.67 &5.726$\pm$0.001 & 5.723$\pm$0.001 & -1.4$\pm$1.9 & $>$3 & $>$149 & 119$\pm$46 \\
\hline
\hline
19 & 600ZD &10:30:33.41& +5:23:41.8 &5.728$\pm$0.001 & 5.726$\pm$0.001 & 3.20$\pm$4.3 & $>$3 & 60$^{+18}_{-13}$ & 71$\pm$64 \\
20 & 600ZD &  10:30:40.80 & +5:27:17.4 & 5.717$\pm$0.001 & 5.715$\pm$0.001 & 0.00$\pm$5.2 & 52$\pm$50 & 57$^{+17}_{-14}$ & 94$\pm$64 \\
21 & 600ZD &  10:29:59.36 & +5:21:55.9 &5.738$\pm$0.001 & 5.736$\pm$0.001 & -1.7$\pm$5.4 & $>$1 & $>$142 & 94$\pm$64 \\
\hline\end{tabular} 
\end{minipage}
\end{table*}

\begin{minipage}{170mm} 
This Appendix contains tables with measurements of spectroscopic
quantities and other estimates for the sample of LAE candidates
observed with \textsc{DEIMOS}. A summary of their photometry is also presented
for completeness. 
The objects are sorted by decreasing weighted skewness S$_{w,10\%}$.
The dashed line separates objects with S$_{w,10\%}{<}3$, which
do not reach the asymmetry condition and are considered 
contaminants. However, all quantities reported are estimated 
assuming that the emission line is \Lya . 

The columns of Tables \ref{t:spec-LAE-J1030} and \ref{t:spec-LAE-J1137} contain:
(1) ID number that follows the sorting criteria of decreasing S$_{w,10\%}$;
(2) name of the grating used in the spectroscopic observations;
(3) right ascension and (4) declination (FK5, J2000.0);
(5) redshift measured at the maximum of the line profile;
(6) redshift measured at the bluest pixel of the line profile (as described in Section \ref{s:redshift-dist});
(7) weighted skewness S$_{w,10\%}$ measured as described 
in Section \ref{s:spectroscopic-LAEs-contamination};
(8) equivalent width measured from the photometry and (9) from the spectra,
as described in Section \ref{s:eqw-measure};
(10) velocity shift of the maximum of the line profile with respect to $z_{\text{BLUE}}$.

The columns of Tables \ref{t:phot-LAE-J1030} and \ref{t:phot-LAE-J1137} contain:
(1) ID number;
(2) absolute magnitude and (3) luminosity in the rest-frame ultraviolet ($\lambda_0{\sim}1350$\AA);
(4) star formation rate (SFR) estimated from the ultraviolet luminosity via 
$L_{\text{UV}}{=}8.0\times10^{27}$SFR$_{\text{UV}}\frac{(\text{erg}\, \text{s}^{-1}\, \text{Hz}^{-1})}{(\text{M}_{\odot}\text{yr}^{-1})}$ \citep{madau1998};
(5), (6), (7) and (8) the aperture photometry in the R$_c$, $i'$, $z'$ and NB\civ\ bands, respectively;
and (9) present the NB excess ($i'-$NB\civ).
\end{minipage}

\FloatBarrier

\begin{table*}
\caption{LAEs in the field J1030+0524.}
\label{t:phot-LAE-J1030}
\begin{tabular}{crrrrcccc}
\hline
  \multicolumn{1}{c}{ID} &
  \multicolumn{1}{c}{M$_{\text{UV}}$} &
  \multicolumn{1}{c}{L$_{\text{UV}}$} &
  \multicolumn{1}{c}{SFR$_{\text{UV}}$} &
  \multicolumn{1}{c}{R$_c$} &
  \multicolumn{1}{c}{$i'$} &
  \multicolumn{1}{c}{$z'$} &
  \multicolumn{1}{c}{NB$_{\text{\civ}}$} &
  \multicolumn{1}{c}{($i'-$NB$_{\text{\civ}}$)} \\
    \multicolumn{1}{c}{} &
  \multicolumn{1}{c}{(mag)} &
  \multicolumn{1}{c}{($\times 10^{28}$ erg s$^{-1}$ Hz$^{-1}$)} &
  \multicolumn{1}{c}{($M_{\odot}$ yr$^{-1}$)} &
   \multicolumn{1}{c}{(2 arcsec)} &
  \multicolumn{1}{c}{(2 arcsec)} &
  \multicolumn{1}{c}{(2 arcsec)} &
  \multicolumn{1}{c}{(2 arcsec)} &
  \multicolumn{1}{c}{(2 arcsec)} \\
\hline
1 & -20.7$\pm$0.5 & 9.1$\pm$4.1 & 11.4$\pm$5.1 & $\simgt$28.35 & 26.24$\pm$0.30& 25.92$\pm$0.53& 24.19$\pm$0.05& 2.0 \\
2 & -21.7$\pm$0.2 & 21.0$\pm$4.7 & 26.2$\pm$5.9 &  $\simgt$28.35 & 25.95$\pm$0.24& 24.92$\pm$0.25& 24.44$\pm$0.08& 1.4 \\
3 & -20.6$\pm$0.7 & 9.20$\pm$5.5 & 11.5$\pm$6.9 &   $\simgt$28.35 & 26.85$\pm$0.486& 26.40$\pm$0.75& 24.35$\pm$0.05& 2.6 \\
4 & -20.7$\pm$0.5 & 9.50$\pm$4.2 & 11.9$\pm$5.2 &  $\simgt$28.35 & $\simgt$28.04  & 25.89$\pm$0.52& 25.11$\pm$0.15& $\simgt$3.0 \\
5 & -19.9$\pm$0.9 & 5.3$\pm$3.8 & 6.6$\pm$4.7 & $\simgt$28.35 & 26.88$\pm$0.50& 26.77$\pm$0.95 & 24.58$\pm$0.09& 2.1 \\
6 & -19.7$\pm$1.0 & 4.8$\pm$3.5 & 6.7$\pm$4.4 &  $\simgt$28.35 & $\simgt$28.04  & 26.9$\pm$1.0 & 25.07$\pm$0.13& $\simgt$3.0  \\
7 & -19.8$\pm$1.0 & 5.2$\pm$3.7 & 6.5$\pm$4.6 &  $\simgt$28.35 & 27.24$\pm$0.642& 26.79$\pm$0.96 & 25.47$\pm$0.21& 1.9 \\
8 & -20.0$\pm$0.9 & 5.7$\pm$3.800 & 7.1$\pm$4.7 &  $\simgt$28.35 & 27.09$\pm$0.58& 26.66$\pm$0.88 & 24.42$\pm$0.07& 2.6 \\
9 & -20.6$\pm$0.6 & 8.5$\pm$4.2 & 10.6$\pm$5.2 & $\simgt$28.35 & 27.27$\pm$0.65& 26.03$\pm$0.58& 25.03$\pm$0.12& 2.1 \\
10 & $>$-19.1 & $<$2 & $<$2.5 &  $\simgt$28.35 & 27.42$\pm$0.72& $\simgt$27.49 & 25.55$\pm$0.21& 1.8 \\
11 & -20.4$\pm$0.6 & 7.6$\pm$4.1 & 9.5$\pm$5.1 &  $\simgt$28.35 & 26.90$\pm$0.50& 26.19$\pm$0.65& 24.80$\pm$0.12& 1.6 \\
12 & -20.9$\pm$0.4 & 11.0$\pm$4.3 & 13.7$\pm$5.3 & $\simgt$28.35 & 25.90$\pm$0.23& 25.69$\pm$0.45& 24.36$\pm$0.07& 1.6 \\
13 & -19.7$\pm$1.0 & 4.7$\pm$3.4 & 5.9$\pm$4.2 &  $\simgt$28.35 & $\simgt$28.04  & 26.9$\pm$1.0 & 25.61$\pm$0.18& $\simgt$2.3 \\
14 & -19.9$\pm$0.9 & 5.4$\pm$3.7 & 6.7$\pm$4.6 &  $\simgt$28.35 & 27.01$\pm$0.54& 26.72$\pm$0.92 & 25.29$\pm$0.16& 1.8 \\
15 & -20.4$\pm$0.6 & 7.7$\pm$4.1 & 9.6$\pm$5.1 &  $\simgt$28.35 & 27.37$\pm$0.70 & 26.17$\pm$0.64& 25.13$\pm$0.20& 2.2 \\
16 & $>$-19.1 & $<$1.9 & $<$2.4 &  $\simgt$28.35 & $\simgt$28.04  & $\simgt$27.49 & 25.42$\pm$0.17& $\simgt$2.7 \\
\hdashline
17 & -20.6$\pm$0.6 & 8.5$\pm$4.1 & 10.6$\pm$5.1 & $\simgt$28.35 & 27.101$\pm$0.58& 26.01$\pm$0.57& 25.79$\pm$0.21& 1.4 \\
18 & $>$-19.1 & $<$1.9 & $<$2.4 &  $\simgt$28.35 & $\simgt$28.04  & $\simgt$27.49 & 25.24$\pm$0.13& $\simgt$2.6 \\
\hline
\hline
19 & -21.2$\pm$0.4 & 13.6$\pm$4.5 & 17.0$\pm$5.6 & 27.67$\pm$0.62 & 26.36$\pm$0.33 & 25.44$\pm$0.37 & 24.12$\pm$0.08 & 1.7 \\
20 & -21.2$\pm$0.4 & 14.3$\pm$4.6 & 17.9$\pm$5.7 &  $\simgt$28.35 &  26.49$\pm$0.37 & 25.38$\pm$0.36 & 24.02$\pm$0.08 & 2.2\\
21 & $>$-19.1 & $<$2.0 & $<$2.5 & $\simgt$28.35 &  $\simgt$28.04 & $\simgt$27.49 & 25.49$\pm$0.16 & 2.5\\
\hline\end{tabular} 
\end{table*}

\begin{table*}
\caption{Same as Table \ref{t:spec-LAE-J1030} for LAEs in the field J1137+3549.}
\label{t:spec-LAE-J1137}
\begin{tabular}{crrrrrrrrr}
\hline
  \multicolumn{1}{c}{ID} &
  \multicolumn{1}{c}{Grating} &
  \multicolumn{1}{c}{RA} &
  \multicolumn{1}{c}{Dec.} &
  \multicolumn{1}{c}{$z_{\text{MAX}}$} &
  \multicolumn{1}{c}{$z_{\text{BLUE}}$} &
  \multicolumn{1}{c}{S$_{w,10\%}$} &
  \multicolumn{1}{c}{EW$_{0,\text{Spec.}}$} &
  \multicolumn{1}{c}{EW$_{0,\text{Phot.}}$} &
  \multicolumn{1}{c}{$\Delta v_{\text{MAX}}$} \\
  \multicolumn{1}{c}{} &
  \multicolumn{1}{c}{} &
  \multicolumn{1}{c}{(J2000)} &
  \multicolumn{1}{c}{(J2000)} &
  \multicolumn{1}{c}{} &
  \multicolumn{1}{c}{} &
  \multicolumn{1}{c}{} &
  \multicolumn{1}{c}{(\AA)} &
  \multicolumn{1}{c}{(\AA)} &
  \multicolumn{1}{c}{(\kms)} \\
\hline
1 & 830G & 11:36:37.373& +35:58:31.63 &5.686$\pm$0.001 & 5.683$\pm$0.001 & 19.9$\pm$1.5 & 176$\pm$158 & 54$^{+17}_{-12}$ & 119$\pm$46 \\
2 & 830G &11:36:37.606& +35:41:03.10&5.708$\pm$0.001 & 5.705$\pm$0.001 & 16.9$\pm$1.3 & 29$\pm$18 & 76$^{+24}_{-19}$ & 136$\pm$46 \\
3 & 600ZD & 11:37:58.70 & +35:56:44.3 &5.695$\pm$0.001 & 5.690$\pm$0.001 & 16.3$\pm$2.5 & 34$\pm$22 & 109$^{+61}_{-42}$ & 190$\pm$65 \\
4 & 830G & 11:36:49.742& +35:57:59.24 &5.683$\pm$0.001 & 5.681$\pm$0.001 & 14.6$\pm$1.5 & $>$4 & 83$^{+46}_{-32}$ & 86$\pm$46 \\
5 & 830G &  11:36:39.089& +35:43:46.44 &5.715$\pm$0.001 & 5.713$\pm$0.001 & 13.7$\pm$0.5 & 62$\pm$41 & 61$^{+17}_{-14}$ & 103$\pm$46 \\
6 & 830G & 11:36:45.269& +35:40:40.37 &5.708$\pm$0.001 & 5.705$\pm$0.001 & 13.5$\pm$0.8 & $>$26 & $>$220 & 136$\pm$46 \\
7 & 830G &11:36:47.088& +35:44:02.22 &5.682$\pm$0.001 & 5.679$\pm$0.001 & 10.9$\pm$0.2 & 193$\pm$141 & 283$^{+200}_{-117}$ & 120$\pm$46\\
8 & 830G & 11:37:18.602& +35:53:00.41 &5.707$\pm$0.001 & 5.704$\pm$0.001 & 10.2$\pm$0.6 & $>$34 & 111$^{+87}_{-47}$ & 120$\pm$46 \\
9 & 600ZD & 11:38:05.27& +35:58:11.5 &5.708$\pm$0.001 & 5.705$\pm$0.001 & 5.30$\pm$2.9 & $>$6 & 30$^{+5}_{-3}$ & 118$\pm$64 \\
10 & 830G & 11:37:01.306& +35:49:17.42 &5.696$\pm$0.001 & 5.694$\pm$0.001 & 3.00$\pm$0.6 & $>$13 & 168$^{+210}_{-105}$ & 86$\pm$46 \\
\hdashline
11 & 600ZD &  11:38:07.69 &+35:54:42.6 & 0.6370$\pm$0.0005 & -- & -1.7$\pm$1.0 & -- & -- & -- \\
12 & 600ZD &11:37:39.82 & +35:52:38.2 &5.713$\pm$0.001 & 5.708$\pm$0.001 & -3.6$\pm$2.4 & $>$7 & 81$^{+64}_{-35}$ & 237$\pm$64 \\ 
\hline\end{tabular} 
\end{table*}

\begin{table*}
\caption{Same as Table \ref{t:phot-LAE-J1030} for LAEs in the field J1137+3549.}
\label{t:phot-LAE-J1137}
\begin{tabular}{crrrrcccc}
\hline
  \multicolumn{1}{c}{ID} &
  \multicolumn{1}{c}{M$_{\text{UV}}$} &
  \multicolumn{1}{c}{L$_{\text{UV}}$} &
  \multicolumn{1}{c}{SFR$_{\text{UV}}$} &
  \multicolumn{1}{c}{R$_c$} &
  \multicolumn{1}{c}{$i'$} &
  \multicolumn{1}{c}{$z'$} &
  \multicolumn{1}{c}{NB$_{\text{\civ}}$} &
  \multicolumn{1}{c}{($i'-$NB$_{\text{\civ}}$)} \\
  \multicolumn{1}{c}{} &
  \multicolumn{1}{c}{(mag)} &
  \multicolumn{1}{c}{($\times 10^{28}$ erg s$^{-1}$ Hz$^{-1}$)} &
  \multicolumn{1}{c}{($M_{\odot}$ yr$^{-1}$)} &
   \multicolumn{1}{c}{(2 arcsec)} &
  \multicolumn{1}{c}{(2 arcsec)} &
  \multicolumn{1}{c}{(2 arcsec)} &
  \multicolumn{1}{c}{(2 arcsec)} &
  \multicolumn{1}{c}{(2 arcsec)} \\
\hline
1 & -21.5$\pm$0.3 & 18.1$\pm$4.9 & 22.6$\pm$6.1 & 27.26$\pm$0.56& 26.02$\pm$0.30 & 25.10$\pm$0.30 & 24.04$\pm$0.17&1.5  \\
2 & -21.1$\pm$0.4 & 13.0$\pm$4.7 & 16.2$\pm$5.9 &  $\simgt$28.04 & 26.77$\pm$0.54 & 25.50$\pm$0.41 & 23.89$\pm$0.11 &2.4 \\
3 & -20.5$\pm$0.6 & 8.10$\pm$4.3 & 10.1$\pm$5.4 & $\simgt$28 & 26.46$\pm$0.43 & 26.12$\pm$0.65 & 24.33$\pm$0.12 & 2.1\\
4 & -20.6$\pm$0.6 & 9.0$\pm$4.4 & 11.2$\pm$5.5 & 27.64$\pm$0.73& 26.37$\pm$0.40 & 25.96$\pm$0.58 & 24.58$\pm$0.13 &1.7 \\
5 & -21.3$\pm$0.3 & 15.4$\pm$4.8 & 19.2$\pm$6.0 & $\simgt$28.04 & 25.95$\pm$0.28 & 25.30$\pm$0.35 & 23.88$\pm$0.06 &2.1 \\
6 & $>$-19.2 & $<$2.1 & $<$2.6 & $\simgt$28.04 &$\simgt$27.61   & $\simgt$27.39 & 24.71$\pm$0.17 &$\simgt$2.9 \\
7 & -20.2$\pm$0.8 & 6.70$\pm$4.1 & 8.4$\pm$5.1 & $\simgt$28.04 &26.16$\pm$0.34  & 26.39$\pm$0.78 & 23.87$\pm$0.06  &2.3 \\
8 & -20.1$\pm$0.8 & 6.40$\pm$4.1 & 8.0$\pm$5.1 &  $\simgt$28.04 &$\simgt$27.61   & $\simgt$27.39 & 24.48$\pm$0.14  &$\simgt$2.9 \\
9 & -22.4$\pm$0.1 & 38.3$\pm$5.3 & 47.9$\pm$6.6 & 26.42$\pm$0.29 & 25.45$\pm$0.19 & 24.26$\pm$0.15 & 23.47$\pm$0.09 & 1.34\\
10 & -19.4$\pm$1.2 & 4.20$\pm$3.4 & 5.2$\pm$4.2 & $\simgt$28.04 &$\simgt$27.61   & 27.2$\pm$1.2 & 24.88$\pm$0.23  &$\simgt$2.8 \\
\hdashline
11 & -- & -- & -- & $\simgt$28.04 & 26.81$\pm$0.55 & 27.2$\pm$1.2 & 24.94$\pm$0.19 & 1.83\\
12 & -20.2$\pm$0.8 & 6.60$\pm$4.1 & 8.20$\pm$5.1 &  $\simgt$28.04 &$\simgt$27.61 & 26.44$\pm$0.80 & 24.72$\pm$0.14 & 2.6 \\
\hline\end{tabular}  
\end{table*}
\floatplacement{figure}{!b}

\section{Spectra of \zlae\ LAEs} \label{app:spec-thum}
This appendix presents individual LAE spectra
observed with \textsc{DEIMOS}.
Two objects have been omitted because they were presented previously.
The first one is LAE 103027+052419 (object ID 4 in the field J1030+0524),
which is presented in Figure \ref{f:LAE04a}, and 
the second one is the \oiii\ emitter at $z{\sim}0.637$ 
(object ID 11 in the field J1137+3549) shown in Figure \ref{f:OIII}.

Each figure presents: the snapshot of the 2D slit spectrum 
in the wavelength range covered by the NB\civ\ filter,
the 1D spectrum of the emission line, the equivalent width 
probability distribution obtained from the photometry and the 1D spectrum
in velocity units with respect to the bluest pixel of the emission.
In the middle panels (1D spectrum), the red dotted line indicates one standard deviation
of the counts in the extraction box (1$\sigma$ error), 
the magenta dot-dashed line is the scaled transmission curve of the NB\civ\ filter, 
the vertical dashed lines indicate the position of the maximum (black line)
and the bluest pixel (blue line), and the grey shaded areas show
the wavelength of the skylines where residuals can be large.
The sky spectrum in an arbitrary scale is given for reference below each object spectrum.
The weighted skewness is measured in the range contained
by the vertical green solid lines that indicate $\lambda_{10\%,blue}$ and 
$\lambda_{10\%,red}$.
In the inset panels (P(EW$_0$)), the solid line and the dashed lines indicate the central value 
of the distribution and the range enclosing a probability of 0.68.
In the bottom panels (velocity units), the vertical solid lines indicate the wavelength range where
the spectroscopic equivalent width was measured. The blue squares 
with error bars at each side of the emission lines show the mean 
counts level of the continuum, where the vertical error bars 
are the standard deviation of the flux in counts and
the horizontal error bars are the wavelength range used to 
estimate the mean. At both sides of the emission, we avoided 
regions of significant skyline residuals (grey areas)
and regions with high contamination from internally reflected light (red-hashed regions).


\begin{figure}
\includegraphics[width=80mm]{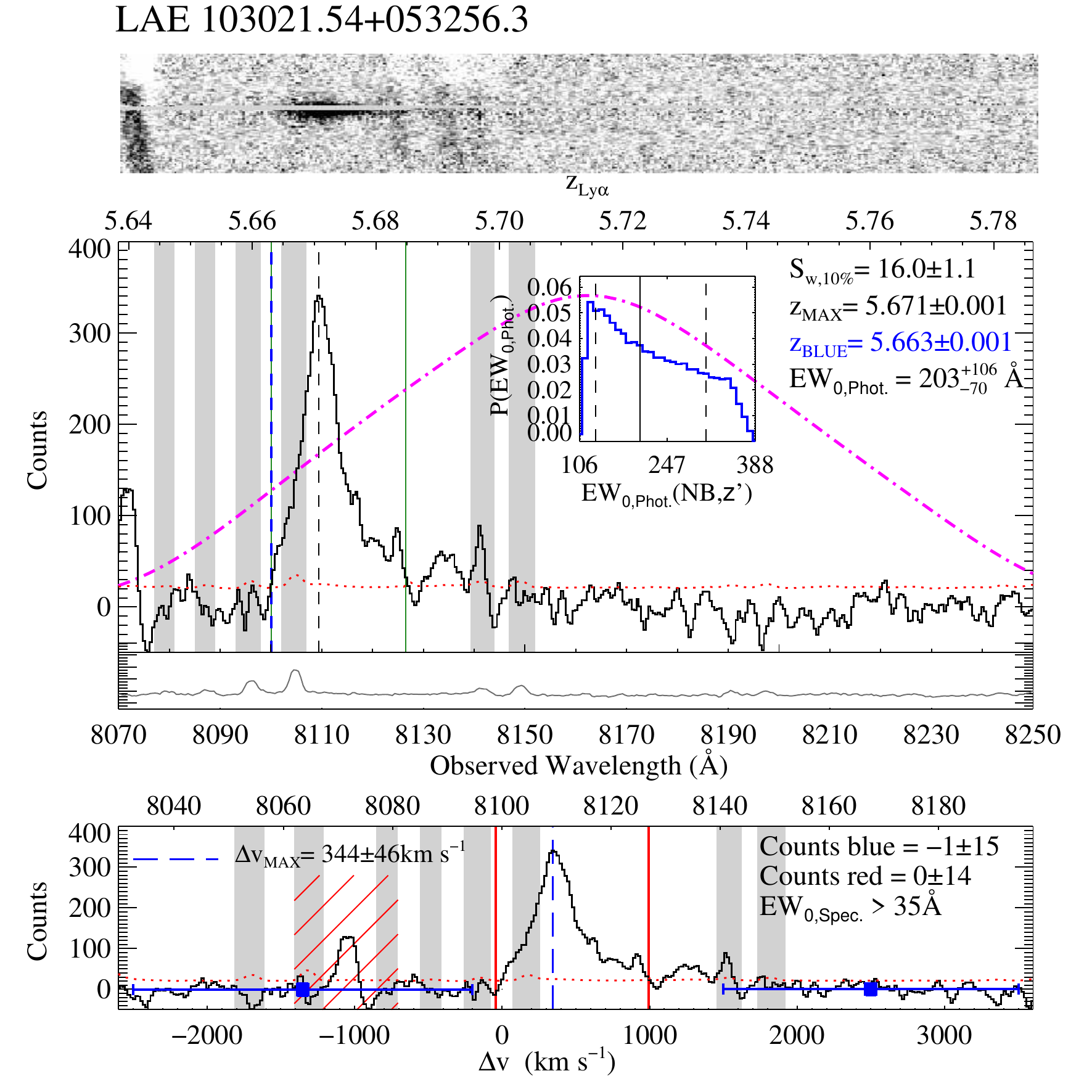} 
\caption{Field J1030+0524. ID 1}
\end{figure}
\begin{figure}
\includegraphics[width=80mm]{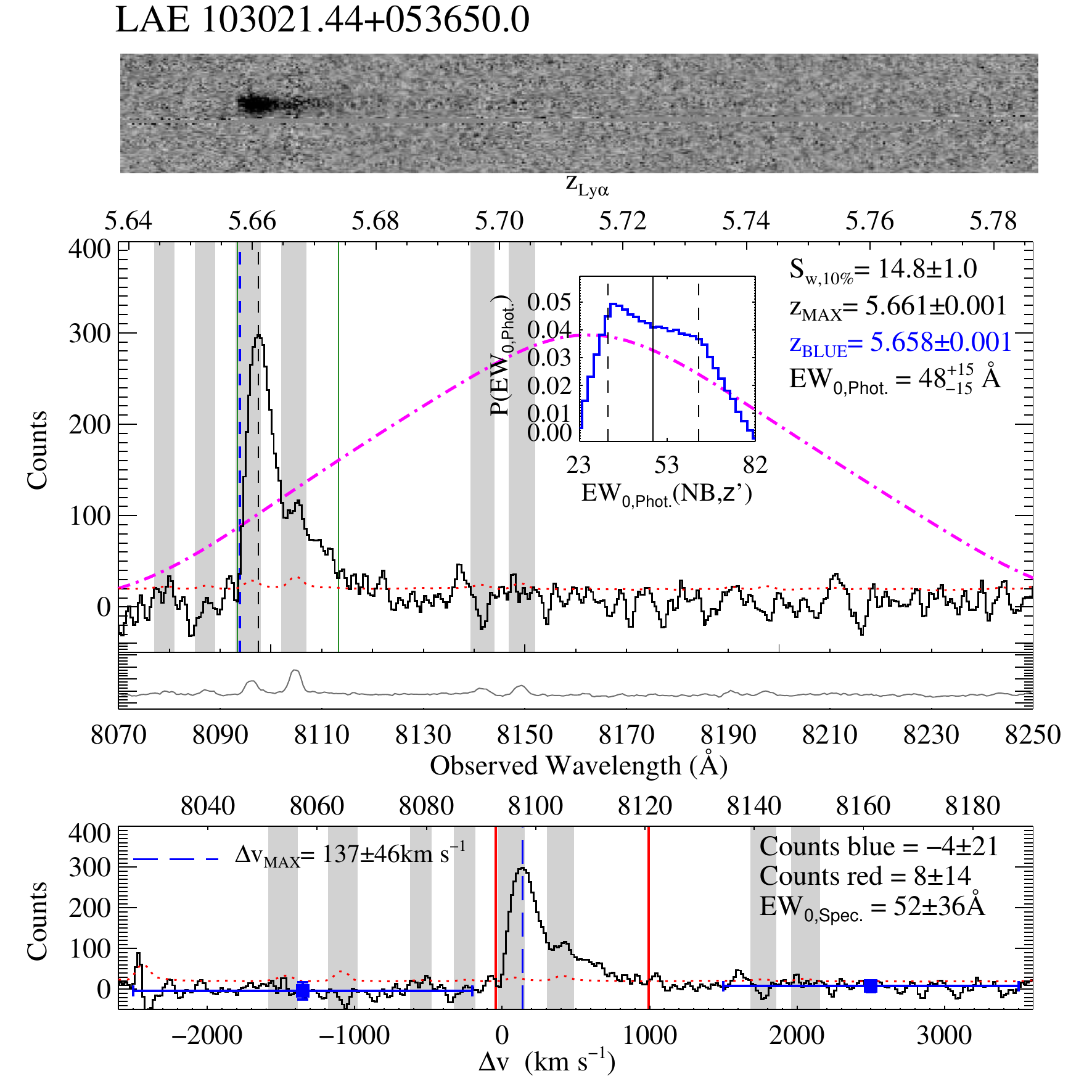} 
\caption{Field J1030+0524. ID 2}
\end{figure}
\begin{figure}
\includegraphics[width=80mm]{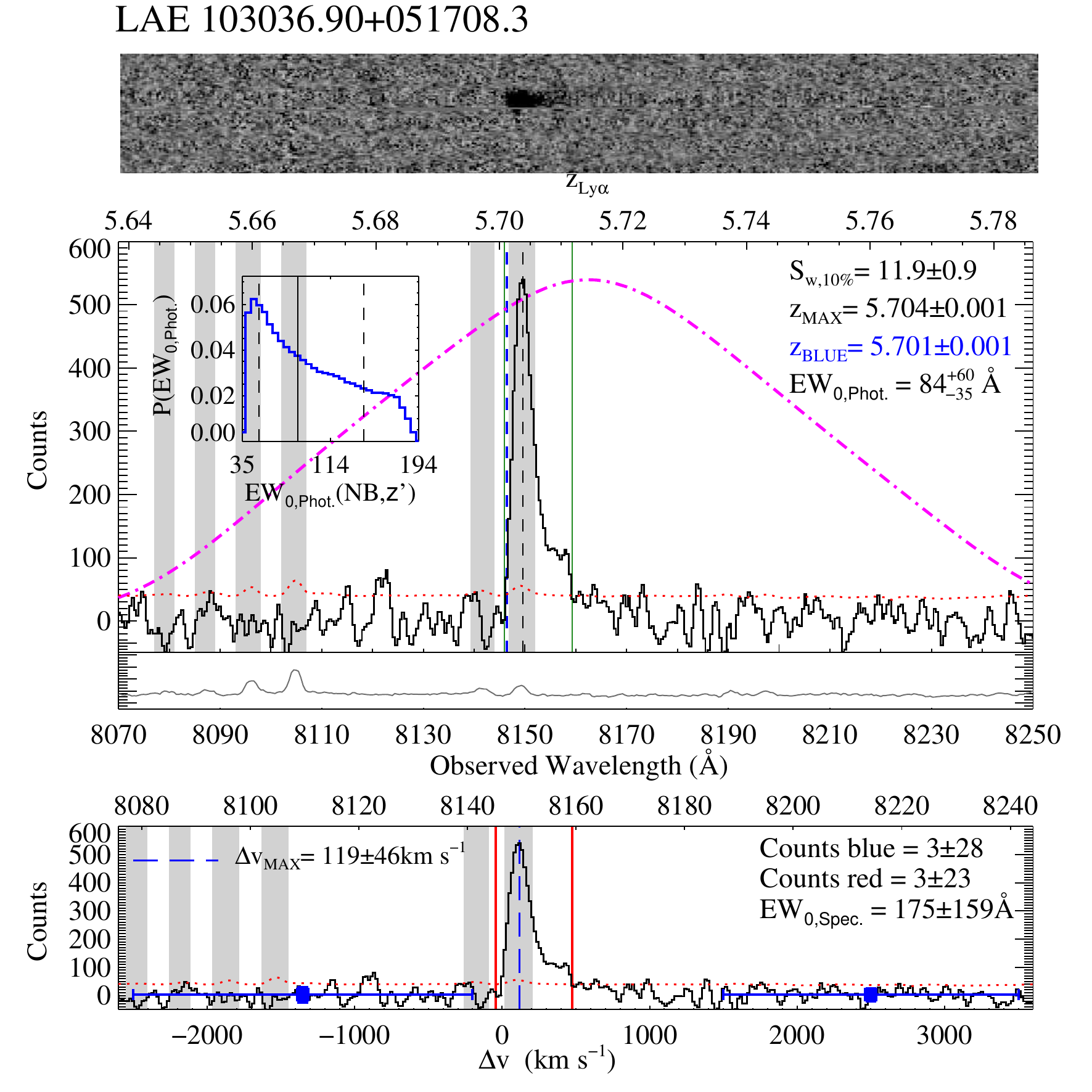} 
\caption{Field J1030+0524. ID 3}
\end{figure}
\begin{figure}
\includegraphics[width=80mm]{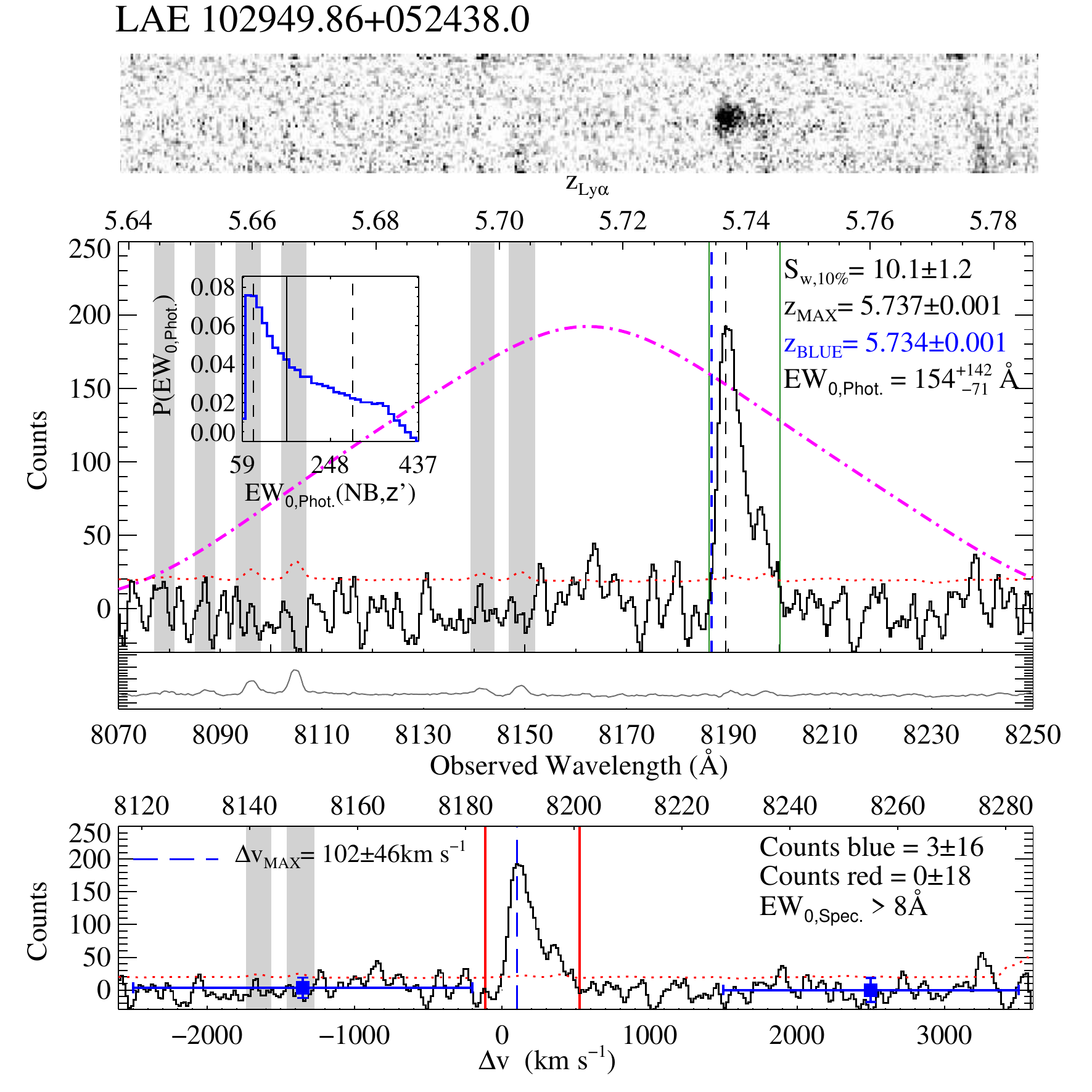} 
\caption{Field J1030+0524. ID 5}
\end{figure}

\begin{figure}
\includegraphics[width=80mm]{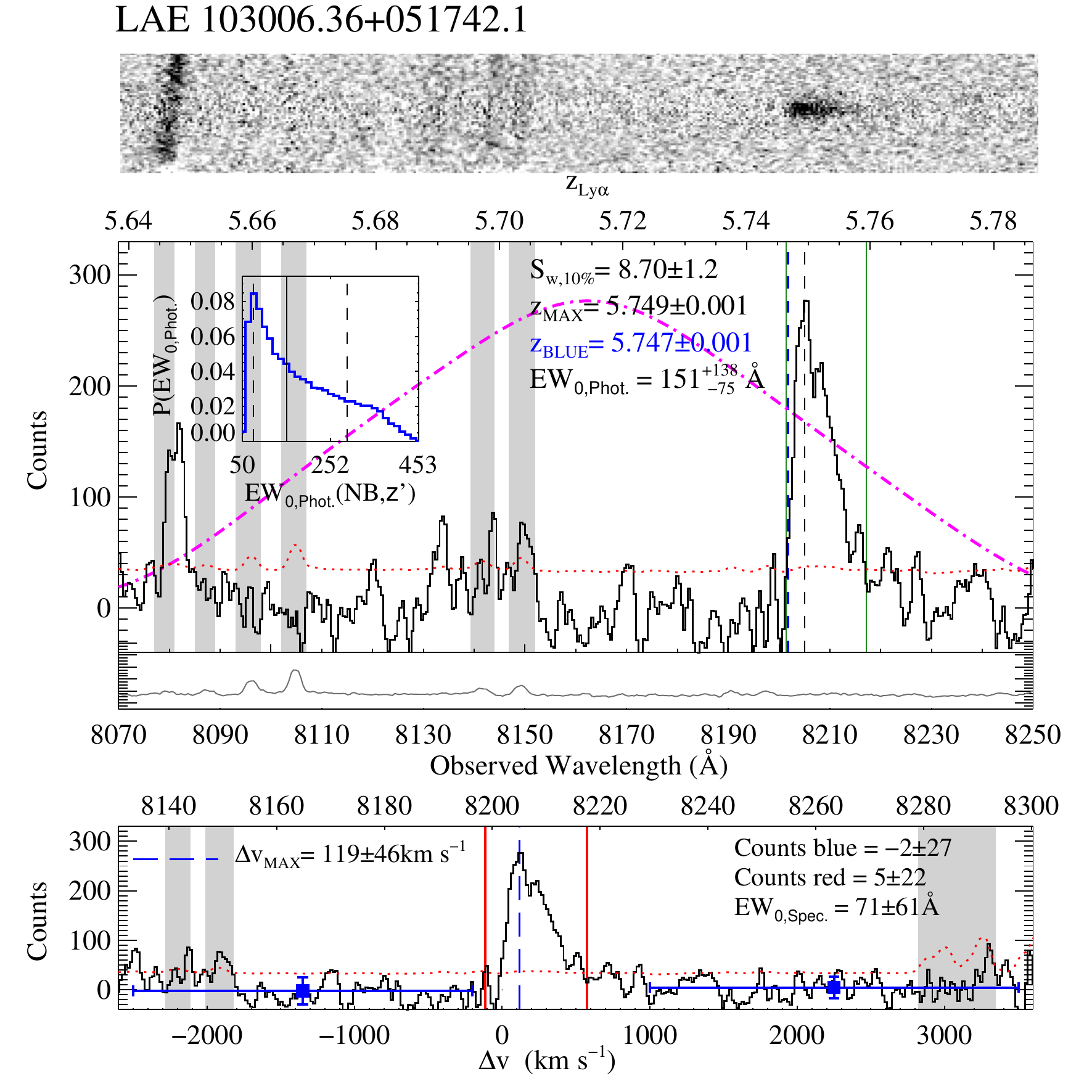} 
\caption{Field J1030+0524. ID 6}
\end{figure}
\begin{figure}
\includegraphics[width=80mm]{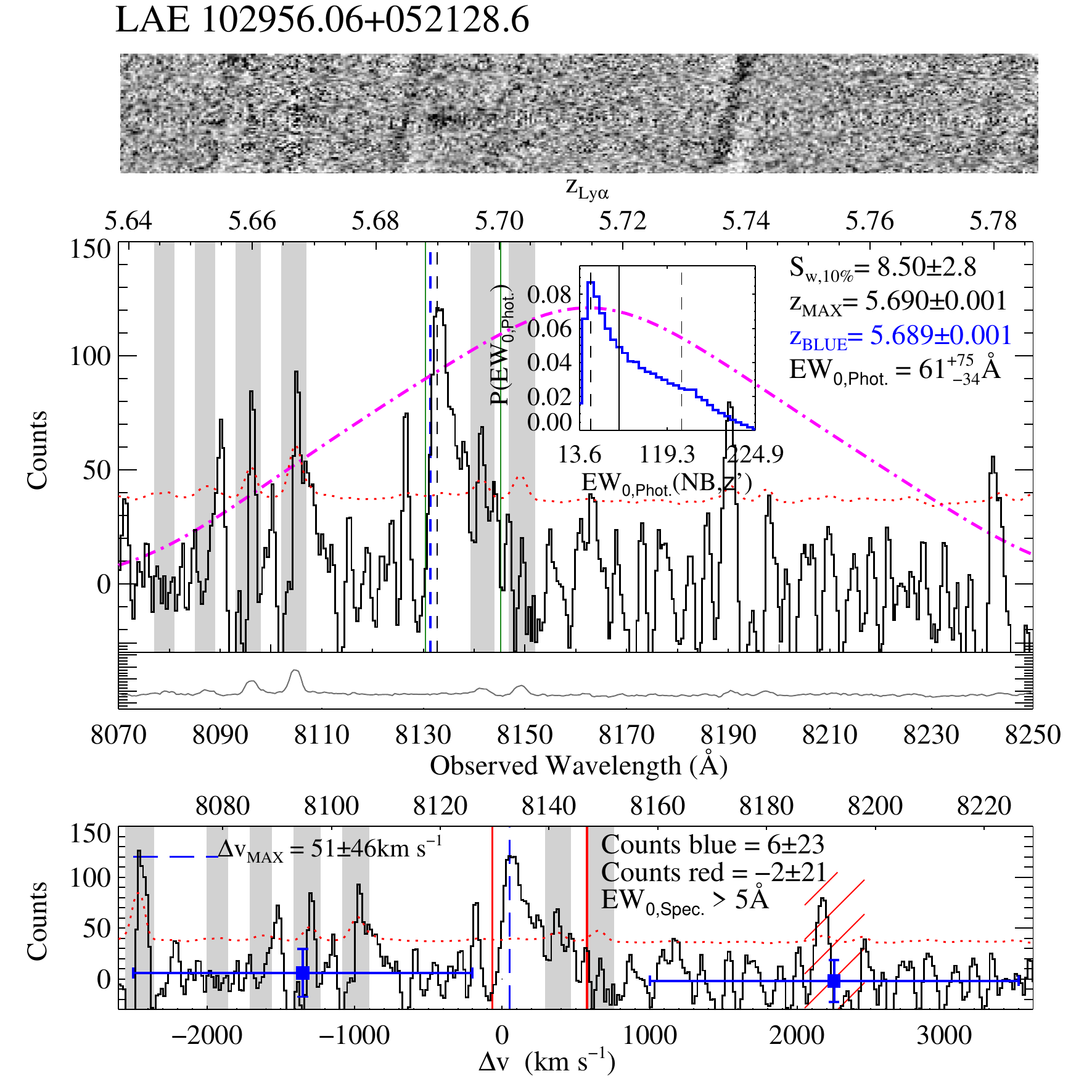} 
\caption{Field J1030+0524. ID 7}
\end{figure}
\begin{figure}
\includegraphics[width=80mm]{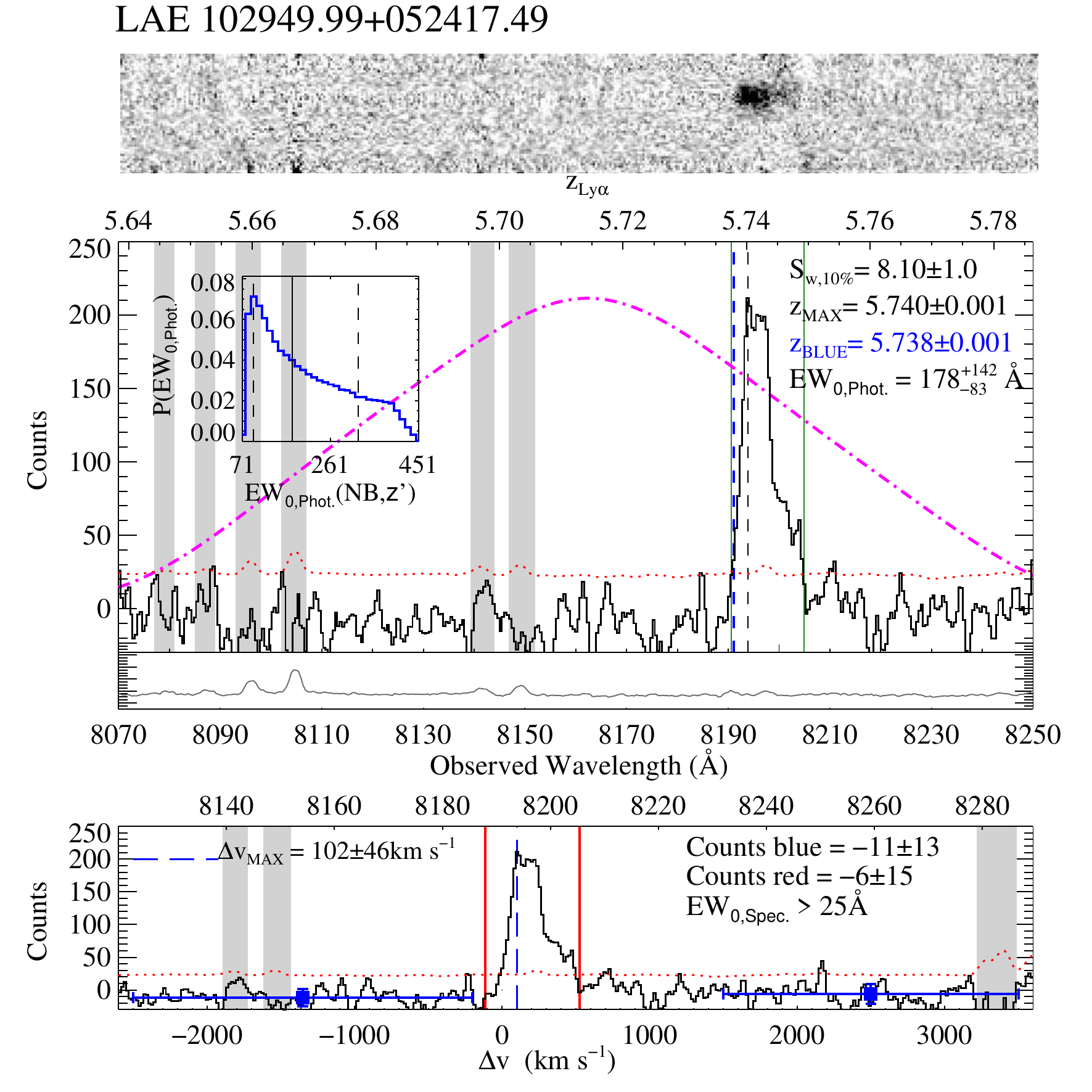} 
\caption{Field J1030+0524. ID 8}
\end{figure}

\begin{figure}
\includegraphics[width=80mm]{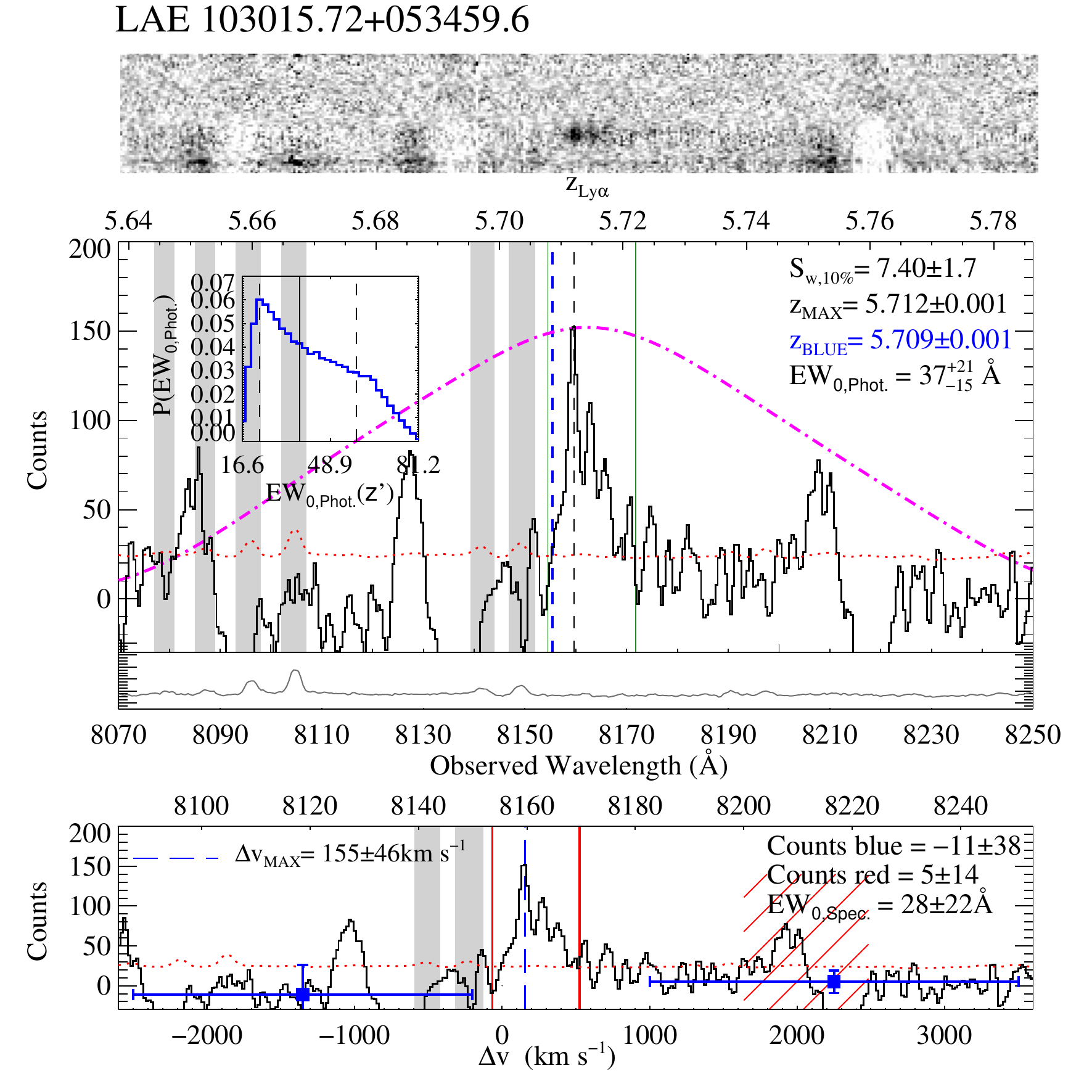} 
\caption{Field J1030+0524. ID 9}
\end{figure}

\begin{figure}
\includegraphics[width=80mm]{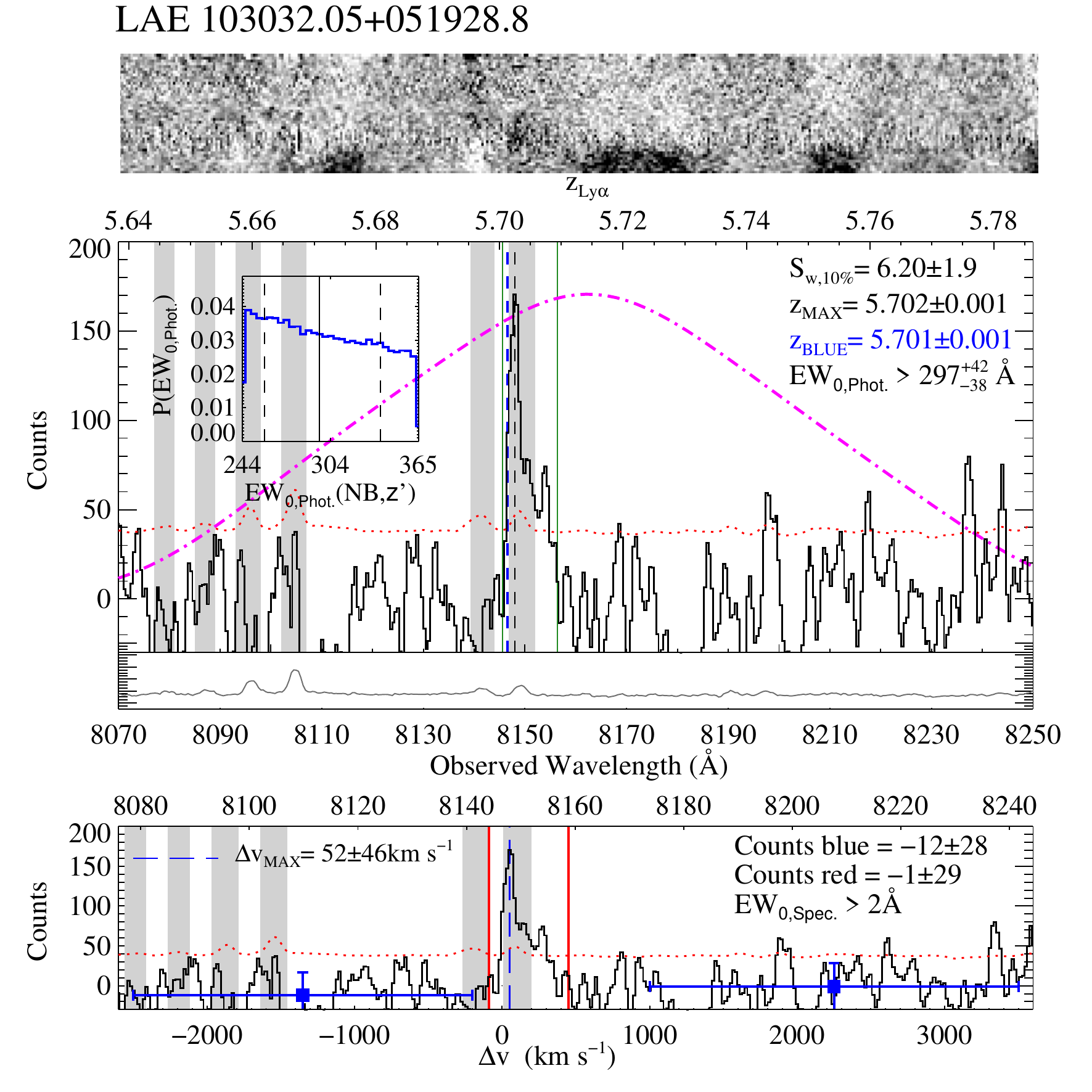} 
\caption{Field J1030+0524. ID 10}
\end{figure}
\begin{figure}
\includegraphics[width=80mm]{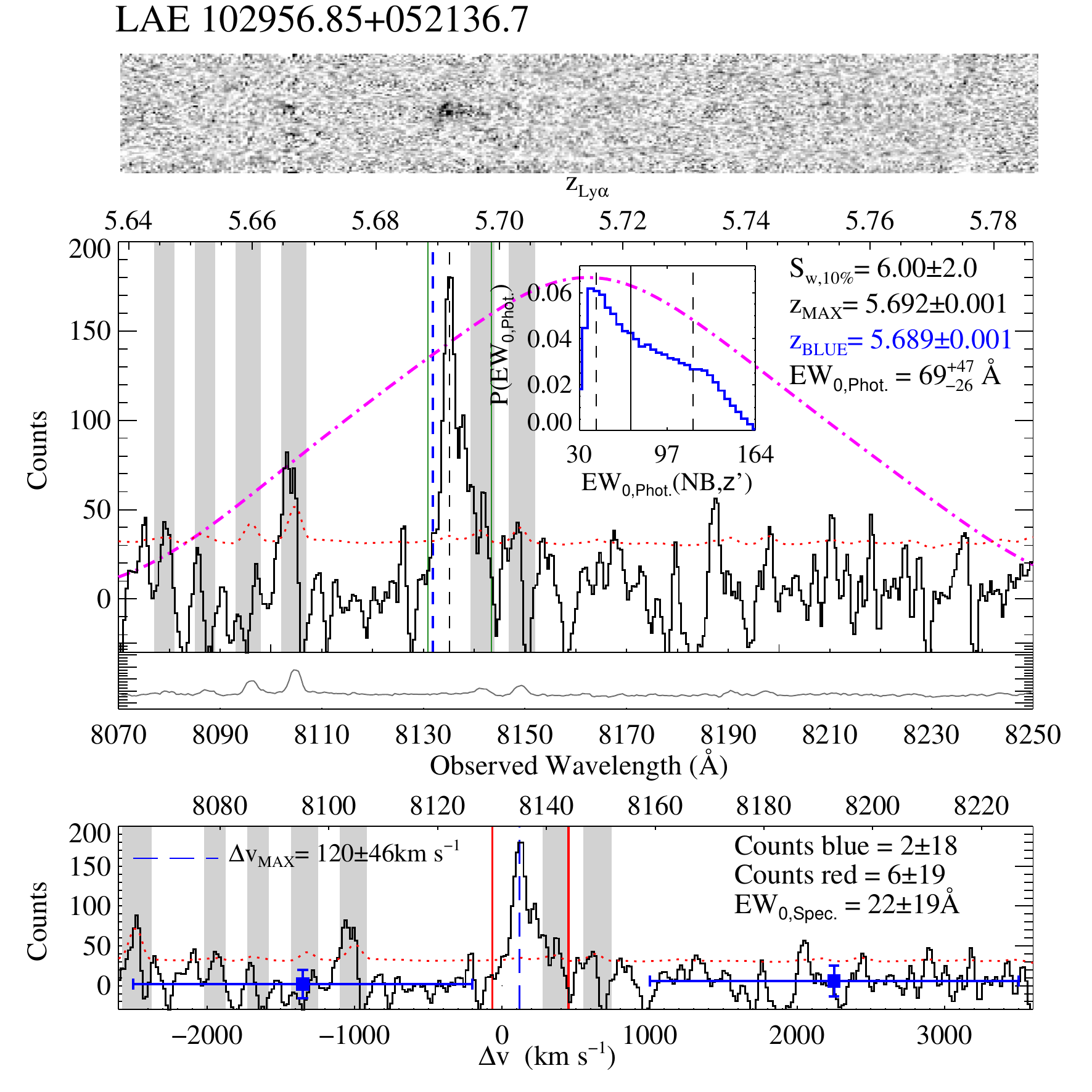} 
\caption{Field J1030+0524. ID 11}
\end{figure}

\begin{figure}
\includegraphics[width=80mm]{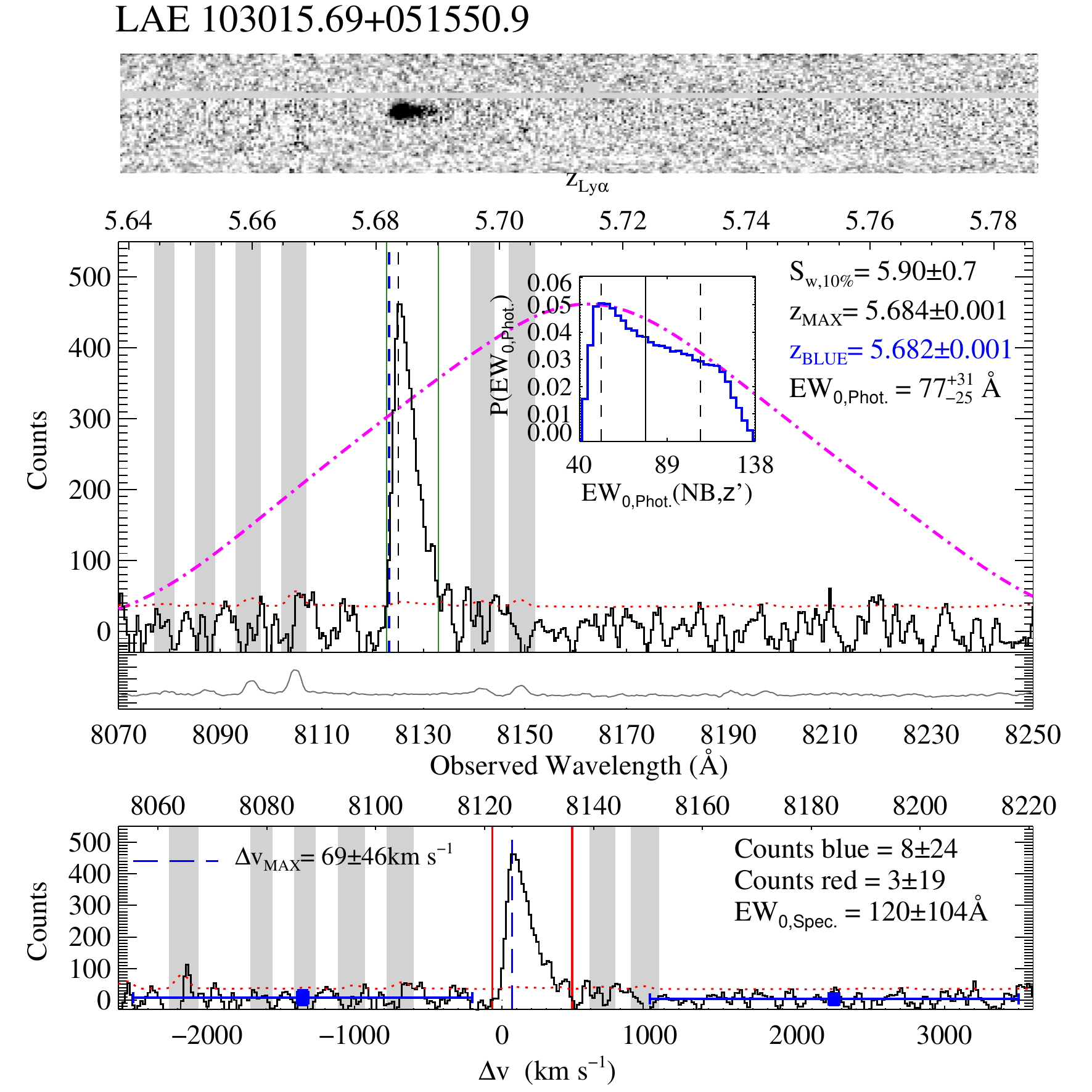} 
\caption{Field J1030+0524. ID 12}
\end{figure}
\clearpage

\begin{figure}
\includegraphics[width=80mm]{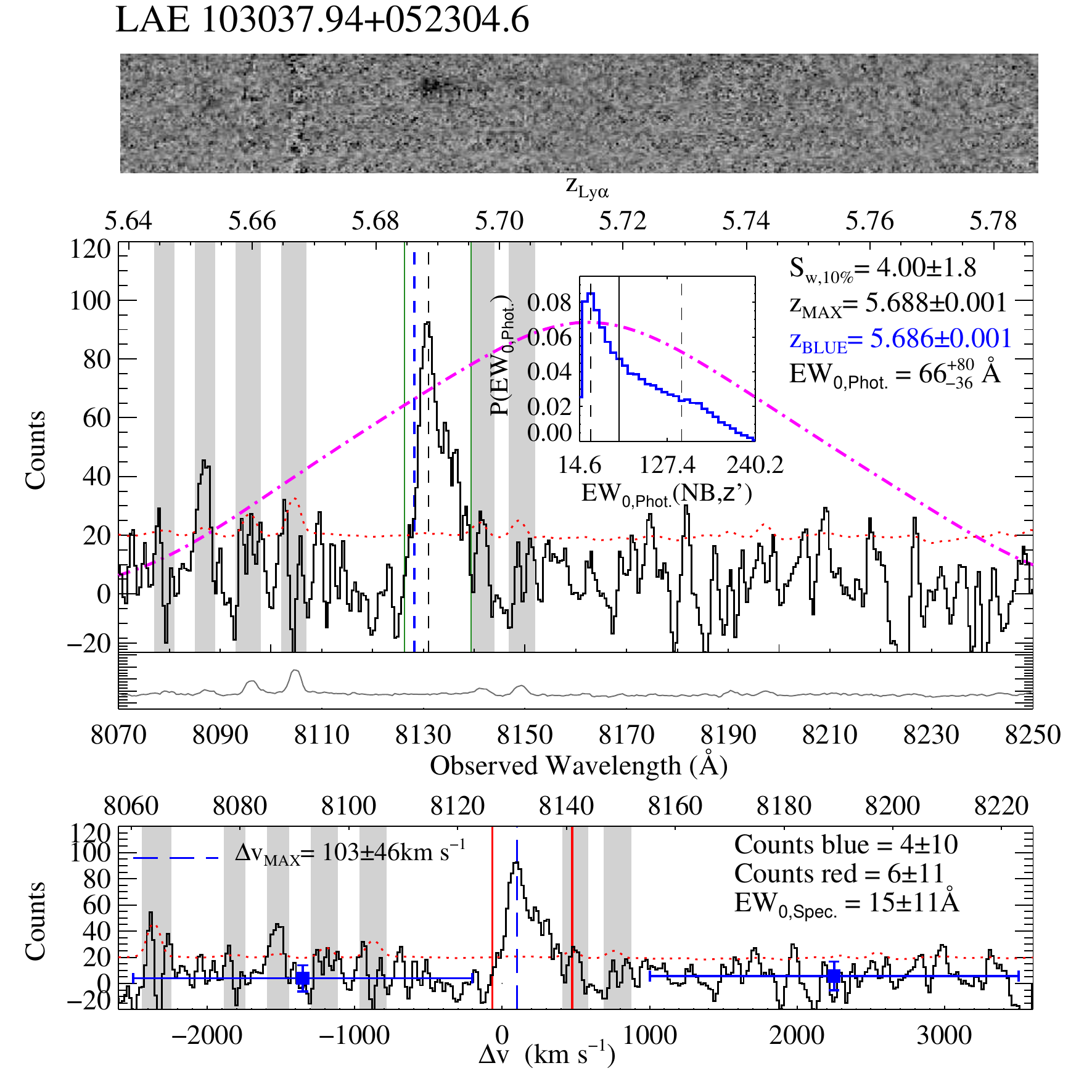} 
\caption{Field J1030+0524. ID 13}
\end{figure}

\begin{figure}
\includegraphics[width=80mm]{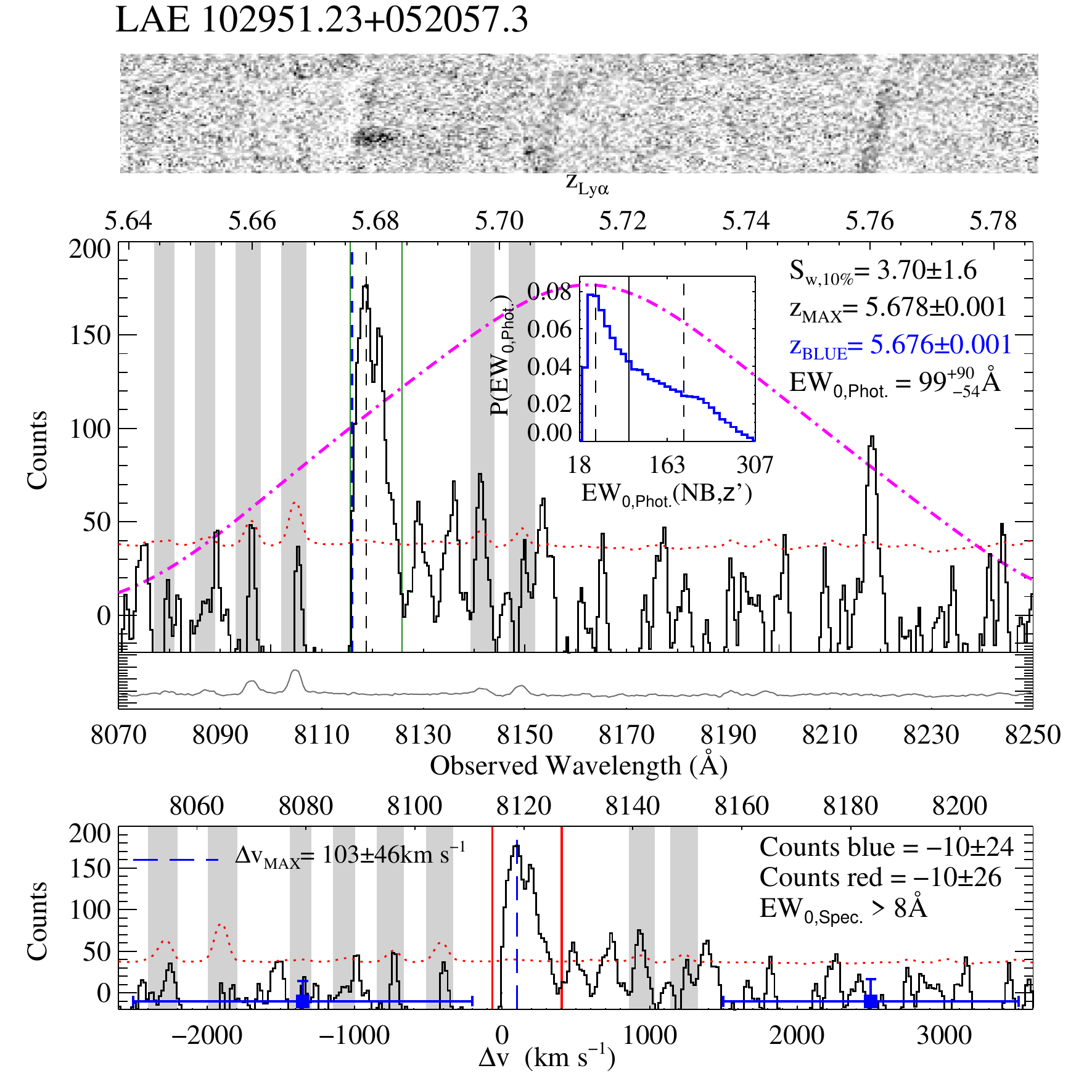} 
\caption{Field J1030+0524. ID14}
\end{figure}
\begin{figure}
\includegraphics[width=80mm]{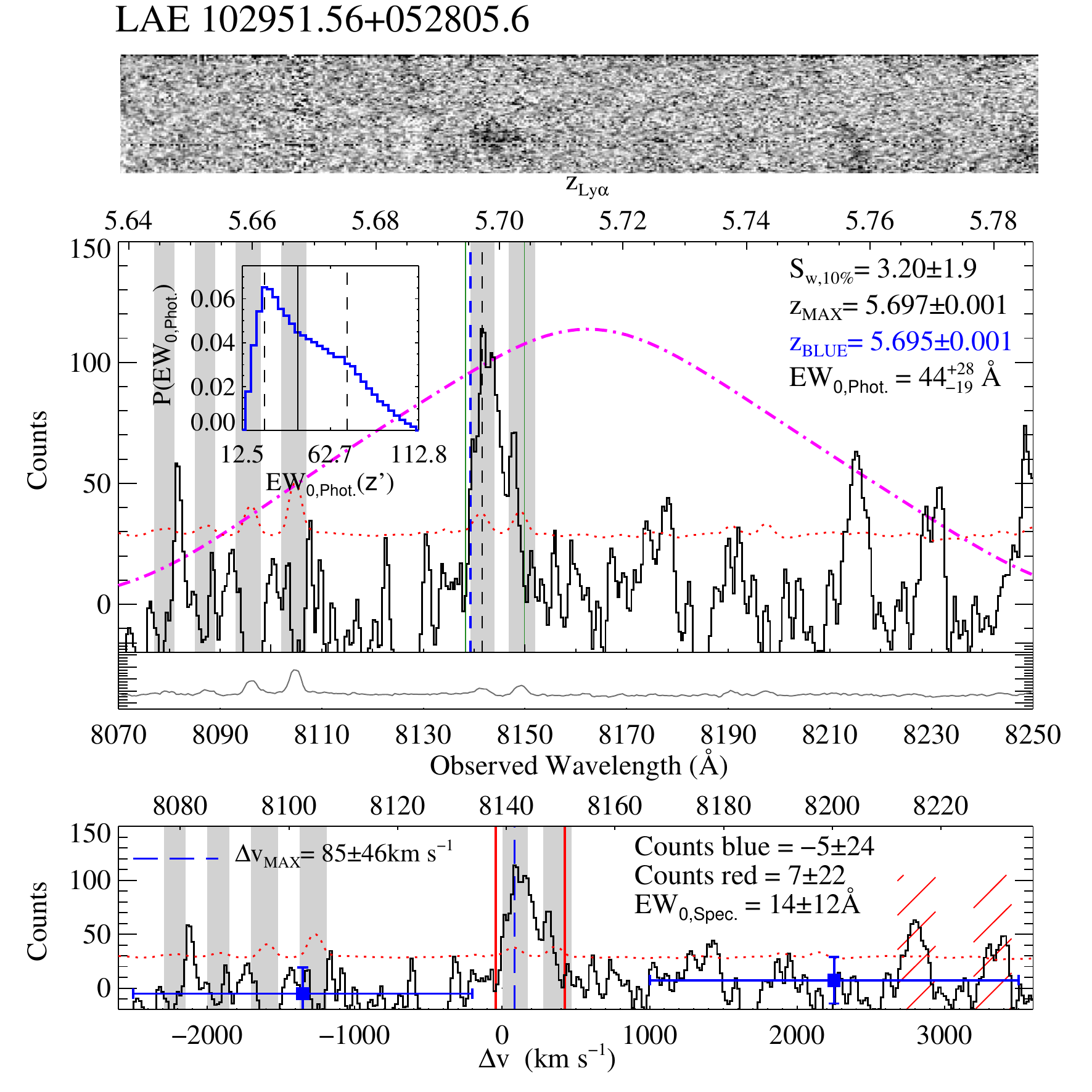} 
\caption{Field J1030+0524. ID 15}
\end{figure}

\begin{figure}
\includegraphics[width=80mm]{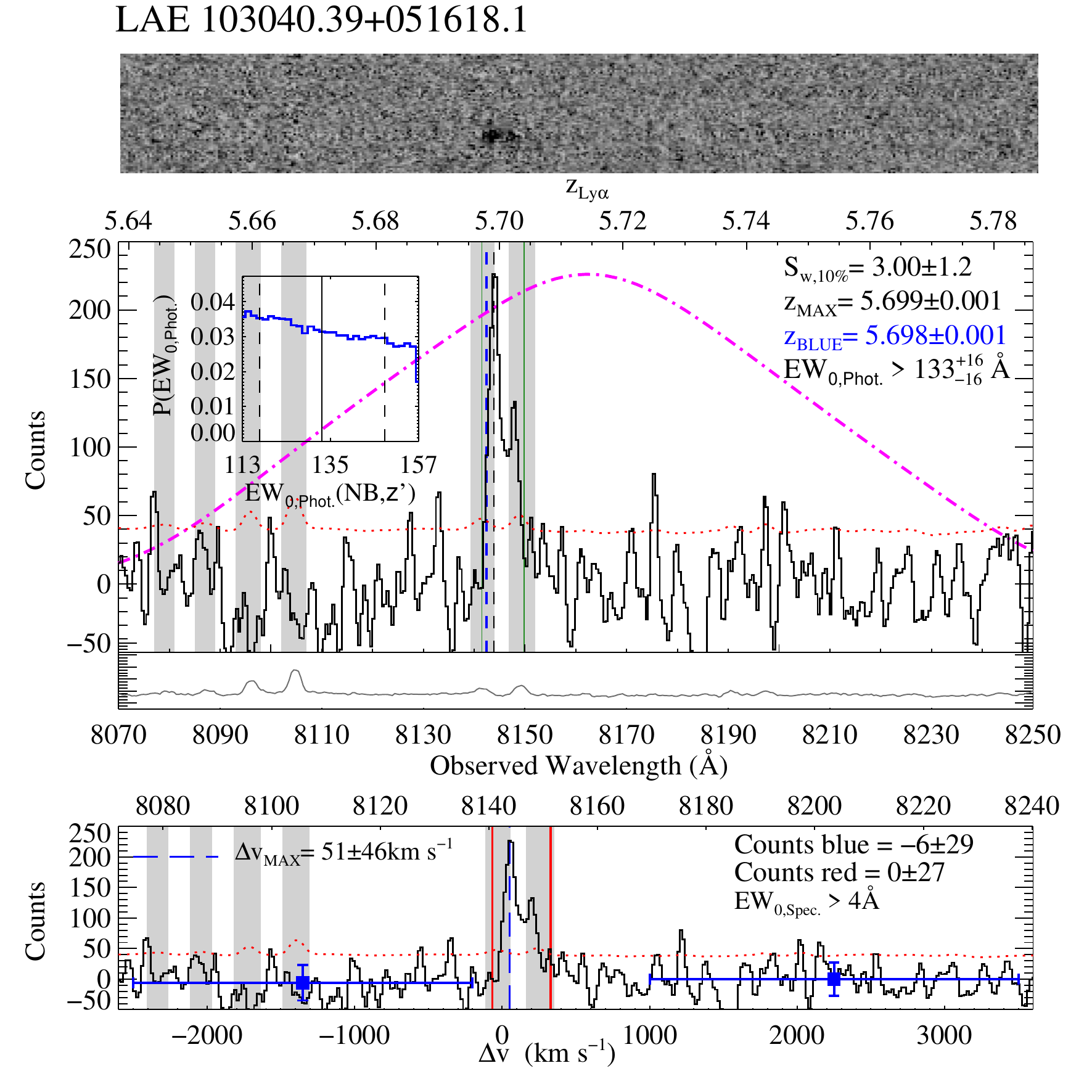} 
\caption{Field J1030+0524. ID 16}
\end{figure}
\clearpage

\begin{figure}
\includegraphics[width=80mm]{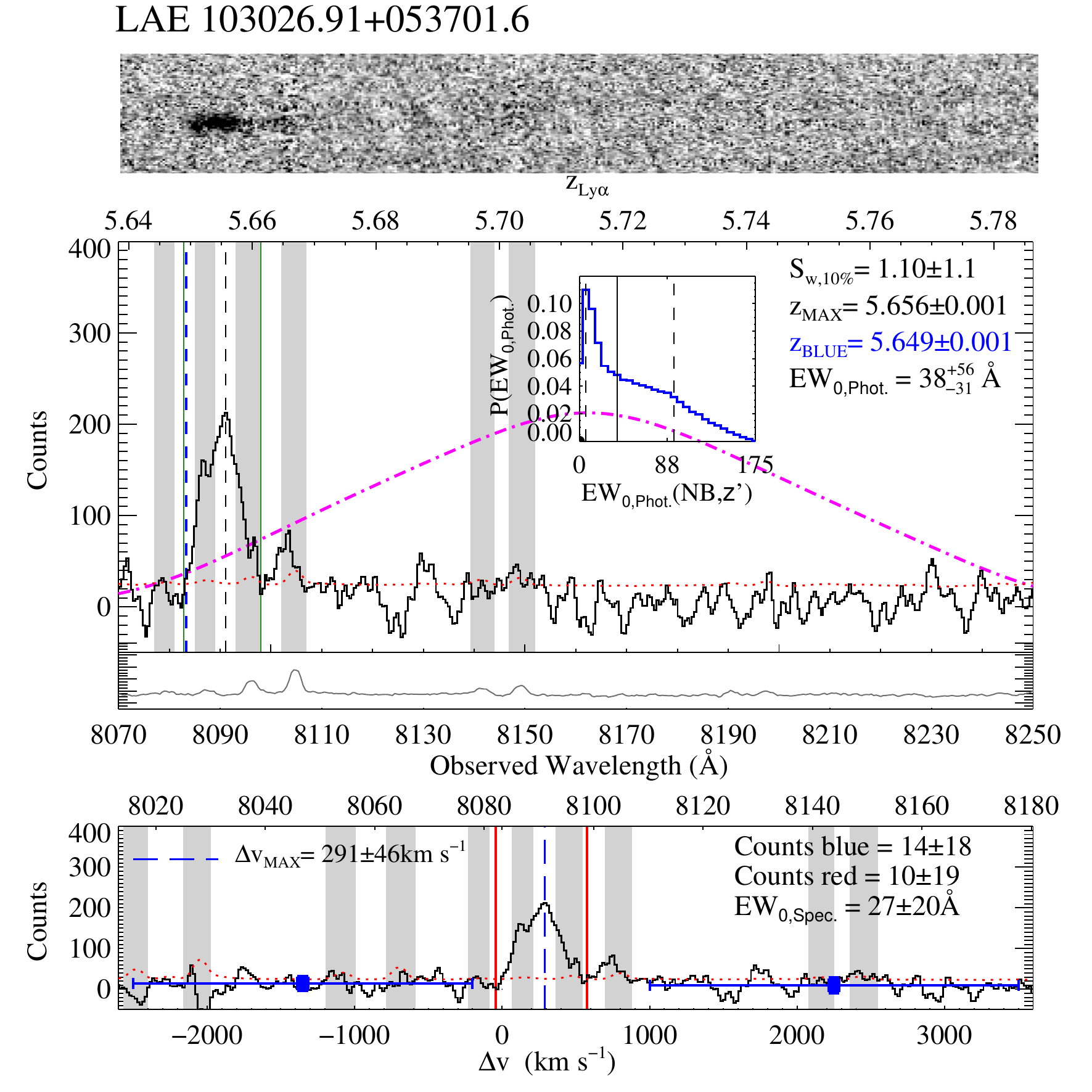} 
\caption{Field J1030+0524. ID 17}
\end{figure}

\begin{figure}
\includegraphics[width=80mm]{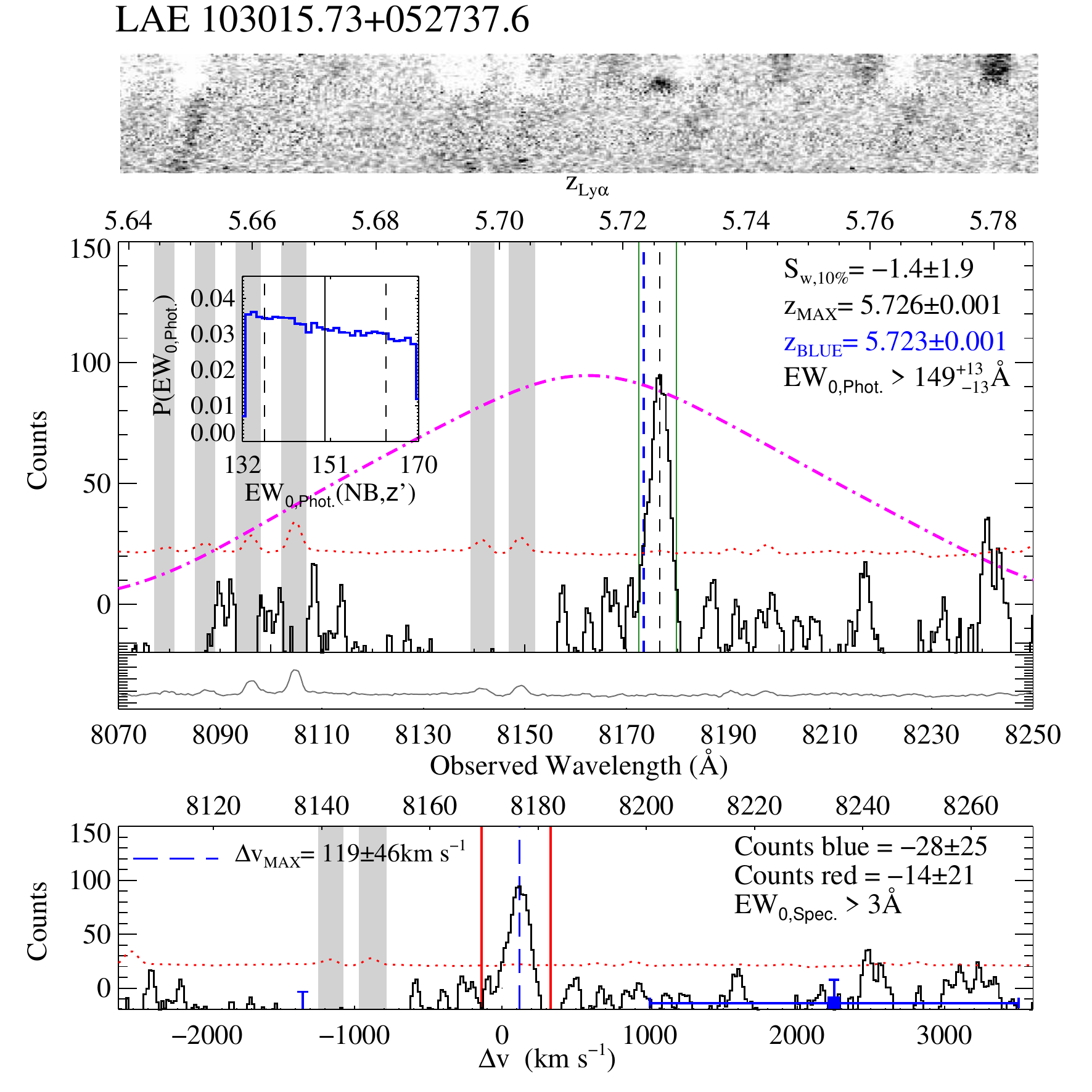} 
\caption{Field J1030+0524. ID 18}
\end{figure}

\begin{figure}
\includegraphics[width=80mm]{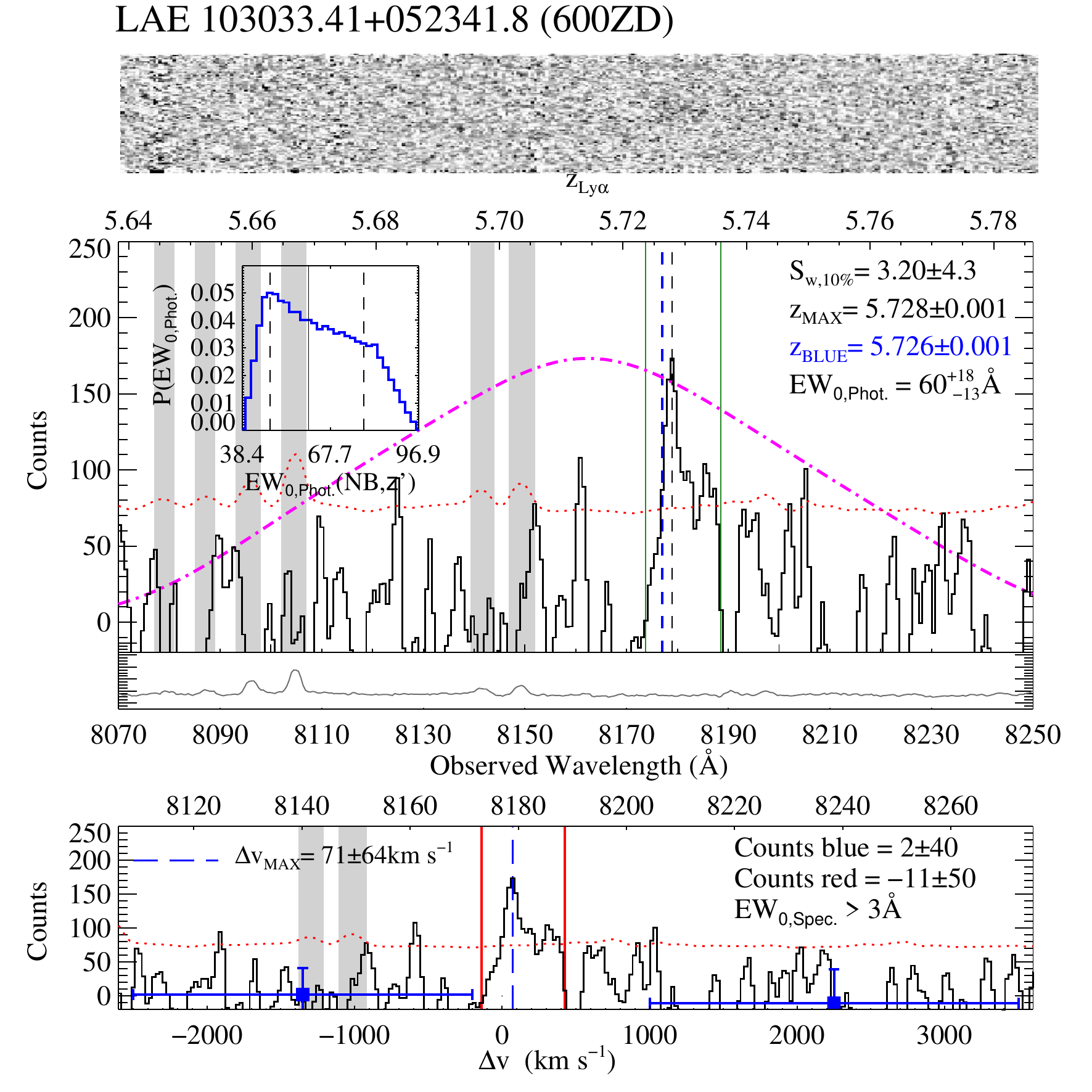} 
\caption{Field J1030+0524. ID 19}
\end{figure}

\begin{figure}
\includegraphics[width=80mm]{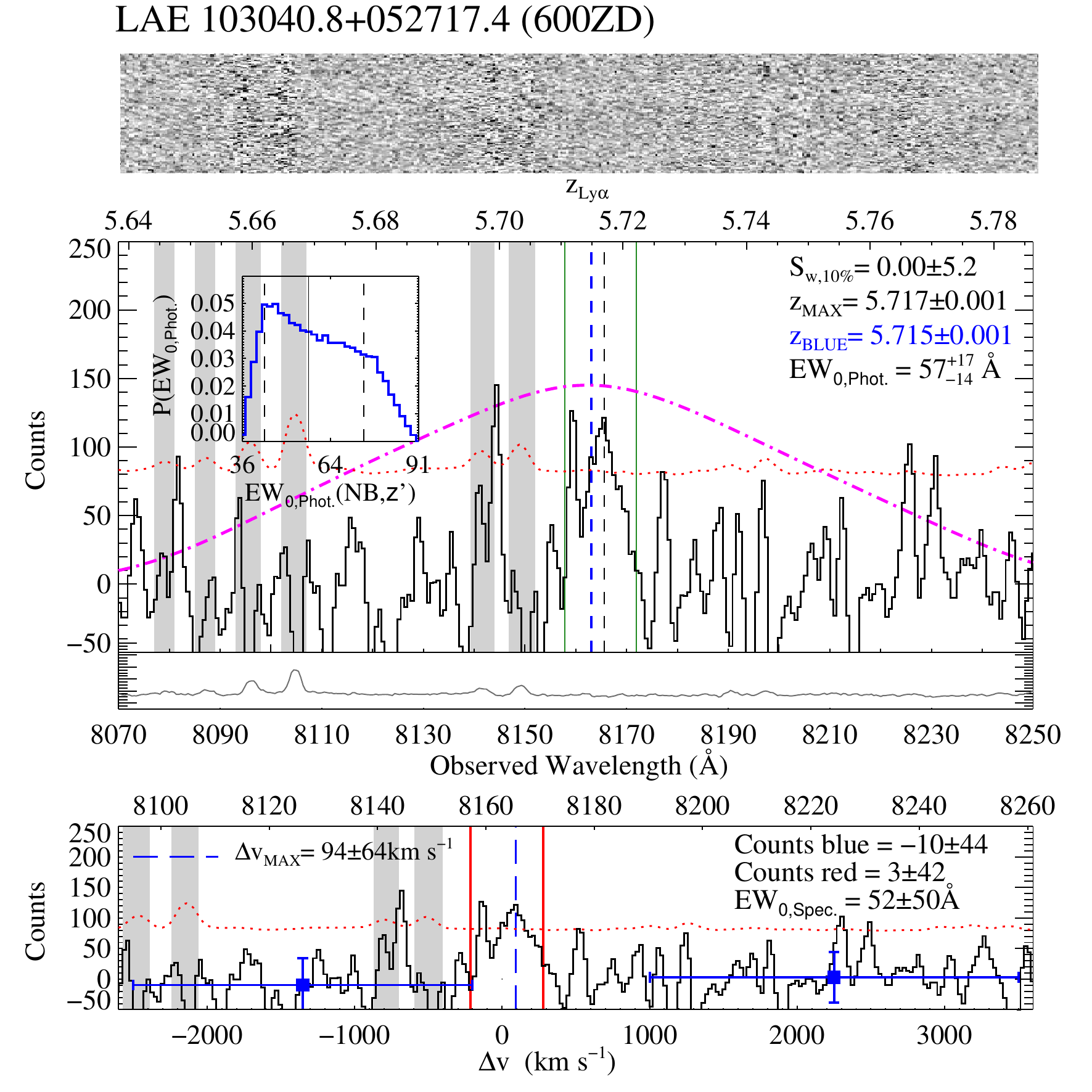} 
\caption{Field J1030+0524. ID 20}
\end{figure}
\begin{figure}
\includegraphics[width=80mm]{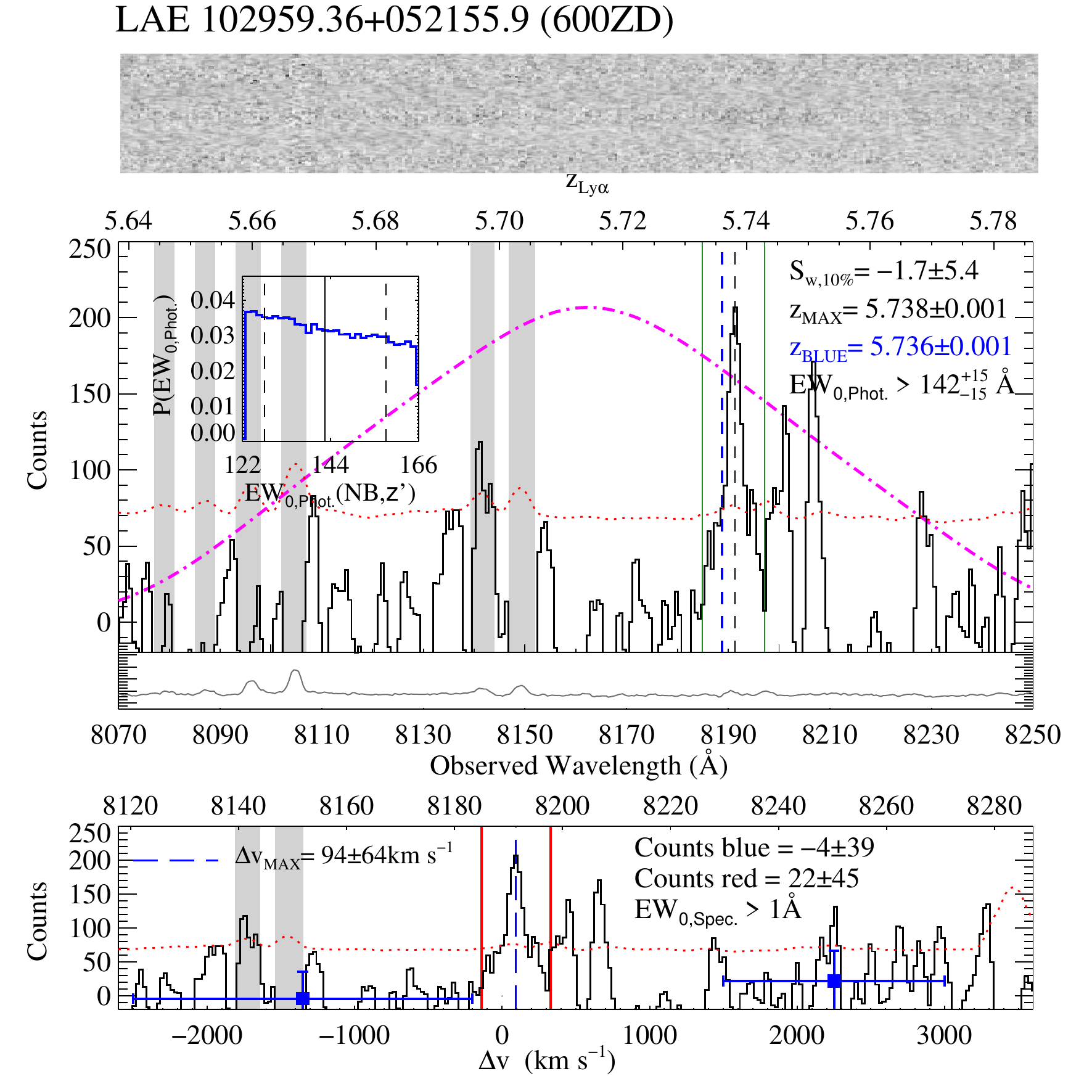} 
\caption{Field J1030+0524. ID 21}
\end{figure}


\begin{figure}
\includegraphics[width=80mm]{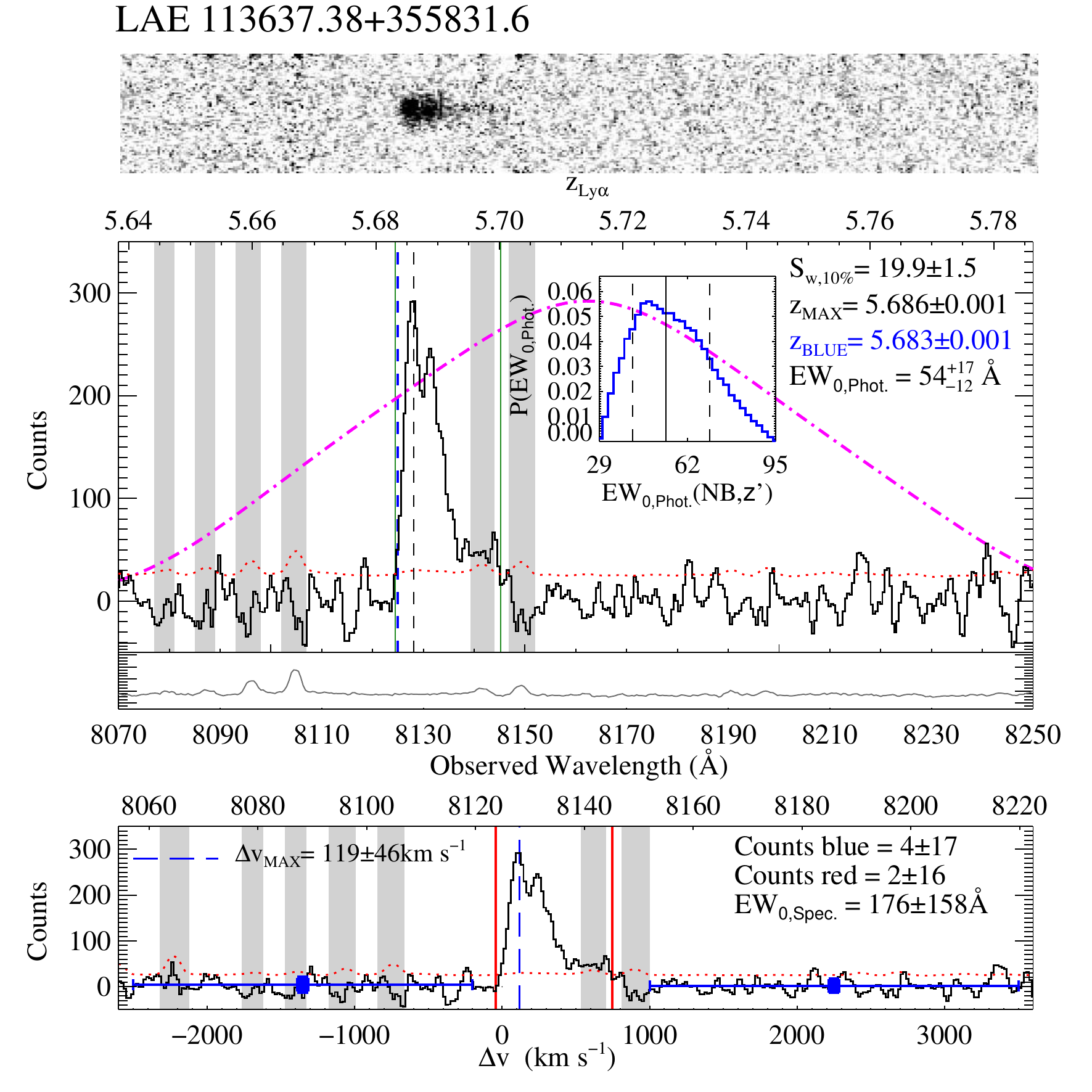} 
\caption{Field J1137+3549. ID 1}
\end{figure}
\begin{figure}
\includegraphics[width=80mm]{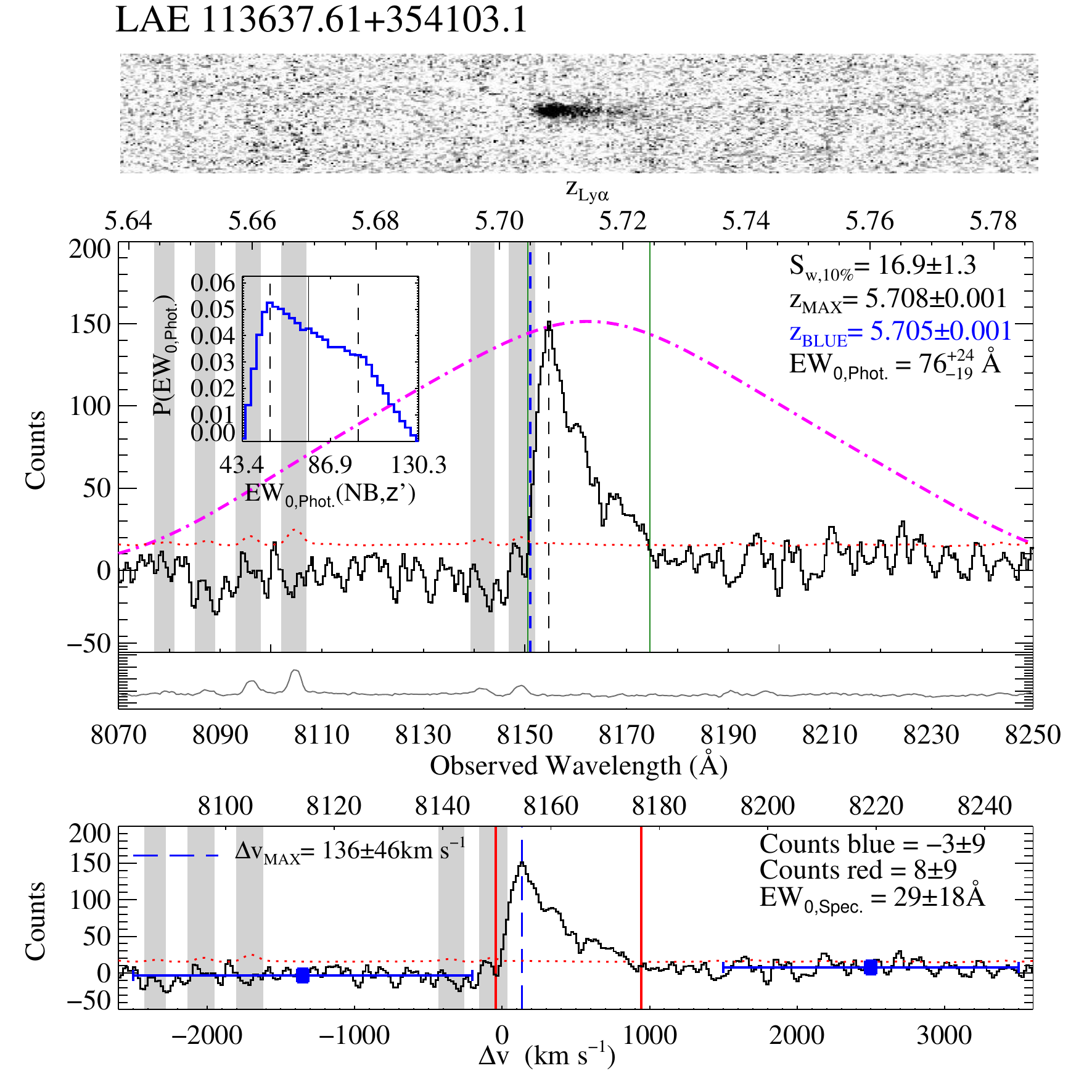} 
\caption{Field J1137+3549. ID 2}
\end{figure}

\begin{figure}
\includegraphics[width=80mm]{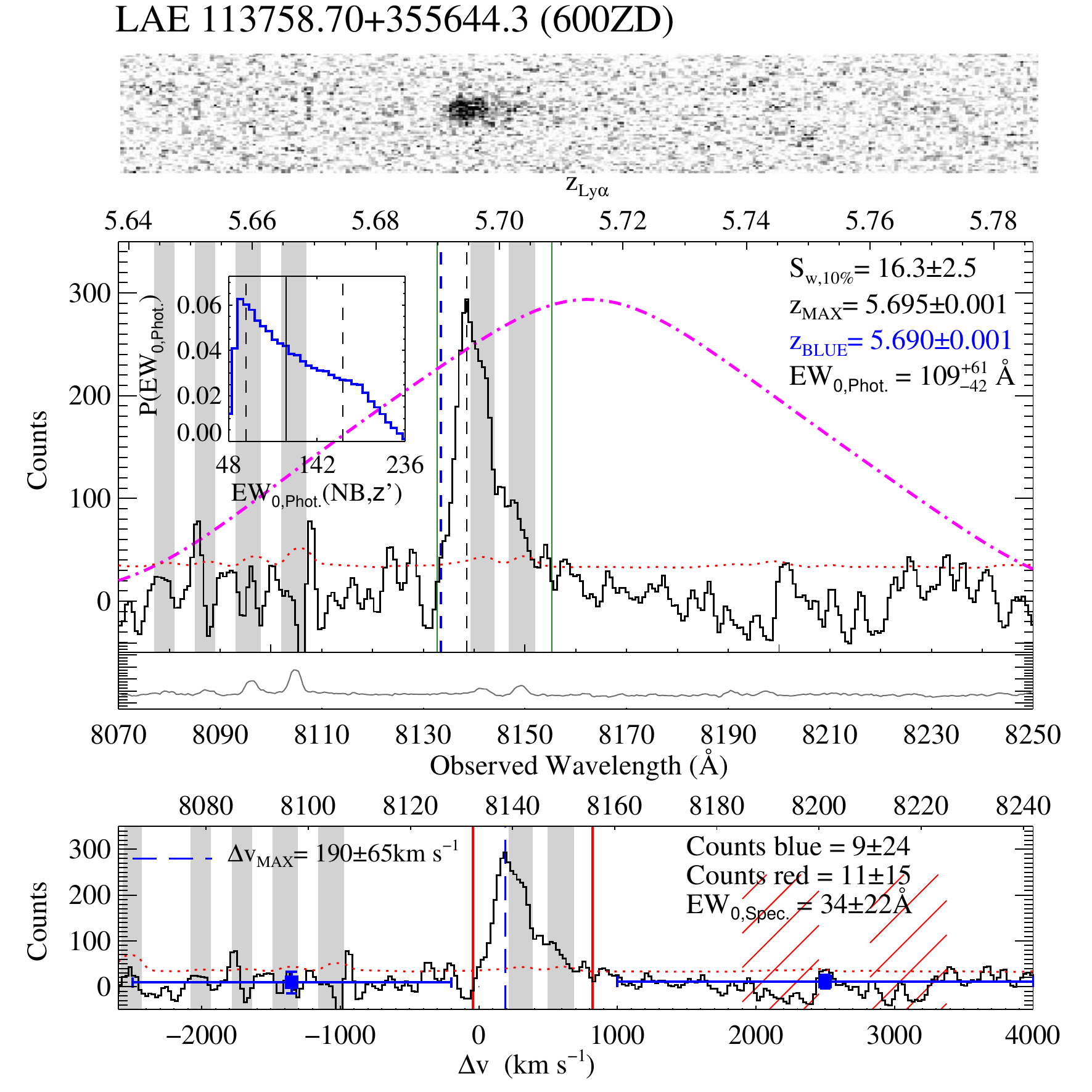} 
\caption{Field J1137+3549. ID 3}
\end{figure}
\begin{figure}
\includegraphics[width=80mm]{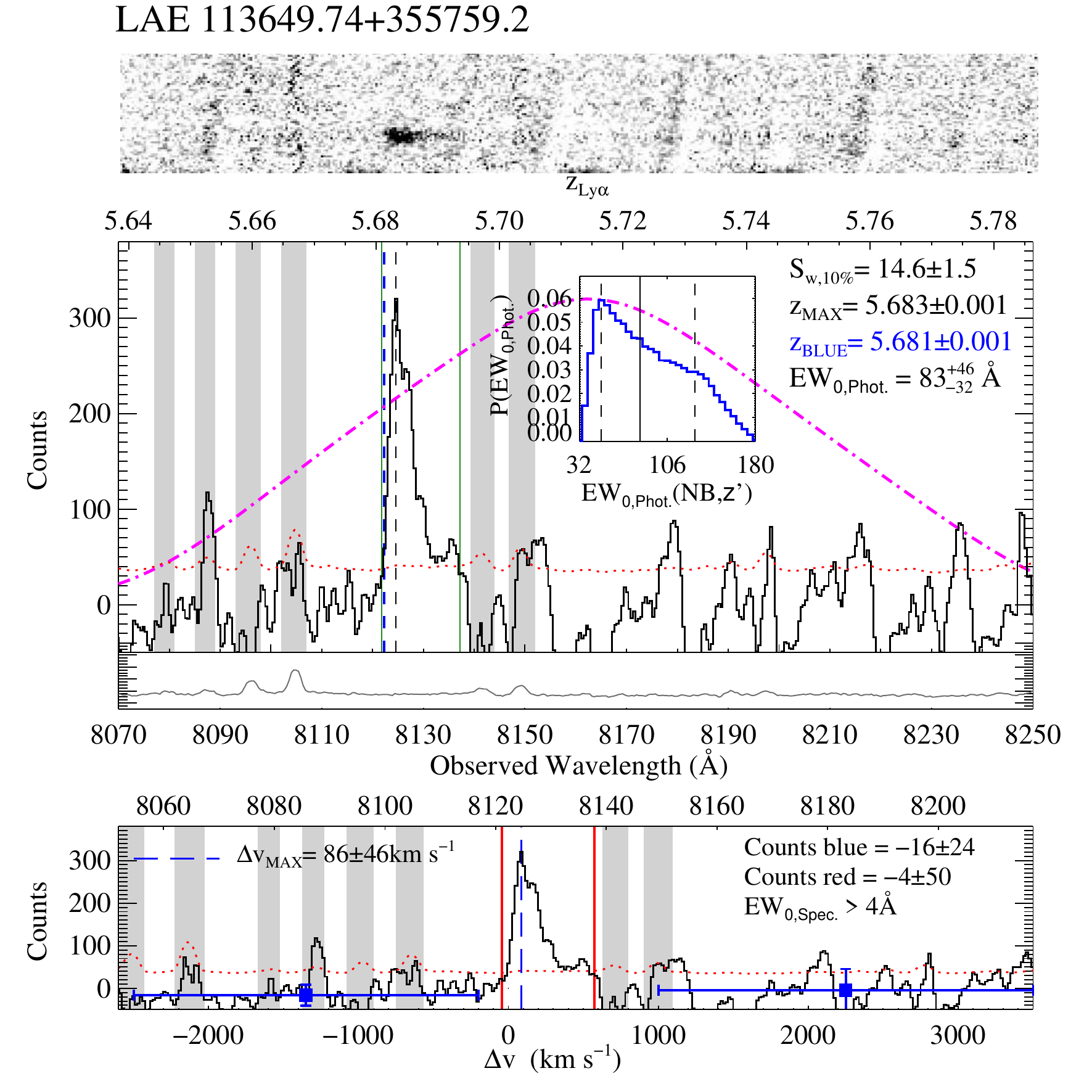} 
\caption{Field J1137+3549. ID 4}
\end{figure}
\begin{figure}
\includegraphics[width=80mm]{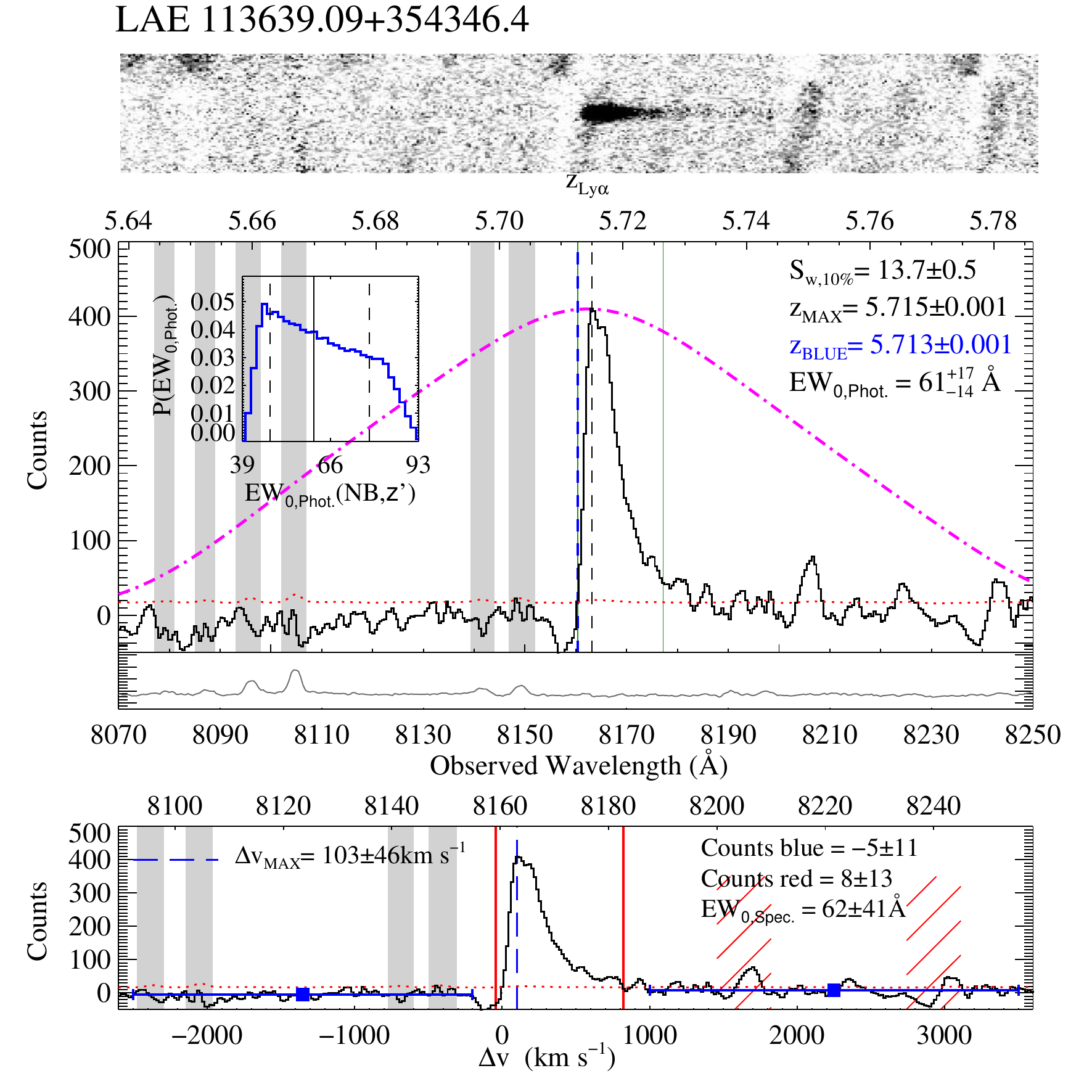} 
\caption{Field J1137+3549. ID 5}
\end{figure}
\begin{figure}
\includegraphics[width=80mm]{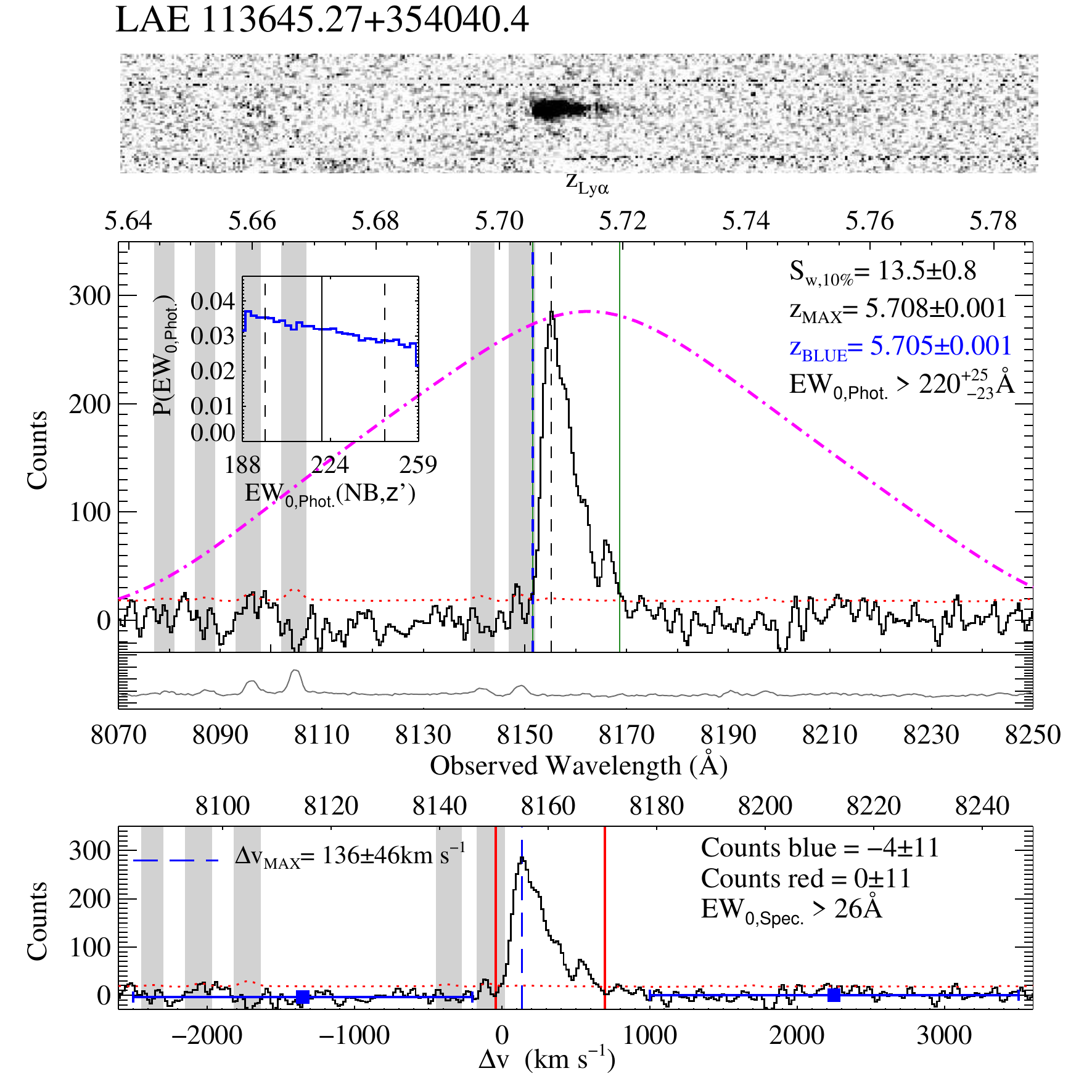} 
\caption{Field J1137+3549. ID 6}
\end{figure}

\begin{figure}
\includegraphics[width=80mm]{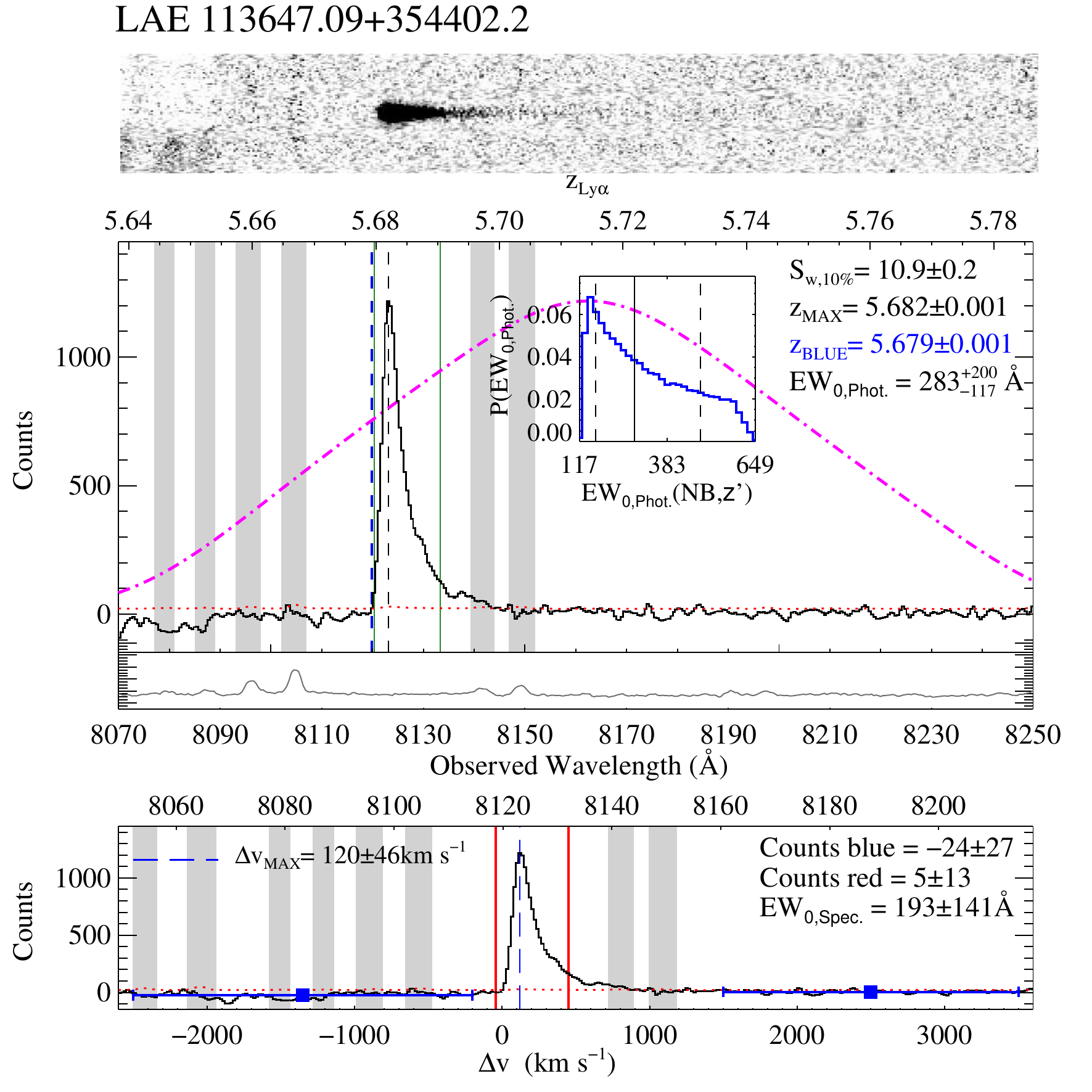} 
\caption{Field J1137+3549. ID 7}
\end{figure}
\begin{figure}
\includegraphics[width=80mm]{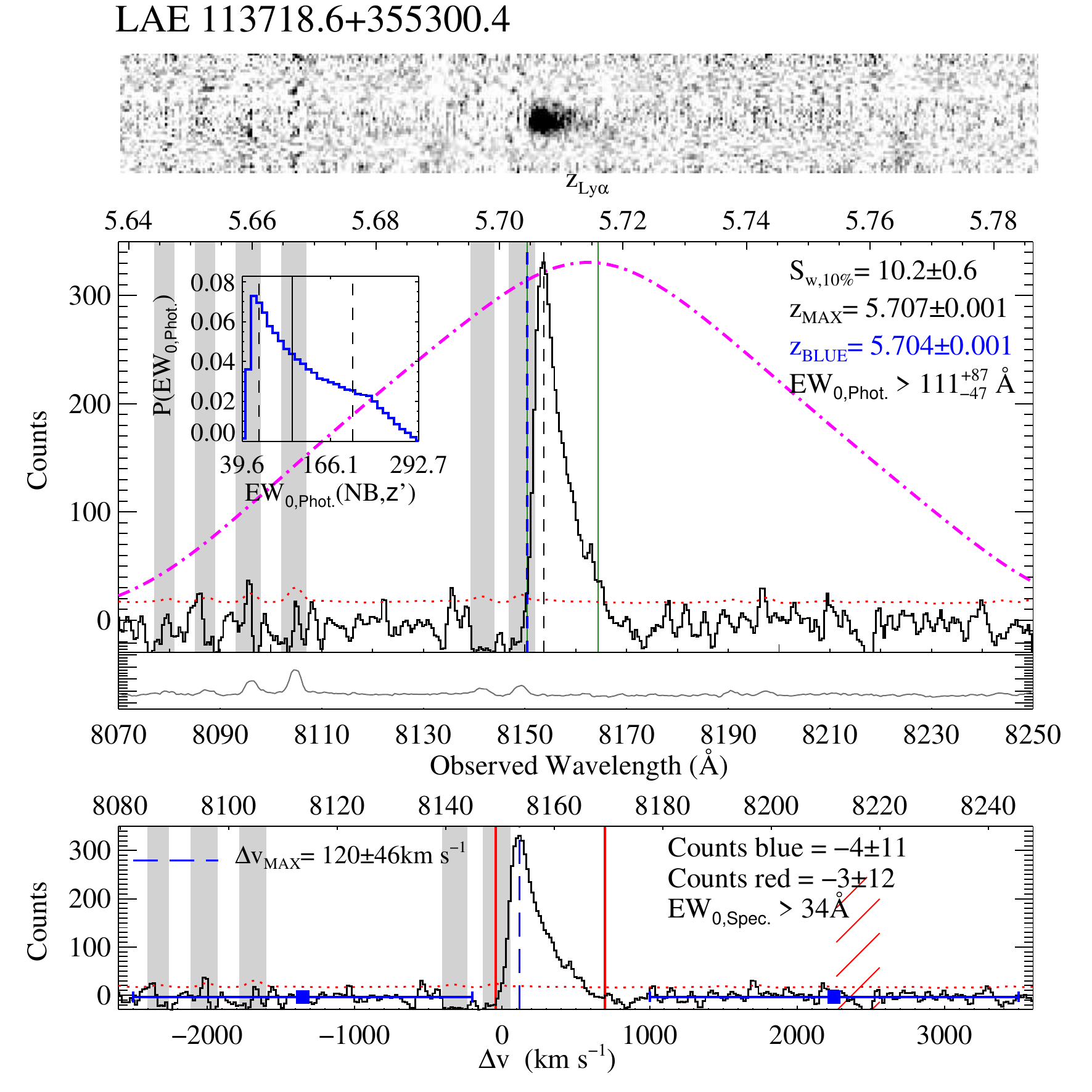} 
\caption{Field J1137+3549. ID 8}
\end{figure}

\begin{figure}
\includegraphics[width=80mm]{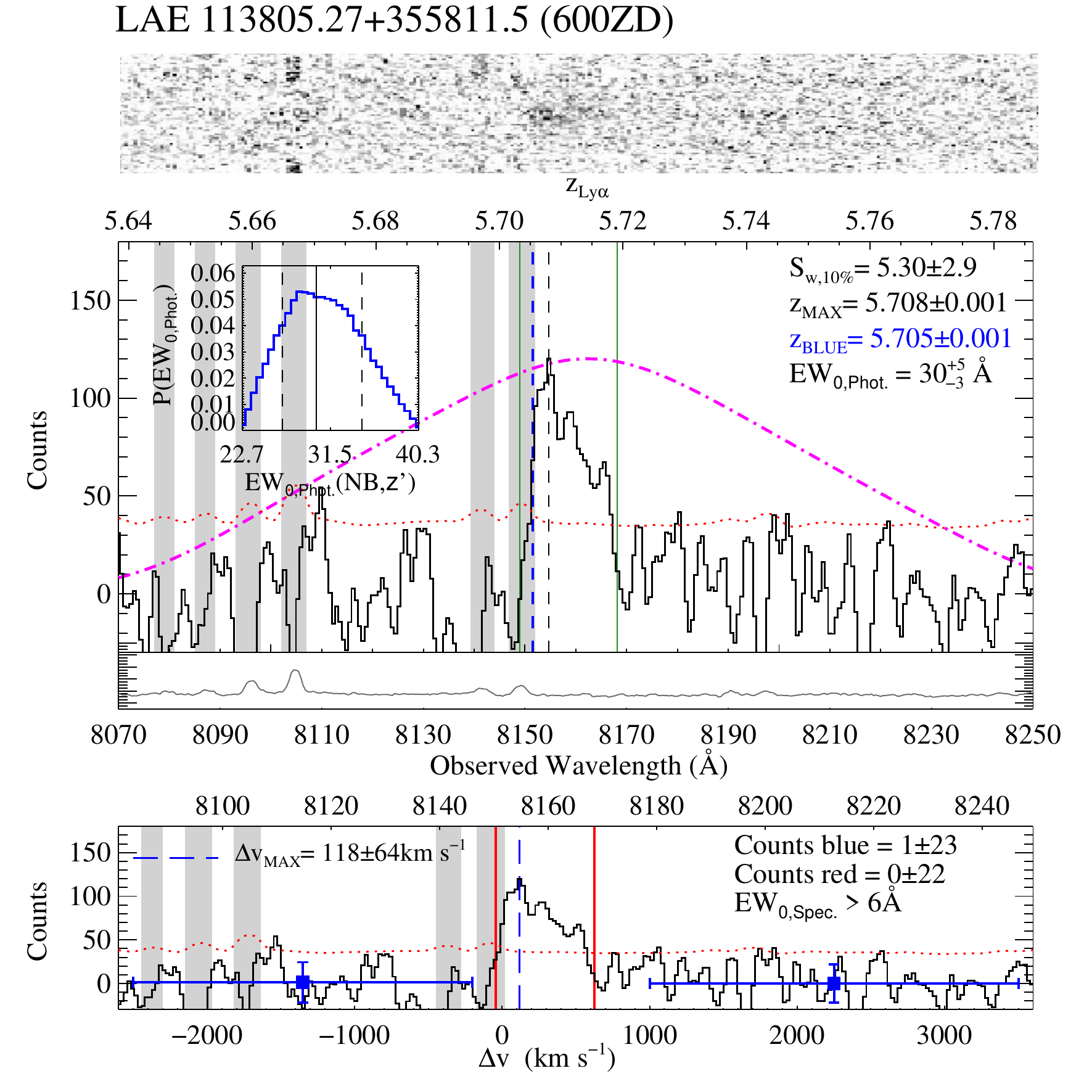} 
\caption{Field J1137+3549. ID 9}
\end{figure}
\begin{figure}
\includegraphics[width=80mm]{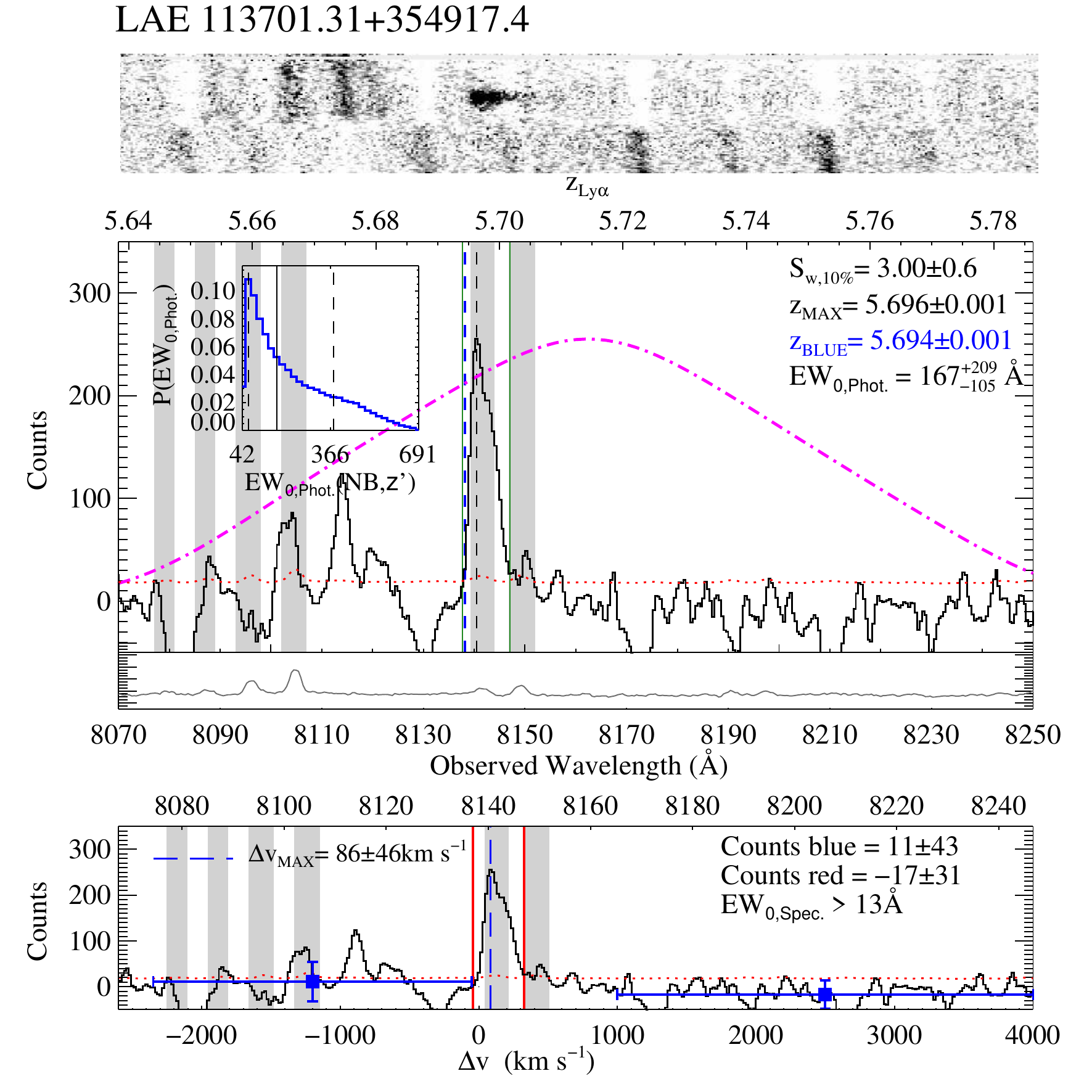} 
\caption{Field J1137+3549. ID 10}
\end{figure}
\begin{figure}
\includegraphics[width=80mm]{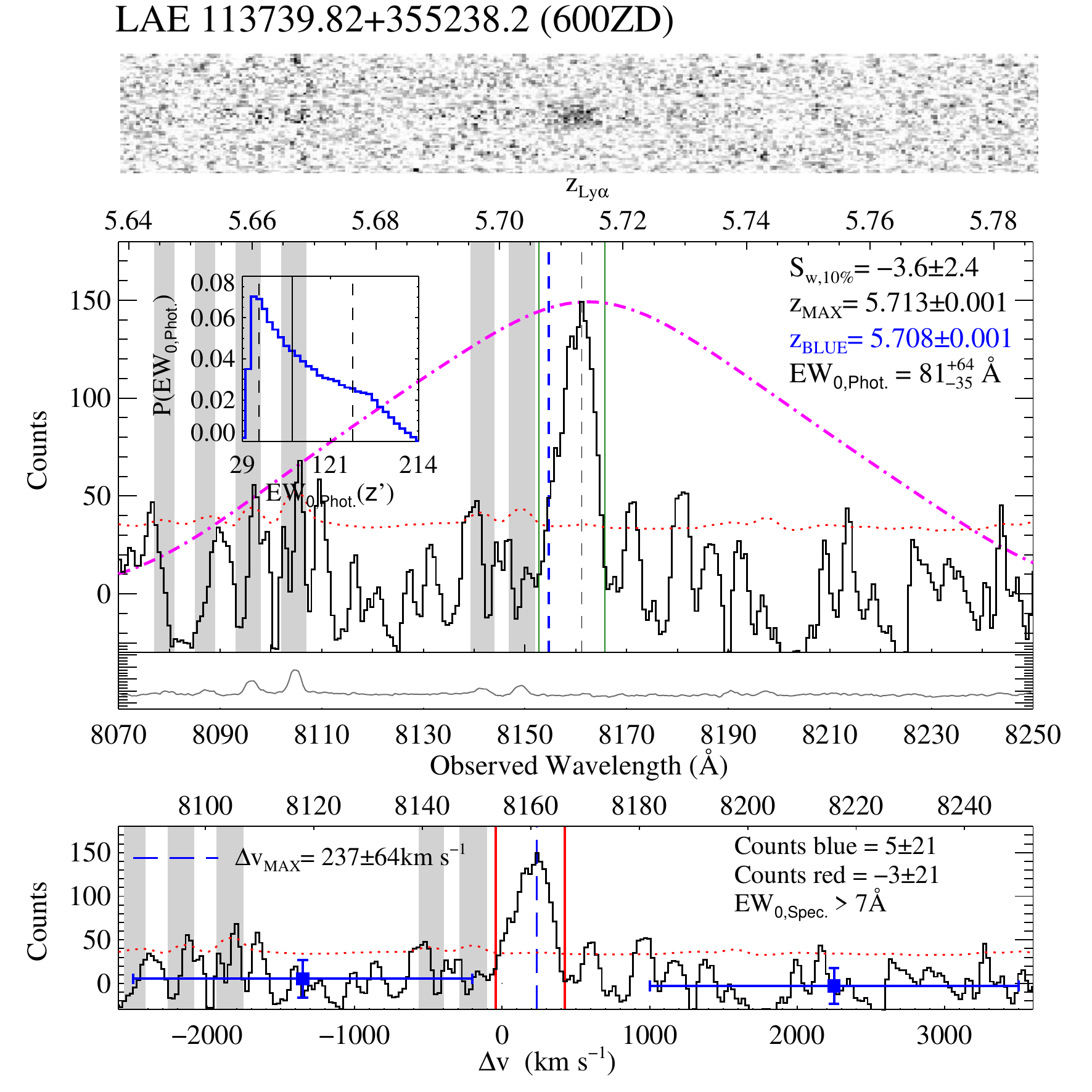} 
\caption{Field J1137+3549. ID 12}
\end{figure}


\end{document}